\documentclass{aa}
\usepackage{graphicx}

\begin{document}

\title{Mass ratios of the components in T~Tauri binary systems and
  implications for multiple star formation
\thanks{Based on observations collected at the German-Spanish Astronomical
 Center on Calar Alto, Spain, and at the European Southern Observatory,
 La Silla, Chile.}}

\author{J.\,Woitas\inst{1, 2}
 \and Ch.\,Leinert\inst{2}
 \and R.\,K\"ohler\inst{3, 2}
 }

\offprints{Jens Woitas, \email{woitas@tls-tautenburg.de}}

\institute{Th\"uringer Landessternwarte Tautenburg, Sternwarte 5,
 07778 Tautenburg, Germany 
 \and Max-Planck-Institut f\"ur Astronomie, K\"onigstuhl 17,
 69117 Heidelberg, Germany
 \and Center for Astrophysics and Space Sciences, University of California,
  San Diego, 9500 Gilman Drive, La Jolla, \newline CA 92093-0424, USA}

\date{Received / Accepted}

\titlerunning{Mass ratios of T Tauri star components}

\abstract{
Using near-infrared speckle interferometry we have obtained
resolved JHK-photometry for the components of 58 young binary systems.
From these measurements, combined with other data taken from literature,
we derive masses and particularly mass ratios of the components.\\
We use the J-magnitude as an indicator for the stellar luminosity and assign
the optical spectral type of the system to the primary. 
On the assumption that the components within a binary are coeval we can then 
place also the secondaries into the HRD and derive masses and mass ratios
for both components by comparison with different sets of current theoretical
pre-main sequence evolutionary tracks. The resulting distribution of mass
ratios is comparatively flat for $M_2/M_1\ge 0.2$, but depends on assumed
evolutionary tracks. The mass ratio is neither correlated with the primary's
mass or the components' separation. These findings are in line with the
assumption that for most multiple systems in T~associations the components'
masses are principally determined by fragmentation during formation and
not by the following accretion processes.\\
Only very few unusually red objects were newly found among the detected
companions.This finding shows that the observed overabundance of binaries
in the Taurus-Auriga association compared to nearby main sequence stars
should be real and not the outcome of observational biases related to
infrared observing.\\
}
\maketitle

\section{Introduction}
Most stars in the solar neighbourhood are members of multiple systems
(e.\,g. Duquennoy \& Mayor \cite{Duquennoy91}, Fischer \& Marcy
\cite{Fischer92}). This raises the question whether most of these systems
were formed as binaries or whether they are the result of later capture
processes.
After high angular resolution techniques in the near infrared (NIR) had been
developed at the beginning of the 1990s this problem became an issue of
observational astronomy. \\
A large number of multiplicity surveys in star forming regions (SFRs) and
clusters has now been done (see Mathieu et al.\,\cite{Mathieu00} and
references therein). It is still a matter of debate if different environmental
conditions of star formation lead to different degrees of multiplicity. One
can however conclude that there is at this time no sample of young stars that
shows a significant binary deficit compared to nearby main sequence stars.
In some SFRs even a strong binary excess is observed. 
The consequence of these results is that multiplicity must be already
established in very early phases of stellar evolution and that star
formation to a large extent has to be considered as formation of multiple
stars. Multiplicity has to be taken into account if one asks for
{\it stellar} properties. If this question is addressed to the systems instead
of the components, misleading results may be obtained.\\
In this paper we will discuss young binary systems in the nearby
SFRs Taurus-Auriga, Upper Scorpius, Chamaeleon\,I and Lupus that have
been detected by Leinert et al.\,(\cite{Leinert93}), Ghez et al.\,
(\cite{Ghez97a}) and K\"ohler et al.\,(\cite{Koehler99}). We present
resolved photometry of the components in the NIR spectral
bands J, H and K (Sect.\,\ref{data}). Based on these data we discuss the
components in a color-color diagram (Sect.\,\ref{ccd}) and in a
color-magnitude diagram (Sect.\,\ref{cmd}). Using the J-band magnitude
as indicator for the stellar luminosity and pulished spectral types for
the primaries, we place the components pairwise 
into the Hertzsprung-Russell diagram (HRD, Sect.\,\ref{hrd}) and derive masses
and in particular mass ratios
from a comparison with theoretical PMS evolutionary models
(Sect.\,\ref{masses}). Implications of the results for theoretical concepts
of multiple star formation and for binary statistics are discussed in
Sect.\,\ref{discussion}.
\section{Observational data}
\label{data}
\subsection{System properties}
\label{sysprops}
If not stated otherwise we take the systems' magnitudes, interstellar
extinction coefficients $A_{\mathrm{V}}$ and spectral types from the
literature. The references are Kenyon \& Hartmann (\cite{Kenyon95}) for
Taurus-Auriga, Walter et al.\,(\cite{Walter94}) and also K\"ohler et
al.\,(\cite{Koehler99}) for Upper Scorpius, Gauvin \& Strom\,(\cite{Gauvin92})
for Chamaeleon\,I and Hughes et al.\,(\cite{Hughes94}) for Lupus.
Possible effects of variability are discussed in Sect.\,\ref{variability}.

\subsection{Spatially resolved photometry}
\label{resolved}
Data for the objects in Taurus-Auriga were obtained during several observing
runs with the NIR camera MAGIC at the 3.5\,m-telescope on Calar Alto from
1993 to 1998. Measurements at this telescope before September 1993 were done
with a device for one-dimensional speckle-interferometry that has been
described
by Leinert \& Haas (\cite{Leinert89}).
Observations of multiple systems in southern SFRs were
carried out in May 1998 at the ESO New Technology Telescope (NTT) on La Silla
that is also a 3.5\,m-telescope, using the SHARP camera of the Max-Planck
Institute for Extraterrestrial Physics (Hofmann et al.\,\cite{Hofmann92}).\\
Since most binaries of our sample have projected separations of less than
$1\arcsec$ a high angular resolution technique is needed to overcome the
effects of atmospheric turbulence and to reach the diffraction limit which is
$\lambda/D = 0\farcs13$ for a 3.5\,m-telescope at
$\lambda = 2.2\,\mu\mathrm{m}$.
We have mostly used two-dimensional speckle interferometry.
Sequences of typically 1000 images with exposure times of
 $\approx 0.1\,\mathrm{sec}$ were taken for the object and a
nearby reference star. After background subtraction, flatfielding and badpixel
correction these data cubes are Fourier-transformed.\\
We determine the modulus of the complex visibility (i.\,e. the Fourier
transform of the object brightness distribution) from power spectrum
analysis. The phase is recursively reconstructed using two different
methods: The Knox-Thompson algorithm (Knox \& Thompson\,\cite{Knox74}) and
the bispectrum analysis (Lohmann et al.\,\cite{Lohmann83}). A detailed
description of this data reduction process has been given by K\"ohler et
al.\,(\cite{Koehler99}, Appendix\,A). Modulus and phase show
characteristic strip patterns for a binary. In the case of a triple
or quadruple star these patterns will be overlayed by similar structures
that belong to the additional companion(s).\\
Fitting a binary model to the complex visibility yields the binary
parameters position angle, projected separation and flux ratio.
The errors of these parameters are estimated by doing this fit for
different subsets of the data. 
The comparison of our position angles and projected separations with those
obtained by other authors that has been done in another paper
(Woitas et al.\,\cite{Woitas2001}) has shown that systematic differences
in relative astrometry are negligible. Differences in resolved K-band
photometry have an order of magnitude that can be explained by the
variability of T~Tauri stars.\\ 
Together with the system brightness that is taken from the literature in most
cases the flux ratio determines the components' magnitudes. We present the
results of our individual measurements in Table\,\ref{new-obs}. To reduce
the effect of the variability of T~Tauri stars we calculate the
mean of all resolved photometric observations in one filter obtained
by us and other authors (see caption of Table\,\ref{jhk-phot} for
references). 

\subsection{Potential of the data set} 

Our sample grew out of surveys for multiplicity in star forming regions.
With a total of 119 individual components
it is of reasonable size. By construction it is 
largely independent of biases due to duplicity. This makes it well suited for
statistical discussions.\\
From the resolved photometry alone we can check for circumstellar excess
emission, search for possible infrared companions, detect contamination
by background stars and have some check on whether the components of a binary
system are coeval. In the last point we encounter the limitations of
our method: the large uncertainty in color, resulting mainly from
variability, seriously degrades possible age determinations.\\
Therefore we use in the HRD known spectral types to derive masses for
the components dominating the visual region, and derived masses for the
companions on the assumption of coevality. The reliability of these
mass determination profits from our explicit knowledge of duplicity.\\
The presentation of the results starts with those resulting from 
resolved photometry alone.

\section{Color-color diagram} 
\label{ccd}

\begin{figure*}  
\resizebox{\hsize}{!}{\includegraphics{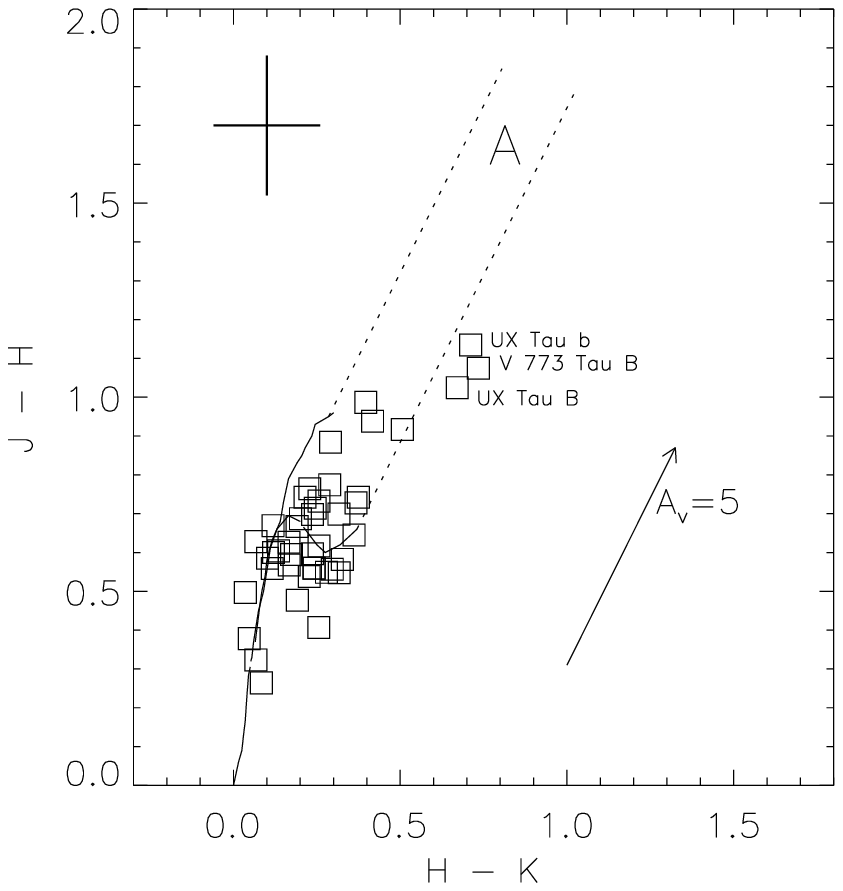}
\includegraphics{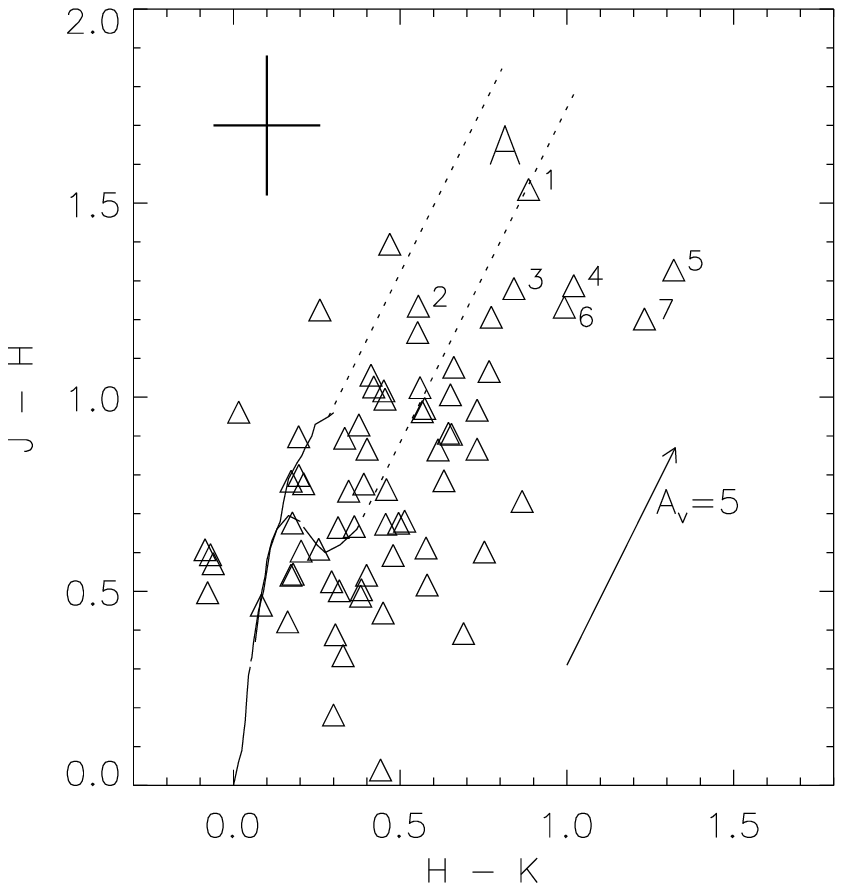}}  
\caption{\label{zweifarb} Color-color diagram for the components of
 weak-lined T~Tauri stars (WTTS, left panel) and classical T~Tauri stars
 (CTTS, right panel). The almost vertical solid line denotes the main
 sequence, with the giant branch to the right (Bessell \& Brett
 \cite{BessellBrett88}). The dotted lines are parallel to the reddening
 vector (indicated for $A_{\mathrm{V}} = 5$ in the lower right of both
 panels). The cross gives the typical error bars of our observations
 (see Sect.\,\ref{errors}). Only in the region labeled `A' and on the 
 main sequence the observed colors are consistent with photospheric emission,
 whereas for objects located right to this region circumstellar excess
 emission is present. The numbers in the right panel denote 
 \object{Haro\,6-28\,B} (1), \object{CZ\,Tau}\,B (2), the infrared companion
 of \object{XZ\,Tau} (3), the components of \object{FS\,Tau} (4, 5),
 \object{HN\,Tau}\,A (6) and the infrared companion of \object{UY\,Aur} (7).}
\end{figure*}

\subsection{Presence of circumstellar excess emission}  
\label{excess}

We correct the resolved JHK-photometry for interstellar extinction using the
reddening law of Rieke \& Lebovsky (\cite{Rieke85}) and applying the
$A_{\mathrm{V}}$ given in Table\,\ref{jhk-phot} to all components of one
system.\\
The resulting dereddened colors are not necessarily
stellar colors, because infrared excess emission caused by circumstellar
disks is a common phenomenon in T~Tauri stars\,(e.\,g. Beckwith et
al.\,\cite{Beckwith90}). For this reason we first consider a subsample of
systems that consist of weak-lined T~Tauri stars (WTTS). The adopted
classification criterion for WTTS is that their H$\alpha$ equivalent width
is less than $10\mathrm{{\AA}}$ (Herbig \& Bell\,\cite{HBC}). WTTS are not
expected to have prominent disks and their mean excess emission in J, H and
K is zero (Hartigan et al.\,\cite{Hartigan95}). Almost all components of these
systems have colors comparable to main sequence stars (Fig.\,\ref{zweifarb},
left panel) as is expected. The only exceptions are the two components
of \object{UX\,Tau}\,B and the companion of \object{V\,773\,Tau}.
For components of systems with classical T~Tauri stars (CTTS) where
significant circumstellar excess emission is expected, the positions in the
color-color diagram are much more spread around(Fig.\,\ref{zweifarb}, 
right panel). These colors cannot be referred to stellar photospheres
in a simple way. 
\subsection{Infrared companion candidates} 
\label{ircs}

It is interesting to look for companions with extreme red colors, because
these can be candidates for infrared companions (IRCs). IRCs are objects
that are very weak or have even not been detected at optical wavelengths,
but dominate the system's brightness in the infrared. They are somewhat
puzzling for star formation theory, because some of them appear to be more
massive than the optical ``primary'' but are at the same time more embedded
and less evolved.\\
At this time 8 IRCs are known (Koresko et al.\,\cite{Koresko97}, Ressler \&
Barsony \cite{Ressler01}). Two of these
objects -- \object{XZ\,Tau}\,B and \object{UY\,Aur}\,B -- do indeed show
unusual red colors in the color-color diagram (Fig.\,\ref{zweifarb},
right panel). Two other known IRCs that belong to our sample -- 
\object{T\,Tau}\,B and \object{Haro\,6-10}~B -- are not discussed here,
because we could not detect those objects in the J-band.
In Fig.\,\ref{zweifarb} we have also indicated \object{CZ\,Tau}\,B,
\object{Haro\,6-28}\,B, both components of \object{FS\,Tau} and
\object{HN\,Tau}\,A as unsually red objects. For these systems additional
spatially resolved observations at longer and shorter wavelengths will be
necessary to decide if they really contain IRCs.\\
The best candidate for a new IRC is the companion of \object{FV\,Tau\,/c}
that we have observed in H and K, but failed to detect in the J-band.
Ghez et al.\,(\cite{Ghez97a}) have proposed \object{HBC\,603}\,B and
\object{VW\,Cha}\,C to be IRCs, because these objects were found by
their K-band survey, but missed at $\lambda = 0.9\,\mu\mathrm{m}$ by
Reipurth \& Zinnecker\,(\cite{Reipurth93}). We have observed these systems
in J and H and did not detect any companion either, which calls
for additional observations at longer wavelengths.\\
Extinction by circumstellar envelopes or edge-on disks is not the
only possible explanation for extremely red colors. The objects mentioned
in this section may also have a very late spectral type and may even be
young brown dwarfs. We will discuss the topic of possible
substellar companions in Sect.\,\ref{substellar}.\\
In any case extremely red objects are not frequent among the companions
detected by Leinert et al.\,\cite{Leinert93} 
in Taurus-Auriga. This indicates
that the observed overabundance of binaries in this SFR compared
to nearby main sequence stars is real and not the result of using
infrared wavelengths for multiplicity surveys among young stars.

\section{Color-magnitude diagram} 
\label{cmd}

\subsection{Conversion of PMS models into the observational plane} 
\label{obsplane}

For comparison with the data, we convert
luminosity $L$ and $T_{\mathrm{eff}}$ for distinct masses and ages from
the theoretical atmosphere models to near-infrared colors
and magnitudes. To this purpose we use relationships
that give the bolometric correction $BC_{\mathrm{V}}$ and several colors
as a function of spectral type or the corresponding $T_{\mathrm{eff}}$. These
relations are tabulated by Bessell\,(\cite{Bessell91}), Bessell \&
Brett\,(\cite{BessellBrett88}) and Schmidt-Kaler\,(\cite{SchmidtKaler82}).
To interpolate between the datapoints given in these tables we use
polynomial fits that have been done by Meyer\,(\cite{Meyer96}).
The apparent J-band magnitude is derived from $L$ and
$T_{\mathrm{eff}}$ using the following equation:
\begin{eqnarray}
\label{j-lum}
J & = & 4.74 - 2.5\log\left(\frac{L}{L_{\odot}}\right) -
  BC_{\mathrm{V}}(T_{\mathrm{eff}}) - (V - J)_{T_{\mathrm{eff}}} \\
 \nonumber & & + (m - M) - 0.282A_{\mathrm{V}}.
\end{eqnarray}
The coefficient before $A_{\mathrm{V}}$ is taken from the interstellar
reddening law of Rieke \& Lebovsky (\cite{Rieke85}).
For the H- and K-band magnitudes there are similar relations.\\

\subsection{Errors of colors and magnitudes} 
\label{errors}

The magnitude errors that are given in Table\,\ref{jhk-phot} are the errors
of the mean magnitude that result from averaging over all spatially resolved
measurements in a given spectral band. If only one observation has been done
the error is calculated from the uncertainties of measured system
photometry and flux ratio, as given in
Table\,\ref{new-obs}. For the comparison of colors and magnitudes to theoretical
PMS models additional error sources have to be taken into account.\\

\subsubsection{Distance}
To obtain the distance modulus $m - M$ used in Eq.\,\ref{j-lum}
we adopt distances to the SFRs that are the mean of all
Hipparcos distances derived for members of the respective association.
The values and references are given in Table\,\ref{distances}.
However, distances to individual objects
may be different from these mean values. To take this into account we
assume that the radial diameters of the SFRs are as large as their
projected diameters on the sky. The latter quantity can be estimated
to be $\approx 20^{\circ}$ for Taurus-Auriga (see Fig.\,1 in K\"ohler
\& Leinert \cite{Koehler98}) as well as for Scorpius-Centaurus (see Fig.\,1
in K\"ohler et al.\,\cite{Koehler99}). Concerning the mean distances
from Table\,\ref{distances} this corresponds to a diameter of 50\,pc.
We assume here $\pm 25\,\mathrm{pc}$ as uncertainty for the
distance of an individual system which is a very conservative
estimate: more than two thirds of the stars will be within $\pm$\,15 pc
for an even spatial distribution.\\

\begin{table}
\caption{\label{distances} Adopted distances to stars in nearby SFRs}
\begin{tabular}{lll}
 & & \\ \hline
SFR & distance\,[pc] & reference \\ \hline
Taurus-Auriga & 142 $\pm$ 14 & Wichmann et al.\,(\cite{Wichmann98}) \\
Upper Scorpius & 145 $\pm$ 2 & de Zeeuw et al.\,(\cite{deZeeuw99}) \\
Chamaeleon\,I & 160 $\pm$ 17 & Wichmann et al.\,(\cite{Wichmann98}) \\
Lupus & 190 $\pm$ 27 &  Wichmann et al.\,(\cite{Wichmann98}) \\
\hline
\end{tabular}
\end{table}

\begin{figure} 
\resizebox{\hsize}{!}{\includegraphics{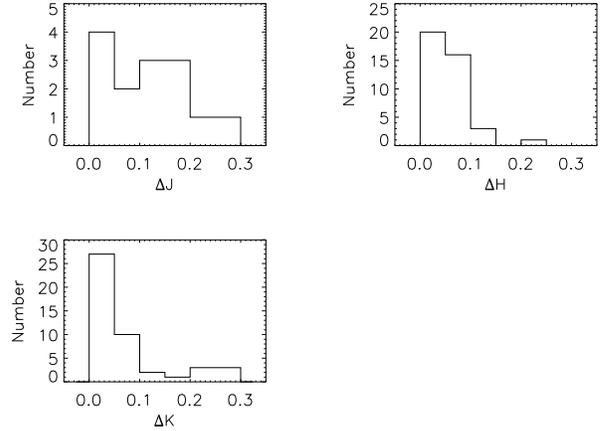}}
\caption{\label{varhist} Distribution of variability amplitudes for components
 of young binary systems in the NIR spectral bands J, H and K. They contain
 all of the systems discussed in this paper for which repeated spatially
 resolved observations in one filter exist (see Table\,\ref{new-obs} for our
 observations and caption of Table\,\ref{jhk-phot} for references to other
 published data).}
\end{figure}

\subsubsection{Variability}
\label{variability}

T~Tauri stars are variable. Although this effect is much less in the
infrared compared to optical wavelengths (Nurmanova
\cite{Nurmanova83}), it cannot be neglected. We use those components for
which there are observations in the same filter at different epochs to
estimate the influence of variability on the magnitudes. The distribution of
variability amplitudes is given in Fig.\,\ref{varhist}. The mean
amplitudes that we consider as variability errors are
$\sigma_{\mathrm{J}} = 0.12\,\mathrm{mag}$,  $\sigma_{\mathrm{H}} =
0.07\,\mathrm{mag}$ and $\sigma_{\mathrm{K}} = 0.09\,\mathrm{mag}$.
These variations, as derived from our data set,
 are similar to those in the tabulation of Rydgren~et~al. (\cite{Rydgren84})
but smaller than those found by Skrutskie~et~al. (\cite{Skrutskie96}).
Therefore our estimate of variability may be somewhat optimistic.

\subsubsection{Dereddening}
Kenyon \& Hartmann (\cite{Kenyon95}) have given for the extinction an error of 
$\sigma (A_{\mathrm{V}}) = 0.3\,\mathrm{mag}$ for the systems in
Taurus-Auriga. We adopt this error also for the systems in other SFRs. 

\begin{figure*} 
\resizebox{\hsize}{!}{\includegraphics{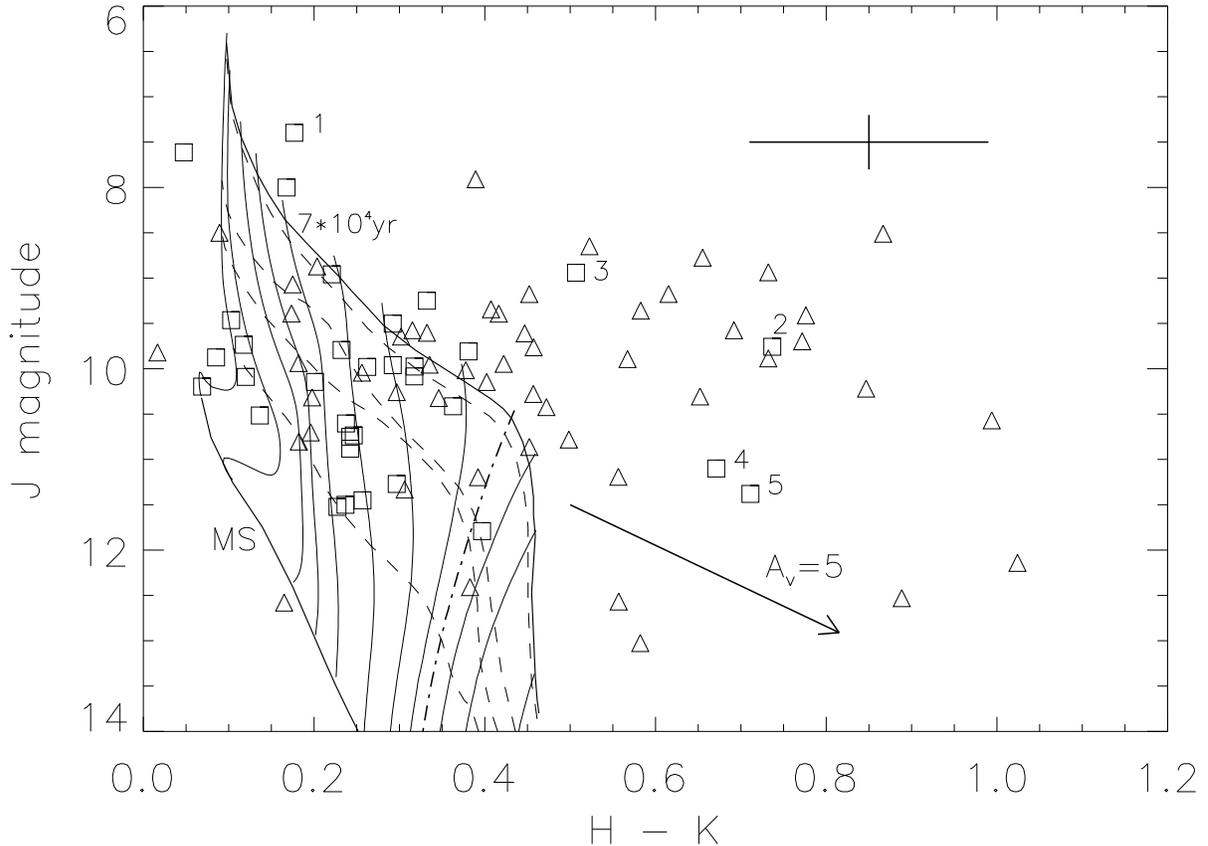}}
\caption{\label{fhd-allcomps.hk} 
 Components of young binary systems placed
 into a NIR color-magnitude diagram together with the theoretical PMS model
 by D'Antona \& Mazzitelli (\cite{dm98}). The solid lines are evolutionary
 tracks for different stellar masses (0.9 to 0.017 $M_{\sun}$), the dashed
 lines are isochrones for ages from $7\times10^4\,\mathrm{yr}$
 to $10^8\,\mathrm{yr}$. The latter age is assigned to the main sequence (MS).
 The crosses denote typical error bars for the stars' positions,
 the arrows are reddening vectors for
 $A_{\mathrm{V}} = 5\,\mathrm{mag}$. CTTS are represented by triangles and
 WTTS by squares. The WTTS indicated with numbers are \object{V\,773\,Tau}\,A
 and B (1, 2) and \object{UX\,Tau}\,A, B and b (3, 4, 5)}
\end{figure*}

\subsubsection{Theoretical Atmosphere Models}
The following two error sources do not affect colors or magnitudes,
but the transformation of theoretical PMS models to observable quantities. 
By this way they also enter the error discussion:

\begin{itemize}
\item For the bolometric correction used in Eq.\,\ref{j-lum} an error
 $\sigma(BC_{\mathrm{V}}) = 0.04\,\mathrm{mag}$ is adopted as given by
 Hartigan et al.\,(\cite{Hartigan94}).
\item As already mentioned in Sect.\,\ref{obsplane} the conversion of
 luminosity and $T_{\mathrm{eff}}$ into NIR magnitudes and colors
 uses polynomial fits to tabulated
 data. This causes an error of $0.07\,\mathrm{mag}$ in J, H and K,
 except for the models of Baraffe~et~al (\cite{Baraffe98}), which directly
 give integrated magnitudes over the near-infrared bands.
\end{itemize}

\subsubsection{Error of colors}

All mentioned errors added in quadrature give the uncertainty of a NIR
{\em magnitude} that has to be considered when placing the components into a
color-magnitude diagram. If a {\em color} is derived from these magnitudes
some errors cancel: The distance is the same for all components and
variability is supposed to be negligible if observations at both wavelenghts
are carried out in the same night. Unfortunately this is not the case
for most of our objects. To obtain a color
from $T_{\mathrm{eff}}$ one does not need a bolometric correction, so
also the respective error can be discarded. 

\begin{figure*} 
\resizebox{88mm}{!}{\includegraphics{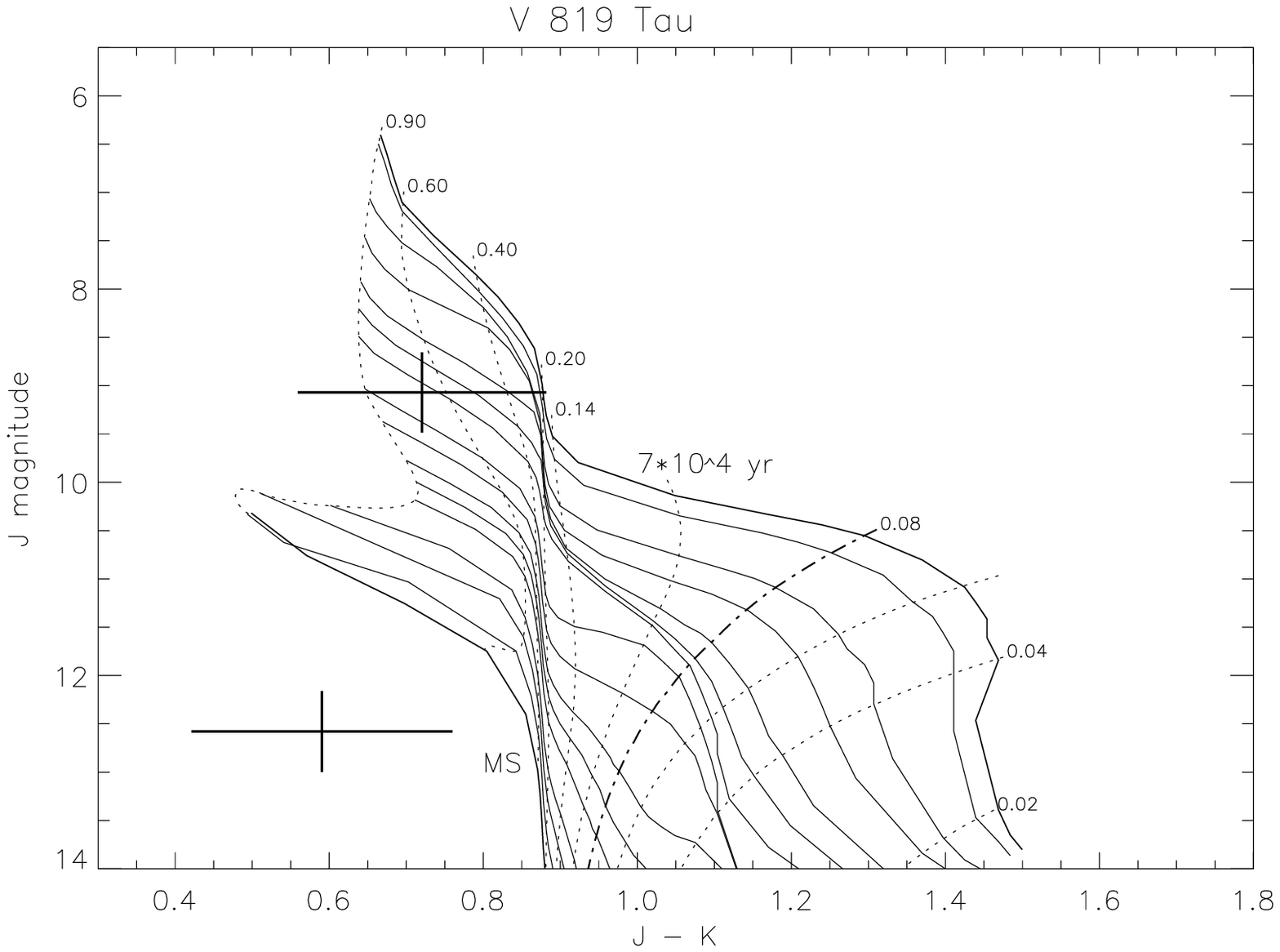}}
\resizebox{88mm}{!}{\includegraphics{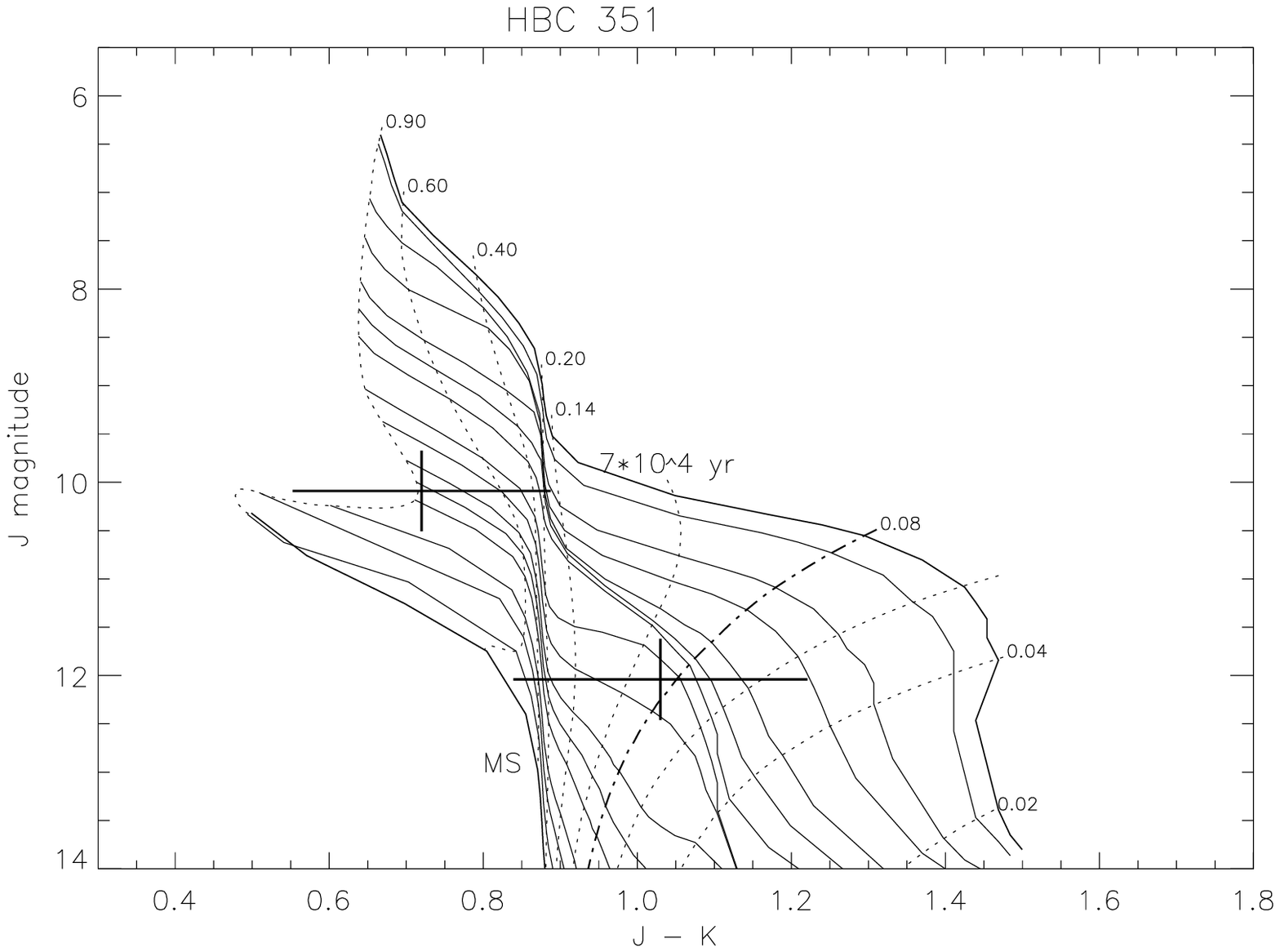}}
\caption{\label{cmd_jjk}
Examples of colour-magnitude diagrams for two WTTS systems.
Left: V819\,Tau, the putative companion appears to be a background object.
Right: HBC\,351: an example for coevality of the components.
The data are overplotted with the PMS model by
    D'Antona \& Mazzitelli\,(\cite{dm98}). The dashed lines denote
    evolutionary tracks for masses from 0.02 to $0.9\,M_{\sun}$, the
    solid lines are isochrones for ages $7\cdot 10^4$, $10^5$, $2\cdot 10^5$,
    $3\cdot 10^5$, $5\cdot 10^5$, $7\cdot 10^5$, $10^6$, $2\cdot 10^6$,
    $3\cdot 10^6$, $5\cdot 10^6$, $7\cdot 10^6$, $10^7$, $2\cdot 10^7$,
    $3\cdot 10^7$, $5\cdot 10^7$ and  $10^8\,\mathrm{yr}$ (MS). The full
figure with all 17 colour-magnitude-diagrams is available 
at CDS as Figure B.1.}
\end{figure*}

\subsection{Placing the components into color-magnitude diagrams} 
\label{allcomps}
In Fig.\,\ref{fhd-allcomps.hk} the placement of the components into a
J/(H~-~K) color-magnitude diagrams is shown. The PMS evolutionary model by
D'Antona \& Mazzitelli\,(\cite{dm98}) is also indicated in
the Figure. The distribution of objects in a J/(J~-~K) colour-magnitude
diagram would look quite similar.\\ 
Almost all components are above the lower main sequence as they are expected
to be. But many stars lie in a region that is not covered by the
evolutionary tracks. This cannot be corrected by further dereddening,
but must be due to the presence of circumstellar color excesses that we have
already mentioned above. E.g., nearly all stars that have an unusually large
H~-~K are CTTS (represented by triangles) and the only WTTS in this region
are the components of \object{V\,773\,Tau} and \object{UX\,Tau} where we
have noticed excess emission already in Sect.\,\ref{excess}.

\subsection{Relative ages of components} 
\label{coeval}

Since there is no resolved spectroscopy of the components, we cannot correct
for the color excesses mentioned in the previous section. For this reason
we restrict the discussion of the components in the color-magnitude diagram
to a subsample of 17 systems that consist of WTTS where no significant
circumstellar excess emission is expected.

The placement of the  individual components into a J/(J-K) color-magnitude
diagram is shown in Fig.\,\ref{cmd_jjk}. Object by object, the positons of
the components in these plots are compared to the PMS evolutionary model by
D'Antona \& Mazzitelli (\cite{dm98}) to check for coevality and validity of
the association. We have also used the sets of PMS
tracks and isochrones given by Swenson et al.\,(\cite{Swenson94}) and
Baraffe et al.\,(\cite{Baraffe98}) and have obtained similar results.\\
With respect to the D'Antona \& Mazzitelli (\cite{dm98}) tracks in 14 out of
17 cases the components appear to be coeval. Unfortunately, this finding
has not much weight because of the large errors in color.
Nevertheless some useful checks can be performed.
The three problematic cases are \object{V\,773\,Tau}, \object{UX\,Tau} and
\object{V\,819\,Tau}. For the first two systems we have already noticed
the presence of excess emission (Sect.\,\ref{excess} and \ref{allcomps}).  
Duch\^{e}ne et al.\,(\cite{Duchene99b}) have proposed
to reclassify \object{UX\,Tau} as a CTTS, because
its H$\alpha$ equivalent width is $9.5\,\mathrm{{\AA}}$ and thus above the
upper limit for WTTS with spectral type K that is $5\,\mathrm{{\AA}}$
according to Mart\'{\i}n\,(\cite{Martin98}). The companion of
\object{V\,819\,Tau} lies far below the main sequence in the color-magnitude
diagram. This indicates that \object{V\,819\,Tau}\,B is in fact not a binary
companion, but a chance projected background star.\\
For HBC 352/353 Fig.\,\ref{fhd-multiplots} yields an age range
of about 2$\times$10$^7$ - 2$\times$10$^8$ yr
which is problematic because this age is larger than the age spread found
for the Taurus SFR (e.\,g. Kenyon \& Hartmann\,\cite{Kenyon95}). 
Also, Mart\'{\i}n et al.\,(\cite{Martin94}) called into question that these
are young stars because they do not show significant Lithium line
absorption.\\
With the mentioned exceptions, our data are compatible with the assumption
that the components of the T Tauri binaries are coeveal.
There are other studies that better prove that this assumption is valid:
Brandner \& Zinnecker (\cite{Brandner97}) have obtained spatially resolved
spectroscopy for the components of eight young binary systems and placed them
into the HRD. In all cases the components are coeval. Particularly well this
has been shown in the case of the quadruple system \object{GG\,Tau} (White
et al. \cite{White99}) from resolved spectroscopy with the HST Faint
Object Spectrograph. Below we will use the assumption of coevality to
estimate masses for the companions and mass ratios for the components in
binary systems.

\section{Placing the components into the HRD} 
\label{hrd}

In this section we will place the components of young binary systems into
the HRD and compare their positions in this diagram with theoretical PMS
evolutionary models to derive masses and mass ratios. 
For this purpose one has to know their luminosities and
spectral types. Since these quantities are not known as directly measured
values we have to make some assumptions to estimate them from our resolved NIR
photometry and the system properties that are given in the literature.

\subsection{Luminosities} 
As described in Sect.\,\ref{obsplane} for the colour-magnitude diagrams,
we transform the theoretical atmosphere models that give $T_{\mathrm{eff}}$
and $L$ as functions of mass and age into a diagram in which the luminosity
is represented by a NIR magnitude.\\
Circumstellar excess emission is minimal around
$\lambda\approx 1\mu\mathrm{m}$
(e.\,g.\,Kenyon \& Hartmann\,\cite{Kenyon95}). For this reason we prefer
the J-band magnitudes of the components to H and K as luminosity indicator.
It is however not clear if this excess emission is negligible in the J-band.
Hartigan et al.\,(\cite{Hartigan95}) have measured the veiling in optical
spectra of T~Tauri stars. Assuming that the excess emission in the J-band
is $r_{\mathrm{J}} = 0.1\times r_{\mathrm{V}}$, where $r_{\lambda} =
F_{\lambda, \mathrm{excess}}/F_{\lambda, \ast}$, they come to the conclusion
that for a sample of 19 CTTS and 10 WTTS the mean value is consistent with
$<r_{\mathrm{J}}> = 0.0$. Folha \& Emerson (\cite{Folha99}) determined the
NIR excess emission directly using infrared spectra of 50 T~Tauri stars.
Their result is that $<r_{\mathrm{J}}> = 0.57$ for the CTTS in their sample
and thus much larger than expected. To take this into account we apply an
excess correction of $0.49\,\mathrm{mag}$ corresponding to this
$<r_{\mathrm{J}}>$, if we use J magnitudes as an indicator for CTTS stellar
luminosities. For the WTTS in their sample Folha \& Emerson (\cite{Folha99})
find values of $r_{\mathrm{J}}$ that are compatible with zero, so for the
components of WTTS systems no excess correction is necessary.

\subsection{Spectral types} 
\label{sp-types}

\begin{figure} 
 \resizebox{\hsize}{!}{\includegraphics{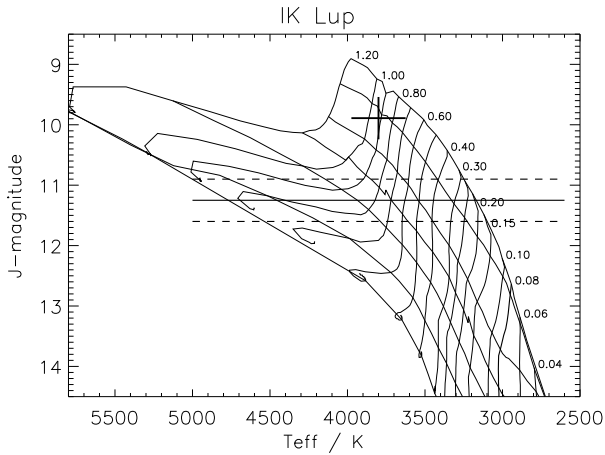}}
 \caption{\label{ik-lup} The components of the young binary system
 \object{IK\,Lup} placed into the HRD as an example for the method 
 described in Sect.\,\ref{hrd}. The cross gives the position of the
 primary, the horizontal dashed lines the locus for the secondary
 and the respective error. The theoretical model is by
 Baraffe et al.\,(\cite{Baraffe98}). The evolutionary tracks are given
 for masses of 0.04, 0.06, 0.08, 0.10, 0.15, 0.20, 0.25, 0.30, 0.40,
 0.50, 0.60, 0.70, 0.80, 0.90, 1.00 and 1.20\,$M_{\odot}$, the isochrones
 denote ages of 1, 2, 5, 10, 20, 50 and 90 Myr. If the components are coeval
 their masses will be $0.9\pm 0.2\,M_{\odot}$ and $0.3\pm 0.1 \,M_{\odot}$.}
\end{figure}

For nearly all of the systems discussed here we know the combined optical 
spectral type from the literature (see Sect.\,\ref{sysprops} and
Table\,\ref{jhk-phot}). We assume that these combined spectra represent
to a good approximation those of the optical primary components,
and we assign the optical spectral type of the system to the brightest
component in the J-band. The spectral type and effective temperature
of the companion is estimated using the assumption that
all components within a system are coeval. 
We are now ready to place the components into the HRD. The procedure is
shown in Fig.\,\ref{ik-lup} using the T~Tauri binary system
\object{IK\,Lup} as an example. For 48 more systems the placement of
the components into the HRD is shown in Fig.\,\ref{hrd-appendix},
available at CDS. The theoretical PMS evolutionary model used is by Baraffe et
al.\,(\cite{Baraffe98}). The position of the primary is determined
by its J-band magnitude and the system's spectral type. For the latter
quantity we assume an error of one spectral subclass as given by
Kenyon \& Hartmann (\cite{Kenyon95}) for the systems in Taurus-Auriga.
The companion's J-band magnitude and the respective error define a locus
for the companion in the HRD. If we assume that both components are coeval
the companion is situated at the point of intersection between this locus
and the isochrone of the primary. In the same way we also defined the loci
of the components in the HRD for the evolutionary tracks of 
Swenson et al.\,(\cite{Swenson94}) and D'Antona \& Mazzitelli
 (\cite{dm98}).

\section{Masses and mass ratios}  
\label{masses} 

The procedure described in the previous section yields the components'
masses. For instance, in the \object{IK\,Lup} system
(Fig.\,\ref{ik-lup}) the components have masses of $0.9\,M_{\odot}$ and
$0.3\,M_{\odot}$ with respect to the Baraffe et al.\,(\cite{Baraffe98})
tracks. The resulting masses derived for the components from all three sets
of PMS tracks used are given in Table\,\ref{mass-table}. For some systems
(indicated with question marks in Table\,\ref{mass-table}) the primary
is located in a region of the HRD that is not covered by the respective
tracks. The Swenson et al.\,(\cite{Swenson94}) model does only cover a
mass range above $0.15\,M_{\odot}$, so for some secondaries only
upper mass limits can be derived from that model. The errors
given in Table\,\ref{mass-table} reflect the range of tracks that is
covered by the stars' locations in the HRD. These uncertainties are 
20~-~30\% for most stars and thus quite large. However, all error sources
discussed so far are random and not systematic. Therefore in a
{\it statistical} analysis of these masses that we will do in
Sects.\,\ref{massfunk} and \ref{massrat-text}, these uncertainties
will partially cancel and have less influence to the results.\\
There are however additional uncertainties
within the PMS models theirselves. One can see from 
Table\,\ref{mass-table} that there are discrepancies in masses obtained
for the same stars from different PMS models that can be much larger
than the indicated errors which trace the uncertainty
of our measurements. The components' mass functions derived from
the three models (see Fig.\,\ref{massfunk-bild}) are different at a 99\%
confidence level which indicates that these mass differences are systematic.\\
There is now some evidence that the Baraffe et al.\,(\cite{Baraffe98})
tracks could be preferrable among the current PMS models:
White et al.\,(\cite{White99}) have placed the four components of
\object{GG\,Tau} into the HRD and compared their positions with different
sets of PMS tracks. They found that the Baraffe et al.\,(\cite{Baraffe98})
model is best consistent with the assumption that all components are
coeval. Simon et al.\,(\cite{Simon2000}) and Steffen et al.
(\cite{Steffen2000}) presented first results of empirical mass
determinations from orbital motion around T~Tauri stars that are also 
comparable with the predictions of the Baraffe et al.\,(\cite{Baraffe98})
model. It would however be premature to consider these results as
a final solution of the problem of inconsistent PMS models, mainly because
the mentioned observations do not cover the whole range of masses and
ages expected for T~Tauri stars. Therefore
in this paper we will -- as we have already done in Sect.\,\ref{coeval} --
rely on the three PMS models given by D'Antona \& Mazzitelli (\cite{dm98}),
Swenson et al.\,(\cite{Swenson94}) and Baraffe et al.\,(\cite{Baraffe98})
and compare the respective results. It can be seen that the uncertainties
inherent in the evolutionary model predictions often surpass the
uncertainties resulting from the measurement errors.\\

\begin{table*}
\caption{\label{mass-table} Masses of the components in young binary systems.
 They are derived placing the components into the HRD as described in
 Sect.\,\ref{hrd} and comparing to the PMS evolutionary tracks from
 D'Antona \& Mazzitelli (\cite{dm98}), Swenson et al.\,(\cite{Swenson94})
 and Baraffe et al.\,(\cite{Baraffe98}). Question marks indicate that for
 these systems the primary's location in the HRD could not be 
 reliably compared to the PMS tracks, because this position falls in a
 region not covered by the respective tracks.}
\begin{tabular}{llllllll}
& & & & \\ \hline
System &
 $M_1[M_{\sun}]$ & $M_2[M_{\sun}]$ &
 $M_1[M_{\sun}]$ & $M_2[M_{\sun}]$ &
 $M_1[M_{\sun}]$ & $M_2[M_{\sun}]$ & \\
 & \multicolumn{2}{c}{DM98} & \multicolumn{2}{c}{Swenson} & 
   \multicolumn{2}{c}{Baraffe} \\
\hline
HBC 351     & 0.85 $\pm$ 0.1 & 0.25 $\pm$ 0.1 & 1.0 $\pm$ 0.1 & 0.4 $\pm$ 0.1 &
 1.0 $\pm$ 0.1 & 0.4 $\pm$ 0.1 \\
HBC 352/353 & 1.0 $\pm$ 0.05 & 0.9 $\pm$ 0.05 & ? & ? & ? &  ? \\
HBC 358\,Aa & 0.35 $\pm$ 0.1 & 0.35 $\pm$ 0.1 &
  0.7 $\pm$ 0.05 & 0.7 $\pm$ 0.05 & 0.4 $\pm$ 0.1 & 0.4 $\pm$ 0.1 \\
HBC 358 B   & 0.4 $\pm$ 0.1 & & 0.7 $\pm$ 0.05 & & 0.5 $\pm$ 0.1 \\
HBC 360/361 & 0.2 $\pm$ 0.05 & 0.2 $\pm$ 0.05 & 0.5 $\pm$ 0.1 &
 0.5 $\pm$ 0.1 & 0.3 $\pm$ 0.1 & 0.3 $\pm$ 0.1 \\
FO Tau & 0.35 $\pm$ 0.05 & 0.14 $\pm$ 0.04 & 0.5 $\pm$ 0.1 & 0.35 $\pm$ 0.1 &
 0.6 $\pm$ 0.1 & 0.3 $\pm$ 0.1 \\
DD Tau & 0.5 $\pm$ 0.1 & 0.4 $\pm$ 0.1 & 0.7 $\pm$ 0.1 & 0.65 $\pm$ 0.1 &
 0.7 $\pm$ 0.1 & 0.6 $\pm$ 0.1 \\
CZ Tau & 0.4 $\pm$ 0.1  & 0.04 $\pm$ 0.04 & 0.7 $\pm$ 0.1 & $< 0.15$ &
 0.6 $\pm$ 0.1 & 0.10 $\pm$ 0.05 \\
FQ Tau & 0.4 $\pm$ 0.05 & 0.4 $\pm$ 0.05 & 0.7 $\pm$ 0.05 & 0.7 $\pm$ 0.05 &
  0.5 $\pm$ 0.1 & 0.5 $\pm$ 0.1 \\
V 819 Tau & 0.45 $\pm$ 0.05 & 0.04 $\pm$ 0.04 & 0.7 $\pm$ 0.05 & $< 0.15$ &
 1.1 $\pm$ 0.1 & 0.08 $\pm$ 0.02 & \\
LkCa 7 & 0.5 $\pm$ 0.1 & 0.2 $\pm$ 0.05 & 0.85 $\pm$ 0.15 & 0.45 $\pm$ 0.1 &
 1.1 $\pm$ 0.1 & 0.6 $\pm$ 0.1 & \\
FS Tau & 0.5 $\pm$ 0.1 & 0.12 $\pm$ 0.04 & 0.75 $\pm$ 0.05 & 0.15 $\pm$ 0.10 &
 0.7 $\pm$ 0.1 & 0.2 $\pm$ 0.05 \\
FV Tau & 0.65 $\pm$ 0.15 & 0.35 $\pm$ 0.15 & 1.15 $\pm$ 0.15 & 0.7 $\pm$ 0.15
 & 1.3 $\pm$ 0.1 & 1.0 $\pm$ 0.1 \\
UX Tau\,AC & 1.2 $\pm$ 0.2 & 0.16 $\pm$ 0.05 & 1.4 $\pm$ 0.2 & 0.25 $\pm$ 0.1&
 1.3 $\pm$ 0.1 & 0.4 $\pm$ 0.1 \\
UX Tau\,Bb & 0.25 $\pm$ 0.05 & 0.20 $\pm$ 0.05 & 0.5 $\pm$ 0.1 &
 0.4 $\pm$ 0.1 & 0.6 $\pm$ 0.1 & 0.5 $\pm$ 0.1 \\
FX Tau & 0.35 $\pm$ 0.05 & 0.35 $\pm$ 0.05 & 0.55 $\pm$ 0.1 &
  0.55 $\pm$ 0.1 & 0.8 $\pm$ 0.2 & 0.8 $\pm$ 0.2 \\
DK Tau & 0.50 $\pm$ 0.05 & 0.15 $\pm$ 0.05 & 0.8 $\pm$ 0.15 &
 0.35 $\pm$ 0.1 & 1.1 $\pm$ 0.1 & 0.4 $\pm$ 0.1 \\
LkH$\alpha$ 331 & 0.16 $\pm$ 0.03 & 0.12 $\pm$ 0.02 & 0.4 $\pm$ 0.05 &
  0.15 $\pm$ 0.05 & 0.20 $\pm$ 0.05 & 0.15 $\pm$ 0.05 \\
HK Tau & 0.5 $\pm$ 0.1 & 0.04 $\pm$ 0.02 & 0.6 $\pm$ 0.1 & $< 0.15$ &
 0.7 $\pm$ 0.1 & 0.08 $\pm$ 0.02 \\
V 710 Tau & 0.4 $\pm$ 0.1 & 0.35 $\pm$ 0.1 & 0.55 $\pm$ 0.1 & 0.45 $\pm$ 0.1 &
 0.65 $\pm$ 0.15 & 0.5 $\pm$ 0.15 \\
HK Tau G2 & 0.35 $\pm$ 0.05 & 0.35 $\pm$ 0.05 & 0.6 $\pm$ 0.1 &
 0.45 $\pm$ 0.1  & 0.9 $\pm$ 0.1 & 0.7 $\pm$ 0.1 \\
GG Tau Aa & 0.4 $\pm$ 0.05 & 0.2 $\pm$ 0.05 & 0.65 $\pm$ 0.05 & 0.4 $\pm$ 0.05
 & 0.9 $\pm$ 0.1 & 0.5 $\pm$ 0.1 \\
GG Tau Bb & 0.09 $\pm$ 0.02 & 0.04 $\pm$ 0.02 & $<0.15$ & $< 0.15$ &
 0.2 $\pm$ 0.05 & 0.04 $\pm$ 0.02 \\
UZ Tau w & 0.20 $\pm$ 0.05 & 0.18 $\pm$ 0.05 & 0.45 $\pm$ 0.1 & 0.35 $\pm$ 0.1
  & 0.35 $\pm$ 0.15 & 0.30 $\pm$ 0.15 \\
GH Tau & 0.35 $\pm$ 0.1 & 0.35 $\pm$ 0.1 & 0.55 $\pm$ 0.1 & 0.55 $\pm$ 0.1 &
 0.6 $\pm$ 0.15 & 0.5 $\pm$ 0.15 \\
Elias 12 & 0.4 $\pm$ 0.05 & 0.3 $\pm$ 0.05 & 0.7 $\pm$ 0.1 & 0.35 $\pm$ 0.05 &
 1.2 $\pm$ 0.1 & 0.9 $\pm$ 0.1 \\
IS Tau & 1.1 $\pm$ 0.1 & 0.4 $\pm$ 0.1 & 1.05 $\pm$ 0.15 & 0.7 $\pm$ 0.1 &
 0.9 $\pm$ 0.1 & 0.7 $\pm$ 0.1 \\
GK Tau / GI Tau & 0.55 $\pm$ 0.1 & 0.4 $\pm$ 0.1 & 0.95 $\pm$ 0.2 &
  0.75 $\pm$ 0.2  & 1.2 $\pm$ 0.1 & 1.0 $\pm$ 0.1 \\
HN Tau & 0.7 $\pm$ 0.1 & 0.35 $\pm$ 0.1 & 0.75 $\pm$ 0.1 & 0.45 $\pm$ 0.1 &
 0.7 $\pm$ 0.1 & 0.4 $\pm$ 0.1 \\
CoKu Tau 3 & 0.4 $\pm$ 0.1 & 0.16 $\pm$ 0.03 & 0.75 $\pm$ 0.05 &
 0.35 $\pm$ 0.1 & 0.6 $\pm$ 0.1 & 0.3 $\pm$ 0.1 \\
HBC 412 & 0.35 $\pm$ 0.1 & 0.3 $\pm$ 0.1 & 0.6 $\pm$ 0.1 & 0.5 $\pm$ 0.1 &
 0.55 $\pm$ 0.15 & 0.50 $\pm$ 0.15 \\
Haro 6-28 & 0.20 $\pm$ 0.05 & 0.04 $\pm$ 0.04  & 0.55 $\pm$ 0.1 & $<0.15$ &
 0.25 $\pm$ 0.15 & 0.06 $\pm$ 0.04 \\
VY Tau & 0.60 $\pm$ 0.1 & 0.25 $\pm$ 0.05 & 0.85 $\pm$ 0.1 & 0.35 $\pm$ 0.1 &
 0.8 $\pm$ 0.1 & 0.4 $\pm$ 0.1 \\
IW Tau & 0.7 $\pm$ 0.1 & 0.6 $\pm$ 0.1 & 0.95 $\pm$ 0.05 & 0.85 $\pm$ 0.05 &
 0.9 $\pm$ 0.1 & 0.8 $\pm$ 0.1 \\
LkH$\alpha$ 332 G1 & 0.4 $\pm$ 0.05 & 0.18 $\pm$ 0.03 & 0.55 $\pm$ 0.1 &
 0.35 $\pm$ 0.1 & 0.8 $\pm$ 0.2 & 0.4 $\pm$ 0.1 \\
LkH$\alpha$ 332 G2 & 0.45 $\pm$ 0.05 & 0.20 $\pm$ 0.05 & 0.75 $\pm$ 0.1 &
 0.3 $\pm$ 0.05 & 1.2 $\pm$ 0.1 & 0.7 $\pm$ 0.1 \\
LkH$\alpha$ 332 & 0.7 $\pm$ 0.1 & 0.6 $\pm$ 0.1 & 0.95 $\pm$ 0.05 &
  0.8 $\pm$ 0.05 & 1.0 $\pm$ 0.1 & 0.8 $\pm$ 0.1 \\
Haro 6-37 Aa & 0.7 $\pm$ 0.1 & 0.09 $\pm$ 0.05 & 1.05 $\pm$ 0.1 &
 0.2 $\pm$ 0.1 & 1.1 $\pm$ 0.1 & 0.3 $\pm$ 0.1 \\
Haro 6-37 /c & 0.35 $\pm$ 0.1 & & 0.65 $\pm$ 0.1 & & 0.7 $\pm$ 0.1 \\
RW Aur & 1.1 $\pm$ 0.1 & 0.4 $\pm$ 0.1 & 1.4 $\pm$ 0.1 & 0.7 $\pm$ 0.1 &
 1.3 $\pm$ 0.1 & 0.9 $\pm$ 0.1 \\
NTTS 155203-2338 & 1.7 $\pm$ 0.1 & 0.7 $\pm$ 0.1 & 1.85 $\pm$ 0.2 &
  0.9 $\pm$ 0.1 & ? & ? \\
NTTS 155219-2314 & 0.16 $\pm$ 0.04 & 0.09 $\pm$ 0.04 & 0.375 $\pm$ 0.075 &
  $<0.15$ & 0.2 $\pm$ 0.05 & 0.1 $\pm$ 0.05 \\
NTTS 160735-1857 & 0.20 $\pm$ 0.05 & 0.16 $\pm$ 0.04 & 0.35 $\pm$ 0.05 &
 0.3 $\pm$ 0.05 & 0.4 $\pm$ 0.1 & 0.3 $\pm$ 0.1 \\
NTTS 160946-1851 & 1.7 $\pm$ 0.1 & 0.6 $\pm$ 0.1 & 1.9 $\pm$ 0.2 &
  0.75 $\pm$ 0.15 & ?  & ? \\
WX Cha & 0.5 $\pm$ 0.1 & 0.1 $\pm$ 0.04 & 0.75 $\pm$ 0.1  & 0.15 $\pm$ 0.1 &
 1.0 $\pm$ 0.1 & 0.35 $\pm$ 0.1 \\
VW Cha AB & 0.6 $\pm$ 0.1 & 0.4 $\pm$ 0.1 & 1.05 $\pm$ 0.2 & 0.7 $\pm$ 0.2
 & ? & ? \\
HM Anon & 1.3 $\pm$ 0.1 & 0.7 $\pm$ 0.1 & 1.35 $\pm$ 0.2 & 0.85 $\pm$ 0.05 &
 1.2 $\pm$ 0.1 & 0.9 $\pm$ 0.1 \\
LkH$\alpha$ 332-17 & 1.75 $\pm$ 0.1 & 0.3 $\pm$ 0.05 & 1.9 $\pm$ 0.2 &
 0.45 $\pm$ 0.1 & ? & ? \\
IK Lup & 0.35 $\pm$ 0.1 & 0.14 $\pm$ 0.04 & 0.5 $\pm$ 0.15 & 0.2 $\pm$ 0.1 &
 0.9 $\pm$ 0.2 & 0.3 $\pm$ 0.1 \\
HT Lup & 1.3 $\pm$ 0.4 & 0.2 $\pm$ 0.15 & 1.9 $\pm$ 0.5 & 0.23 $\pm$ 0.1
  & ? & ? \\
HN Lup & 0.25 $\pm$ 0.05 & 0.25 $\pm$ 0.05 & 0.4 $\pm$ 0.1 & 0.4 $\pm$ 0.1 &
  0.7 $\pm$ 0.1 & 0.6 $\pm$ 0.1 \\
HBC 604 & 0.11 $\pm$ 0.02 & 0.05 $\pm$ 0.02 & 0.23 $\pm$ 0.05  & $<0.15$
 & ? & ? \\
HO Lup & 0.35 $\pm$ 0.05 & 0.14 $\pm$ 0.02 & 0.5 $\pm$ 0.1 & 0.2 $\pm$ 0.1 &
 0.7 $\pm$ 0.1 & 0.3 $\pm$ 0.1 \\
\hline
\end{tabular}
\end{table*}

\begin{table*}
\caption{\label{massrat-table} Mass ratios for the binary pairs from
 Table\,\ref{mass-table} derived with respect to the sets of PMS
 evolutionary tracks by D'Antona \& Mazzitelli (\cite{dm98}), Swenson et
 al.\,(\cite{Swenson94}) and Baraffe et al.\,(\cite{Baraffe98}).}
\begin{tabular}{lllll}
& & & \\ \hline
System \hspace{2.0cm}  & DM98 \hspace{1.5cm}  & Swenson \hspace{1.5cm} &
 Baraffe \hspace{1.5cm} & d[arcsec] \\ \hline
HBC\,351 & 0.29 $\pm$ 0.15 & 0.40 $\pm$ 0.14 & 0.40 $\pm$ 0.14 &
  0.61 $\pm$ 0.03 \\
HBC\,352/353 & 0.90 $\pm$ 0.10 & & & 8.6 $\pm$ 0.8 \\
HBC\,358\,Aa & 1.0 $\pm$ 0.57 & 1.0 $\pm$ 0.14 & 1.0 $\pm$ 0.5 & 
  0.15 \\
HBC\,360/361 & 1.0 $\pm$ 0.5 & 1.0 $\pm$ 0.4 & 1.0 $\pm$ 0.67
 & 7.2 $\pm$ 0.8 \\
FO\,Tau & 0.40 $\pm$ 0.17 & 0.70 $\pm$ 0.34 & 0.50 $\pm$ 0.25
 & 0.165 $\pm$ 0.005 \\
DD\,Tau & 0.80 $\pm$ 0.36 & 0.93 $\pm$ 0.28 & 0.86 $\pm$ 0.27
 & 0.57 $\pm$ 0.03 \\
CZ\,Tau & 0.10 $\pm$ 0.13 & $<0.21$ & 0.17 $\pm$ 0.11
 & 0.33 $\pm$ 0.01 \\
FQ\,Tau & 1.00 $\pm$ 0.25 & 1.0 $\pm$ 0.14 & 1.0 $\pm$ 0.4 
 & 0.79 $\pm$ 0.01 \\
V\,819\,Tau & 0.09 $\pm$ 0.10 & $<0.21$ & 0.07 $\pm$ 0.02 &
 10.5 $\pm$ 0.3 \\
LkCa\,7 & 0.40 $\pm$ 0.18 & 0.53 $\pm$ 0.21 & 0.55 $\pm$ 0.14 & 
 1.05 $\pm$ 0.01 \\
FS\,Tau & 0.24 $\pm$ 0.13 & 0.20 $\pm$ 0.15 & 0.29 $\pm$ 0.11 & 
 0.265 $\pm$ 0.005 \\
FV\,Tau & 0.54 $\pm$ 0.36 & 0.61 $\pm$ 0.21 & 0.77 $\pm$ 0.14 & 
 0.72 $\pm$ 0.10 \\
UX\,Tau\,AC & 0.13 $\pm$ 0.06 & 0.18 $\pm$ 0.10 & 0.31 $\pm$ 0.10 & 
 2.7 $\pm$ 0.1 \\
UX\,Tau\,Bb & 0.80 $\pm$ 0.36 & 0.80 $\pm$ 0.36 & 0.83 $\pm$ 0.31 & 0.138 \\
FX\,Tau & 1.00 $\pm$ 0.29 & 1.00 $\pm$ 0.36 & 1.00 $\pm$ 0.50 & 
 0.91 $\pm$ 0.01 \\
DK\,Tau & 0.30 $\pm$ 0.13 & 0.44 $\pm$ 0.21 & 0.36 $\pm$ 0.13 & 
 2.8 $\pm$ 0.3 \\
LkH$\alpha$\,331 & 0.75 $\pm$ 0.27 & 0.38 $\pm$ 0.17 & 0.75 $\pm$ 0.44 &
 0.30 $\pm$ 0.01 \\
HK\,Tau & 0.08 $\pm$ 0.06 & $<0.25$ & 0.11 $\pm$ 0.05 & 2.4 $\pm$ 0.1 \\
V\,710\,Tau & 0.88 $\pm$ 0.47 & 0.82 $\pm$ 0.33 & 0.77 $\pm$ 0.41 & 
 3.24 $\pm$ 0.10 \\
HK\,Tau\,G2 & 1.00 $\pm$ 0.29 & 0.75 $\pm$ 0.29 & 0.78 $\pm$ 0.20 & 
 0.18 $\pm$ 0.01 \\
GG\,Tau\,Aa & 0.50 $\pm$ 0.19 & 0.62 $\pm$ 0.12 & 0.56 $\pm$ 0.17 & 
 0.26 $\pm$ 0.01 \\
GG\,Tau\,Bb & 0.44 $\pm$ 0.32 & & 0.20 $\pm$ 0.15 & 1.4 $\pm$ 0.2 \\
UZ\,Tau w & 0.90 $\pm$ 0.48 & 0.78 $\pm$ 0.40 & 0.86 $\pm$ 0.80 & 
 0.34 $\pm$ 0.06 \\
GH\,Tau & 1.00 $\pm$ 0.57 & 1.00 $\pm$ 0.36 & 0.83 $\pm$ 0.46 & 
 0.35 $\pm$ 0.01 \\
Elias\,12 & 0.75 $\pm$ 0.22 & 0.50 $\pm$ 0.14 & 0.75 $\pm$ 0.15 & 
 0.41 $\pm$ 0.01 \\
IS\,Tau & 0.36 $\pm$ 0.12 & 0.67 $\pm$ 0.19 & 0.78 $\pm$ 0.20 & 
 0.21 $\pm$ 0.02 \\
GK\,Tau / GI\,Tau & 0.73 $\pm$ 0.31 & 0.79 $\pm$ 0.37 & 0.83 $\pm$ 0.15 & 
 12.2 $\pm$ 0.2 \\
HN\,Tau & 0.50 $\pm$ 0.21 & 0.60 $\pm$ 0.21 & 0.57 $\pm$ 0.22 & 
 3.1 $\pm$ 0.1 \\
CoKu\,Tau\,3 & 0.40 $\pm$ 0.18 & 0.47 $\pm$ 0.16 & 0.50 $\pm$ 0.25 & 
 2.04 $\pm$ 0.07 \\
HBC\,412 & 0.86 $\pm$ 0.53 & 0.83 $\pm$ 0.31 & 0.91 $\pm$ 0.52 & 
 0.70 $\pm$ 0.01 \\
Haro\,6-28 & 0.20 $\pm$ 0.25 & $<0.27$ & 0.24 $\pm$ 0.30 & 
 0.66 $\pm$ 0.02 \\
VY\,Tau & 0.42 $\pm$ 0.15 & 0.41 $\pm$ 0.17 & 0.50 $\pm$ 0.19 & 
 0.66 $\pm$ 0.02 \\
IW\,Tau & 0.86 $\pm$ 0.27 & 0.89 $\pm$ 0.10 & 0.89 $\pm$ 0.21 & 
 0.27 $\pm$ 0.02 \\
LkH$\alpha$\,332\,G1 & 0.45 $\pm$ 0.13 & 0.64 $\pm$ 0.30 & 0.50 $\pm$ 0.25 &
 0.23 $\pm$ 0.02 \\
LkH$\alpha$\,332\,G2 & 0.44 $\pm$ 0.16 & 0.40 $\pm$ 0.12 & 0.58 $\pm$ 0.13 &
 0.30 $\pm$ 0.01 \\
LkH$\alpha$\,332 & 0.86 $\pm$ 0.27 & 0.84 $\pm$ 0.10 & 0.80 $\pm$ 0.18 &
 0.33 $\pm$ 0.03 \\
Haro\,6-37\,Aa & 0.13 $\pm$ 0.09 & 0.19 $\pm$ 0.11 & 0.27 $\pm$ 0.12 &
 0.331 $\pm$ 0.005 \\
RW\,Aur & 0.36 $\pm$ 0.12 & 0.50 $\pm$ 0.11 & 0.69 $\pm$ 0.13 & 
 1.50 $\pm$ 0.01 \\
NTTS\,155203-2338 & 0.41 $\pm$ 0.08 & 0.49 $\pm$ 0.11 & & 0.758 $\pm$ 0.007 \\
NTTS\,155219-2314 & 0.56 $\pm$ 0.39 & $<0.40$ & 0.50 $\pm$ 0.38 &
 1.485 $\pm$ 0.003 \\
NTTS\,160735-1857 & 0.80 $\pm$ 0.40 & 0.86 $\pm$ 0.27 & 0.75 $\pm$ 0.44 & 
 0.299 $\pm$ 0.003 \\
NTTS\,160946-1851 & 0.35 $\pm$ 0.08 & 0.39 $\pm$ 0.12 & & 0.203 $\pm$ 0.006 \\
WX\,Cha & 0.20 $\pm$ 0.12 & 0.20 $\pm$ 0.16 & 0.35 $\pm$ 0.14 & 
 0.79 $\pm$ 0.04 \\
VW\,Cha\,AB & 0.67 $\pm$ 0.28 & 0.67 $\pm$ 0.32 & & 0.66 $\pm$ 0.03 \\
HM\,Anon & 0.54 $\pm$ 0.12 & 0.63 $\pm$ 0.13 & 0.75 $\pm$ 0.15 & 
 0.27 $\pm$ 0.03 \\
LkH$\alpha$\,332-17 & 0.17 $\pm$ 0.04 & 0.24 $\pm$ 0.08 & & 5.3 $\pm$ 0.2 \\
IK\,Lup & 0.40 $\pm$ 0.23 & 0.40 $\pm$ 0.32 & 0.33 $\pm$ 0.19 & 6.5 $\pm$ 0.3\\
HT\,Lup & 0.15 $\pm$ 0.16 & 0.12 $\pm$ 0.08 & & 2.8 $\pm$ 0.1 \\
HN\,Lup & 1.00 $\pm$ 0.40 & 1.00 $\pm$ 0.50 & 0.86 $\pm$ 0.27 &
  0.24 $\pm$ 0.01 \\
HBC\,604 & 0.45 $\pm$ 0.26 & $<0.65$ & & 1.99 $\pm$ 0.09 \\
HO\,Lup & 0.40 $\pm$ 0.11 & 0.40 $\pm$ 0.28 & 0.43 $\pm$ 0.20 &
 1.49 $\pm$ 0.07\\ 
\hline
\end{tabular}
\end{table*}

\subsection{Candidates for substellar companions} 
\label{substellar}
In six of our systems the mass determination from the D'Antona \& Mazzitelli
(\cite{dm98}) tracks leads to companion masses
that are below the hydrogen burning mass limit of
$M\approx 0.08\,M_{\odot}$ (see Oppenheimer et al.\,\cite{Oppenheimer00} and
references therein). This is the case for \object{CZ~Tau}~B,
\object{V~819~Tau}~B, \object{HK~Tau/c}, \object{GG~Tau}~b,
\object{Haro~6-28}~B and \object{HBC~604}~B. With respect to the
Swenson et al.\,(\cite{Swenson94}) model that does not cover the
region close above and below the hydrogen burning mass limit all six
mentioned objects have masses below $0.15\,M_{\odot}$. The Baraffe
et al.\,(\cite{Baraffe98}) tracks yield masses of $M\le 0.08\,M_{\odot}$
for \object{V\,819\,Tau}~B, \object{HK\,Tau/c}, \object{GG\,Tau}~B
and \object{Haro\,6-28}~B. The primary of \object{HBC\,604} could not
be reliably compared to the Baraffe et al.\,(\cite{Baraffe98}) tracks,
so we cannot give a Baraffe mass for the secondary.\\
We emphasize that a definitive classification of a companion as
a substellar object is not possible on the basis of our data
and requires spatially resolved spectra of the components.
It has already been mentioned (Sect.\,\ref{ircs}) that based on NIR colors
we cannot distinguish between stars with very late spectral types and deeply
embedded objects. \object{HK~Tau/c} definitely belongs to the latter
class of objects, because it has an edge-on seen disk detected by
Stapelfeldt et al.\,(\cite{Stapelfeldt98}). For two of the other mentioned
objects, namely the companions of \object{CZ\,Tau} and \object{Haro\,6-28},
we have detected unusually large NIR color excesses by placing them into a
color-color diagram (see Fig.\,\ref{zweifarb}) which makes them good
candidates for heavily extincted objects.\\
\object{V\,819\,Tau}~B may be a chance projected background star as has
been mentioned in Sect.\,\ref{coeval}. The apparent low luminosity would
in this case be the result of underestimating its distance.\\
Substellar companions to young stars probably do exist. \object{GG Tau}~b has
been placed into the HRD based on spatially resolved spectroscopy by White
et al.\,(\cite{White99}). They derived a mass of $0.044\pm 0.006\,M_{\odot}$
which is in line with our mass estimate of $0.04\,M_{\odot}$ for this
object derived from the D'Antona \& Mazzitelli (\cite{dm98}) and the
Baraffe et al.\,(\cite{Baraffe98}) models. Meyer et al.\,(\cite{Meyer97})
have estimated a mass of $\approx 0.06\,M_{\odot}$ for the companion of
\object{DI~Tau} that has been detected by Ghez et al.\,(\cite{Ghez93}).
This system is not within our object list because its projected separation of
0\farcs12 is below the diffraction limit of a 3.5\,m telescope in the K-band.
There are no strong substellar companion candidates among the
components covered by our study.

\subsection{The mass function for the components in Taurus-Auriga}
\label{massfunk} 
\begin{figure} 
 \resizebox{\hsize}{!}{\includegraphics{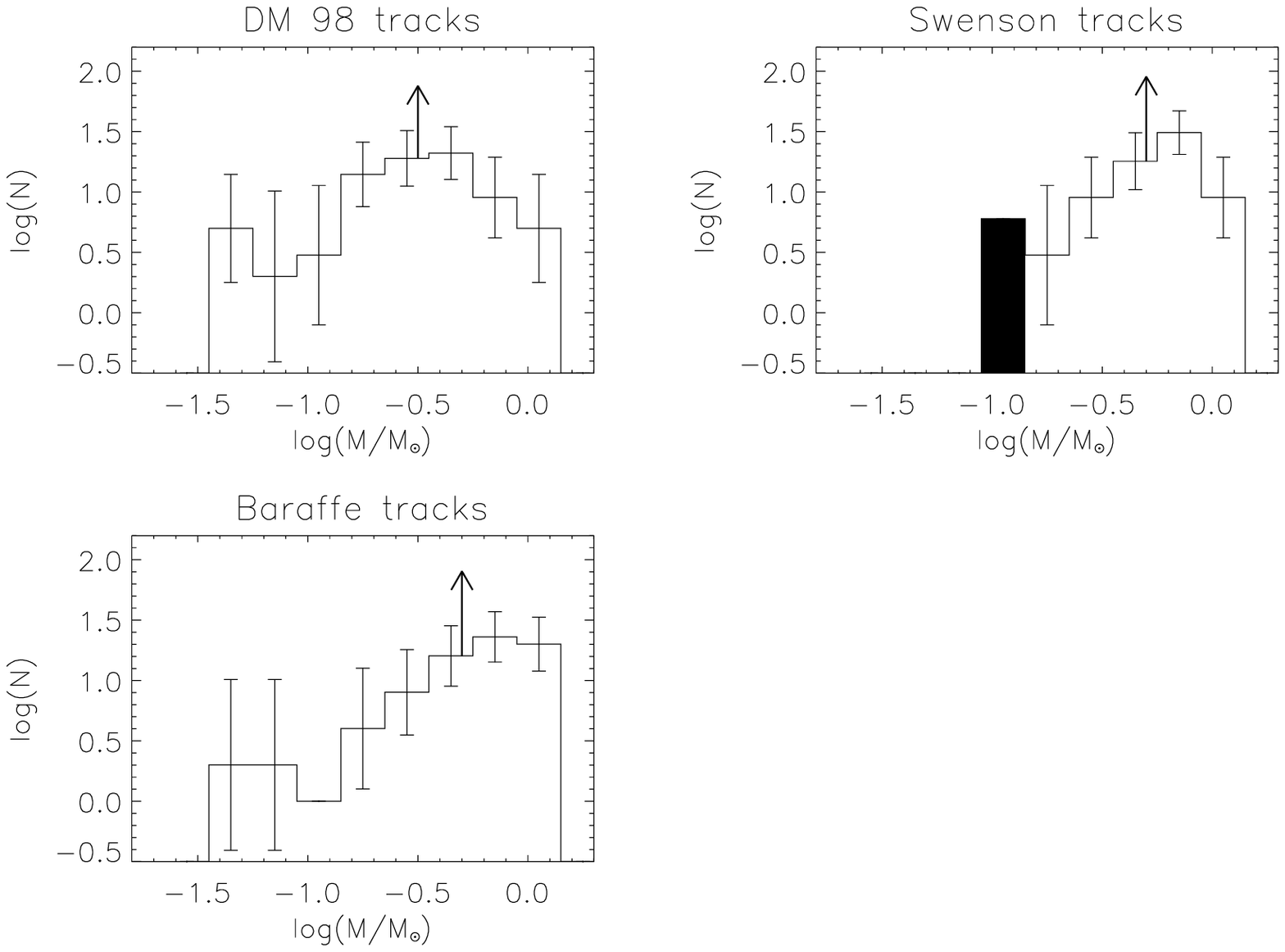}}
 \caption{\label{massfunk-bild} Mass function for the components of
 young binary systems in Taurus-Auriga from Table\,\ref{mass-table}
 derived using the PMS evolutionary models from D'Antona \& Mazzitelli
 (\cite{dm98}), Swenson et al.\,(\cite{Swenson94}) and Baraffe et al.\,
 (\cite{Baraffe98}). The bins indicated with arrows and all bins left of
 them only represent lower limits, because of the incompleteness of our
 sample (see Sect.\,\ref{massfunk}). The objects represented by the
 shaded bin the upper right panel can also have lower masses than
 indicated here, because the Swenson et al.\,(\cite{Swenson94}) model
 does only cover masses with $M\ge 0.15\,M_{\odot}$.}
\end{figure}

Among T~Tauri stars in the Taurus-Auriga association there is a significant
overabundance of binaries compared to main sequence stars in the solar
neighbourhood (see K\"ohler \& Leinert \cite{Koehler98} and references
therein). If the binary excess detected with lunar occultation observations,
speckle interferometry and direct imaging is extrapolated towards the
whole range of projected separations one comes to the conclusion that nearly
all stars in this SFR belong to multiple systems. It is therefore
interesting to derive the components' mass function for the Taurus-Auriga
association, because this should be a better representation of the mass
function in this SFR than the systems' mass function - including
unresolved binaries -  that has been given by Kenyon \& Hartmann
(\cite{Kenyon95}).\\
The mass functions for the components of young multiple systems in 
Taurus-Auriga for which we have given masses in Table\,\ref{mass-table} are
plotted in Fig.\,\ref{massfunk-bild} for the three sets of PMS tracks used.
We have now to ask to what degree these mass functions can be representative
of the whole binary population in this SFR. Our sample is taken from
Leinert et al.\,(\cite{Leinert93}). It is restricted to systems with projected
separations from 0\farcs13 to
13\arcsec and apparent magnitudes $K_{\mathrm{sys}} \le 9.5\,\mathrm{mag}$.
For the first restriction one has to assume that the components' masses
are not a function of their separation. The latter restriction means that we
 can detect all primaries with $K_{\mathrm{sys}} \le 10.25\,\mathrm{mag}$
(for a flux ratio $I_2/I_1 = 1$) while the secondaries are complete
to a magnitude of K = 12 mag. After transforming the theoretical evolutionary
model from the HRD to a diagram where the luminosity is indicated by the
K-band magnitude (see Sect.\,\ref{obsplane}) one can determine which mass
range is completely above the K = 10.25 mag brightness limit 
for ages less than $10^7\,\,\mathrm{yr}$.
This leads to completeness limits of $M\ge 0.4 M_{\odot}$
(D'Antona \& Mazzitelli \cite{dm98} tracks), $M\ge 0.5 M_{\odot}$ (Baraffe
et al.\,\cite{Baraffe98} tracks) and $M\ge 0.55 M_{\odot}$ (Swenson et al.
\cite{Swenson94} tracks). The first incomplete bins
are indicated with arrows in Fig.\,\ref{massfunk-bild}. It is thus not
possible based on our data to answer the question if and where the 
components' mass function has a maximum and how it continues into the
substellar regime. For this purpose deep imaging surveys for low luminosity
young stars and follow up high angular resolution observations will be
necessary. Concerning mass ratios, our sample is much less subject
to incompleteness.\\  

\subsection{Mass ratios} 
\label{massrat-text}

\begin{figure}  
 \resizebox{\hsize}{!}{\includegraphics{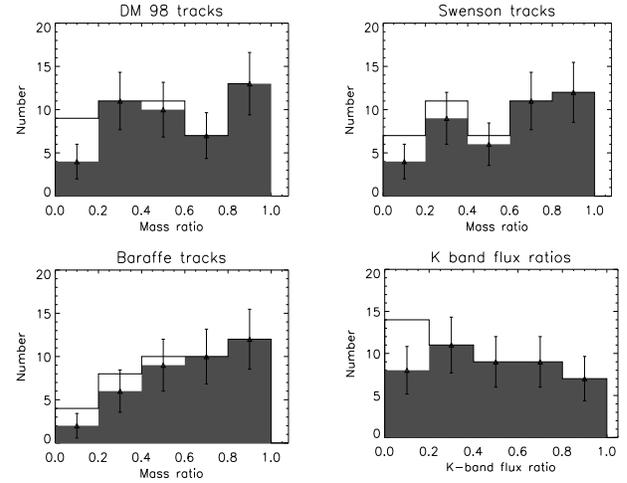}}
 \caption{\label{massrat-bild} Distribution of mass ratios for the
 young binary systems from Table\,\ref{mass-table} derived with respect
 to three sets of PMS evolutionary models. The open histograms
 represent all 51 binaries from that table, the shaded histograms show a
 restricted sample (45 binaries) for which all companions can be detected
 with the applied observational techniques (see Sect.\,\ref{massrat-text}).
 In the lower right panel also the distribution of K band flux ratios
 is given.}
\end{figure}

The distribution of mass ratios of binary components and their dependence on
other parameters are of special interest, because they can be compared
to predictions of theoretical models for multiple star formation
(see Sect.\,\ref{discussion}). To place reliable constraints on this quantity
one has to define a complete sample that is unaffected by
biases caused by the observational techniques. K\"ohler et
al.\,(\cite{Koehler99}) have come to the conclusion that using speckle
interferometry at a 3.5\,m-telescope in the K-band\footnote{This discussion
can be restriced to the K-band, because all companions examined here
have been detected in J, H and K.} will detect any companion
with a projected separation above 0\farcs13 and a magnitude difference
to the primary of $\Delta K \le 2.5\,\mathrm{mag}$. The first condition
is fulfilled for all components in Table\,\ref{mass-table}. By applying
the second restriction and obtaining a homogeneous data set we exclude six
systems from the following discussion, namely \object{V\,819\,Tau},
\object{UX\,Tau\,AC}, \object{HK\,Tau}, \object{HN\,Tau}, \object{WX\,Cha}
and \object{LkH$\alpha$\,332-17}. They have very faint companions that have
been detected using direct imaging.

\subsubsection{Mass ratio distribution} 

The mass ratios derived from three sets of PMS tracks are shown
in Table\,\ref{massrat-table}. For triple and quadruple systems we have only
given them for close pairs in hierarchical systems.
The errors indicated in Table\,\ref{massrat-table} are
formally derived from the mass errors given in Table\,\ref{mass-table}.
This is a conservative estimate, since distance errors affect the components 
of a binary in a similar way and therefore do not fully influence the
mass ratio.
To take the uncertainties that are within the PMS models into account
we compare the mass ratio distributions derived from the 
D'Antona \& Mazzitelli\,(\cite{dm98}), Swenson et al.\,(\cite{Swenson94})
and Baraffe et al\,(\cite{Baraffe98}) models (Fig.\,\ref{massrat-bild}).
The shaded histograms show the restricted sample, the open histograms
represent all pairs from Table\,\ref{mass-table}. The distributions 
derived from the D'Antona \& Mazzitelli\,(\cite{dm98}) and
Swenson et al.\,(\cite{Swenson94}) tracks are flat for $M_2/M_1 \ge 0.2$
within the uncertainties. The apparent deficit in the first bin is
probably due to incomplete detections in this regime. Actually we {\it have}
found more pairs with mass ratios below $0.2$.
The Baraffe et al.\,(\cite{Baraffe98})
model suggests a rising of the mass ratio distribution towards unity,
but this may also be caused by incompletes of the sample at low mass
ratios and is not very significant. The distribution for
$M_2/M_1 \ge 0.2$ is still compatible with being flat on a 57\%
confidence level. In discussing the mass ratio distributions given
in Fig.\,\ref{massrat-bild} one has to take into account
that the mean error of the individual mass ratios (Table\,\ref{massrat-table})
is about one binsize. This effect will cause a flattening of any
given distribution. We will conclude this discussion with the statement
that our data does not support the preference of any mass ratios. On the
other hand we admit that it would be premature to say that the mass
ratio distribution is definetely flat taking into account the large
differences between the distributions shown in Fig.\,\ref{massrat-bild}.\\
A flat distribution of mass ratios is supported by the distribution of K-band
flux ratios (Fig.\,\ref{massrat-bild}, lower right panel). For low-mass PMS
stars there is a K-band mass-luminosity relation of about $L\sim M$ (e.\,g.
Simon et al.\,\cite{Simon92}), so the distribution of this quantity should in
a good approximation resemble that of the mass ratios. There is again no
clustering towards $M_2/M_1 = 1$. If also the systems with $\Delta K > 2.5$,
i.\,e. $F_2(K)/F_1(K) < 0.1$ are included there seems to appear even
a slight overabundance of low mass~ratios.\\
\begin{figure} 
 \resizebox{\hsize}{!}{\includegraphics{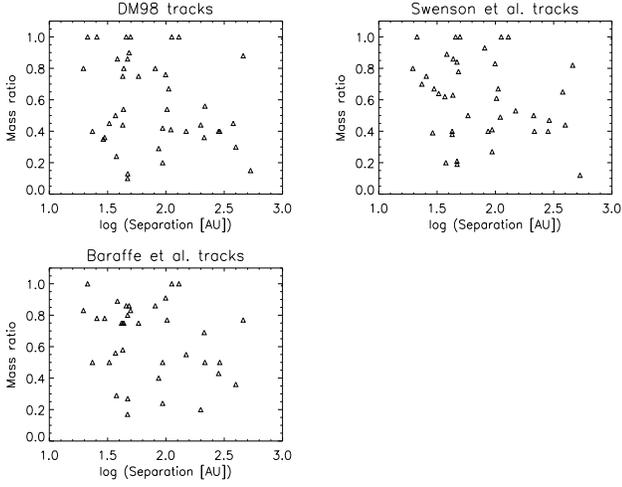}}
 \caption{\label{massep-bild} Distribution of mass ratios as a function of
  the projected separation for the three sets of PMS models used in this
  paper.}
\end{figure}
\begin{table} 
\caption{\label{massep-table} Dependence of mass ratios
 with projected separation}
\begin{tabular}{lll}
& & \\ \hline
& $d_{\mathrm{proj}} < 100\,\mathrm{AU}$ &
 $d_{\mathrm{proj}}\ge 100\,\mathrm{AU}$ \\ \hline
\multicolumn{3}{l}{D'Antona \& Mazzitelli (\cite{dm98})}\\
$M_2/M_1 > 0.5$ & 14 $\pm$ 3.7 & 9 $\pm$ 3.0\\
$M_2/M_1 \le 0.5$ & 12 $\pm$ 3.5 & 10 $\pm$ 3.2 \\ 
\multicolumn{3}{l}{Swenson et al. (\cite{Swenson94})}\\
$M_2/M_1 > 0.5$ & 16 $\pm$ 4.0 & 9 $\pm$ 3.0\\
$M_2/M_1 \le 0.5$ & 10 $\pm$ 3.2 & 8 $\pm$ 2.8 \\ 
\multicolumn{3}{l}{Baraffe et al. (\cite{Baraffe98})}\\
$M_2/M_1 > 0.5$ & 17 $\pm$ 4.1 & 8 $\pm$ 2.8\\
$M_2/M_1 \le 0.5$ & 8 $\pm$ 2.8 & 6 $\pm$ 2.5\\ 
\hline
\end{tabular}
\end{table}

\subsubsection{Correlations between mass ratios and other binary parameters}

First, we want to examine a possible 
correlation of mass ratios and the components'
separations. For this purpose it is useful to convert the measured
projected angular separations to physical separations. This allows us
to discuss binaries in SFRs with different distances
(see Table\,\ref{distances}) simultaneously.
Moreover this yields a comparison between mass ratios and characteristic
length scales like the radii of circumstellar disks.
Applying this conversion one has to consider that values for the projected
distance obtained with ``snap-shot'' observations can be very different
from the semimajor axis $A$. This cannot be corrected for individual systems. 
Leinert et al.\,(\cite{Leinert93}) have however calculated that the
relationship between the projected separation $d_{\mathrm{proj}}$ and the
semimajor axis $A$ is {\it on average}
\begin{equation}
d_{\mathrm{proj}} = 0.95\,A.
\end{equation}
So for a large sample the distribution of $d_{\mathrm{proj}}$ will
resemble the distribution of semimajor axes in a good approximation.
In Fig.\,\ref{massep-bild} the mass ratios derived from three sets
of PMS evolutionary tracks are plotted as a function of
$d_{\mathrm{proj}}$. To present the results more clearly we have divided
these scatter plots into four fields (Table\,\ref{massep-table}) and have
particularly discriminated between separations
that are larger or lower than a typical circumstellar disk radius that
is $\approx 100\,\mathrm{AU}$ (e.\,g. Hartmann\,\cite{Hartmann98}).\\
With respect to the D'Antona \& Mazzitelli\,(\cite{dm98}) and the
Swenson et al.\,(\cite{Swenson94}) models in both regions the numbers of
systems with low and high mass ratios are comparable within the uncertainties.
The result obtained from the Baraffe et al.\,(\cite{Baraffe98}) model
suggests a slight preference of larger mass ratios at lower separations,
but this is significant only at the 1$\sigma$ level.

\begin{figure} 
 \resizebox{\hsize}{!}{\includegraphics{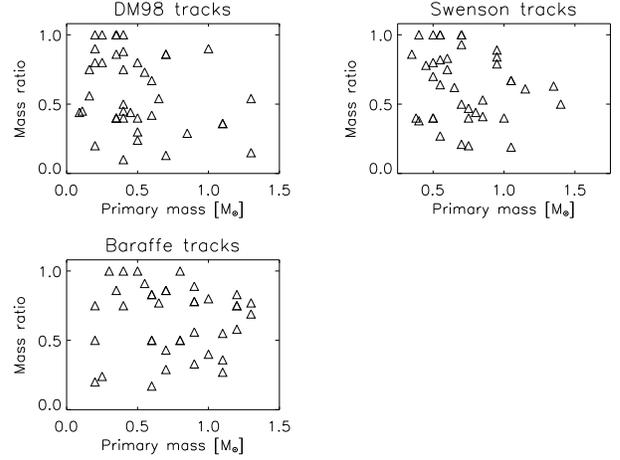}}
 \caption{\label{primmass-bild} Distribution of mass ratios as function
 of the primary mass}
\end{figure}
\begin{table} 
\caption{\label{primmass-table} Dependence of mass ratio
 from primary mass}
\begin{tabular}{lll}
 & & \\ \hline
\multicolumn{3}{l}{D'Antona \& Mazzitelli (\cite{dm98})} \\
& $M_1 < 0.5\,M_{\odot}$ & $M_1 \ge 0.5\,M_{\odot}$ \\ 
$M_2/M_1 > 0.5$   & 15 $\pm$ 3.9 & 8 $\pm$ 2.8 \\  
$M_2/M_1 \le 0.5$ & 11 $\pm$ 3.3 & 11 $\pm$ 3.3 \\ 
\multicolumn{3}{l}{Swenson et al.\,(\cite{Swenson94})} \\
& $M_1 < 0.7\,M_{\odot}$ & $M_1 \ge 0.7\,M_{\odot}$ \\ 
$M_2/M_1 > 0.5$   & 13 $\pm$ 3.6 & 11 $\pm$ 3.3 \\  
$M_2/M_1 \le 0.5$ & 5 $\pm$ 2.2 & 13 $\pm$ 3.6 \\ 
\multicolumn{3}{l}{Baraffe et al.\,(\cite{Baraffe98})} \\
& $M_1 < 0.7\,M_{\odot}$ & $M_1 \ge 0.7\,M_{\odot}$ \\ 
$M_2/M_1 > 0.5$   & 10 $\pm$ 3.2 & 15 $\pm$ 3.9 \\  
$M_2/M_1 \le 0.5$ & 6 $\pm$ 2.5 & 8 $\pm$ 2.8 \\ 
\hline
\end{tabular}
\end{table}
Finally we ask for a correlation between the mass ratio and the primary
mass. The result is shown in Fig.\,\ref{primmass-bild} and
Table\,\ref{primmass-table}. The threshold between ``low'' and ``high''
primary masses is arbitrarily chosen to have a comparable number of objects
in both groups for each theoretical PMS model considered. With regard to
the Baraffe et al.\,(\cite{Baraffe98}) model the behaviour of both groups
is the same. The D'Antona \& Mazzitelli (\cite{dm98}) and Swenson et 
al.\,(\cite{Swenson94}) models suggest a preference of higher mass
ratios for lower primary masses. The latter result has also been mentioned
by Leinert et al.\,(\cite{Leinert93}) on the basis of K band magnitudes
and flux ratios.\\
We conclude that correlations between mass ratio and other binary
parameters are weak if they exist at all.

\section{Discussion} 
\label{discussion}
\subsection{Theoretical models for multiple star formation} 
\label{theory}
It is now widely believed that fragmentation during protostellar collapse
is the major process for forming multiple stars in low-density SFRs
as discussed here (e.\,g.\, Clarke et al.\,\cite{Clarke2000}).
Our results generally are in line with this assumption:

\begin{itemize}
\item We have some additional support from our data for coeval formation of
 the components (Sect.\,\ref{coeval}). Capture processes of independently
 formed single stars (that would produce a widespread distribution of
 relative ages) should not play a dominant role in the formation of the
 binaries discussed here.
\item There are only a few companions with masses $M~\le~0.1~\,~M_{\odot}$
 which is a typical mass of a T Tauri stars' disk (Beckwith et al.\,
 \cite{Beckwith90}). Furthermore these low mass companions do not
 preferentially occur at separations $d\le 100\,\mathrm{AU}$ that
 are comparable to typical disk radii. This is difficult to reconcile with
 the idea that a large number of companions is formed from disk
 instabilities.
\item Fragmentation during protostellar collapse does not lead to a 
 preference of distinct mass ratios or to a dependence of mass ratio from
 other binary parameters like the components' separation
 (e.\,g. Ghez et al.\,\cite{Ghez2000}). We have indeed not seen any such
 correlations (Sect.\,\ref{massrat-text}).
\end{itemize}

Regarding the last item one has however to consider that there is
a large time span between the end of numerical simulations of fragmentation
during protostellar collapse and the state of dynamically stable T~Tauri
multiple systems as observed by us. Particularly, at the end of the
simulations presented e.\,g. by Burkert et al.\,(\cite{Burkert96})
only $\approx 10\%$ of the parent cloud's mass has condensed into fragments.
Therefore it is a highly important question how subsequent accretion
processes influence the properties of young multiple systems. Since it is not
possible today to cover the entire evolution of a molecular cloud core into a
binary system with {\it one} simulation, theory has to take a different
approach.\\
For this purpose Bate (\cite{Bate97}, \cite{Bate98}) and Bate \& Bonnell
(\cite{BateBonnell97}) have simulated the behaviour of a ``binary'' formed
out of two point masses that are situated in a cavity within a surrounding
gas sphere. The mass ratio of the binary and the angular momentum of the
infalling material are variable initial conditions. The result of these
simulations is that in the course of the accretion process the system's mass
ratio increases and approaches unity if the total cloud mass is accreted.
Mass ratios close to unity should be more probable in close systems than for
wide pairs.\\
This is not in agreement with our result (Sect.\,\ref{massrat-text}) that
there is no significant preference of mass ratios close to unity and that the
mass ratio is probably independent of the components' separation. So it seems
that the initial conditions used by Bate \& Bonnell are somehow unrealistic
and that the ``final'' stellar masses are more dependent on the fragment
masses than on the following accretion processes. One explanation for this
could be that in the course of protobinary evolution most of the initial mass
is condensed into fragments, before accretion becomes important. Another idea
is that there is some process that halts accretion before a large amount of
the remaining cloud mass is accreted onto the fragments. In any case our
knowledge about this issue is preliminary and additional theoretical and
observational effort is necessary to decide what physical processes determine
stellar masses.
 
\subsection{Implications for binary statistics} 
\label{binstat}
Multiplicity surveys by Leinert et al.\,(\cite{Leinert93}) and Ghez et
al.\,(\cite{Ghez93}) led to the surprising result that there is a significant
overabundance of binaries in the Taurus-Auriga SFR compared to main sequence
stars in the solar neighbourhood. This result was further proved by the
follow-up studies done by Simon et al.\,(\cite{Simon95}) and K\"ohler \&
Leinert (\cite{Koehler98}). Although this paper is not directly concerned
with binary statistics, we can draw some conclusions that further support the
idea that this binary excess is real and not a result of observational
biases.\\
If one compares the binary frequency among young and evolved stars one
has to take into account that due to evolutionary effects companions can
be relatively bright in their PMS phase, but invisible on the main sequence
stage. This is particularly the case for substellar companions.
One has further to consider that the multiplicity surveys were done at
infrared wavelengths in SFRs, but in the optical range for main sequence
stars. So there might be a bias that supports the detection of very red
``infrared companions'' (IRCs, see Sect.\,\ref{ircs}) in the vicinity of
PMS stars. Another problem is that the surveys in Taurus-Auriga and
the solar neighbourhood could be not directly comparable if they were
sensitive to a different range of mass ratios and thus stellar masses.\\
If the presence of substellar companions or IRCs in the vicinity of T~Tauri
stars in Taurus-Auriga were a common phenomenon this could at least partially
explain the observed binary excess in this SFR. Our results presented in
Sect.\,\ref{ircs} and\,\ref{substellar} show that this is not the case:
K\"ohler \& Leinert (\cite{Koehler98}) have found that after applying
a statistical correction for chance projected background stars there
are $48.9\pm 5.3$ companions per 100 primaries (including single stars)
in Taurus-Auriga. We have denoted only 5 out of 40 companions (for which we
have given masses in Table\,\ref{mass-table}) as candidates
for substellar objects based on masses derived from the D'Antona \&
Mazzitelli (\cite{dm98}) PMS evolutionary tracks (Sect.\,\ref{substellar}).
With respect to the Baraffe et al.\,(\cite{Baraffe98}) model this number
is even lower. If we take the mentioned 5 out of 40 companions as an estimate
for the real number of brown dwarf companions in Taurus-Auriga
and subtract this from the companion frequency given by K\"ohler \&
Leinert (\cite{Koehler98}) this value diminishes to $42.8\pm 6.0$. 
This is still far above the value of $25.3\pm 3.9$ that was given by
Duquennoy \& Mayor (\cite{Duquennoy91}) for G-dwarfs in the solar
neighbourhood.
Furthermore we have found that only 3 out of 51 companions in Taurus-Auriga
are detectable in the H-band and at longer wavelengths, but were missed
at $1.25\,\mu\mathrm{m}$, so IRCs are probably not a frequent phenomenon.
The binary frequency does not have to be corrected for IRCs, because their
successors in the main sequence phase will be ``normal'' stellar companions
(see Koresko et al.\,\cite{Koresko97} for estimates of IRCs' masses).\\
Duquennoy \& Mayor (\cite{Duquennoy91}) have claimed that their sample
is complete for mass ratios $M_2/M_1\ge 0.1$. It has already been mentioned
by K\"ohler \& Leinert (\cite{Koehler98}, Sect.\,5.2) that the
completeness limit of the binary surveys in Taurus-Auriga is in any case
not lower, the actual value dependent on the mass-luminosity relation used.
We can further prove this result, because there are only 2 out of 51 systems
with mass ratios less than 0.1 with respect to the D'Antona \& Mazzitelli
(\cite{dm98}) PMS model and 1 out of 50 considering the Baraffe et 
al.\,(\cite{Baraffe98}) tracks (Table\,\ref{massrat-table}). \\  
We conclude that the observed binary excess in Taurus-Auriga compared
to nearby main sequence stars is neither the result of a higher sensitivity
in mass ratio nor a consequence of a large frequency of substellar or
infrared companions. The strange overabundance of binaries in
Taurus-Auriga remains a fact also after this more detailed analysis of the
systems found by Leinert et al.\,(\cite{Leinert93}).

\section{Summary} 
\label{summary}
From speckle interferometry and direct imaging we have derived resolved
JHK photometry for the individual components of T Tauri binary systems
in nearby star forming regions. These measurements are combined with other
data taken from literature (resolved JHK photometry from other authors,
system magnitudes and spectral types, extinction coefficients) to study
properties of the components in young binary systems. The main results are:

\begin{itemize}
\item We have found only very few unusually red objects that may be young
 substellar objects or infrared companions. Their number is too small
 to have significantly influenced the binary statistics in Taurus-Auriga.
\item The placement of the components into NIR color magnitude diagrams
 is affected by large errors and thus allows no precise determination of
 stellar ages from PMS evolutionary models. We can however detect problematic
 cases and find that \object{V~819~Tau}~B is probably an unrelated background
 object. The locations of the components of the 16 other
 WTTS systems into the the CMDs are in line with the assumption that all 
 components within a system are coeval.
\item The determination of {\it masses} from the HRD has been performed
 using the following procedure: We derive stellar luminosities from
 the J-band magnitudes, assign the optical system spectral type to the
 primary and use the assumption that all components within one system are
 coeval.
\item The use of three different sets of PMS tracks then yields mass functions
 that are different at the 99\% confidence level. For this reason we discuss
 the results of all three models used separately. These differences tend to
 be larger than the uncertainties resulting from observational errors.
 In addition, the latter mass errors are random and thus partially cancel
 in a statistical discussion.
\item Within the uncertainties the distribution of mass ratios is flat
 for $M_2/M_1\ge 0.2$. There are no significant correlations between
 mass ratio and projected separation or mass ratio and primary mass.
 These results are in line with the wideley accepted idea that binaries
 are formed by fragmentation during protostellar collapse processes.
 Moreover, they suggest that the final masses of the components are
 largely determined by fragmentation itself and not by subsequent accretion.
\end{itemize}

\begin{acknowledgements}
We are grateful to Andreas Eckart and Klaus Bickert for their support
in observing with the SHARP camera. We also thank the staff at ESO La Silla
and Calar Alto for their support during several observing runs. The authors
appreciate fruitful discussions with Michael Meyer, Monika Petr, Matthew
Bate and Coryn Bailer-Jones, and they thank an anonymous referee for
pointed and productive comments. This research has made use of the SIMBAD
database, operated at CDS, France.
\end{acknowledgements}

%

\appendix

\section{Near-infrared photometry for the components of young 
	binary systems}                                     

The spatially resolved observations of T Tauri binary systems
obtained by the authors are listed in Table~A1.
Column 1 gives the most common name of the object, columns 2 and 3
the wavelength band and the date of observation, column 4 the
brightness ratio (secondary/primary, usually but not always $<$ 1),
columns 5 through 7 the magnitude of system, primary and
secondary components in the respective wavelength band. 

Table~A2 contains the adopted near-infrared magnitudes of the
components of the young binary systems which compose our sample.
The values given here are the mean of all spatially resolved
 photometric observations, i.\,e. our data from Table \ref{new-obs} and 
 measurements published by Chelli et al.\,(\cite{Chelli95}),
 Duch\^{e}ne\,(\cite{Duchene99a}), Ghez et al.\,(\cite{Ghez93},
 \cite{Ghez97a}, \cite{Ghez97b}), Haas et al.\,(\cite{Haas90}),
 Hartigan et al.\,(\cite{Hartigan94}), K\"ohler et al.\,(\cite{Koehler99}),
 Moneti \& Zinnecker\,(\cite{Moneti91}), Richichi et al.\,(\cite{Richichi99}),
 Roddier et al.\,(\cite{Roddier96}) and  Simon et al.\,(\cite{Simon92}).
The values for visual extinction $A_{\mathrm{V}}$ and the 
spectral type given with the primaries,
 belong to the systems (see Sect.\,\ref{sysprops} for references).
 The numbers in the last column denote the star forming region:
Taurus-Auriga (1), Upper Scorpius (2),
 Chamaeleon\,I (3) and Lupus (4).


\clearpage





\setcounter{table}{0}
\begin{table*}   
\caption{\label{new-obs} Spatially resolved observations of young
 T~Tauri binary systems obtained by the authors.}
\begin{tabular}{llllllll}
& & & & & & & \\ \hline
System & Filter & Date & $I_2/I_1$ & $m_{\mathrm{Sys}}$ &
 $m_{\mathrm{Prim}}$ & $m_{\mathrm{Sec}}$ \\ \hline
HBC 351     & J & 30\,Sep\,1993 & 0.165 $\pm$ 0.004 & 9.92 $\pm$ 0.05 & 
            10.09 $\pm$ 0.05 & 12.04 $\pm$ 0.07 \\ 
            & H & 8.\,Jan\,1993 & 0.190 $\pm$ 0.015 & 9.30 $\pm$ 0.02 &
            9.49 $\pm$ 0.03 & 11.29 $\pm$ 0.09 \\
            & K & 16.\,Feb\,1992 & 0.22 $\pm$ 0.02 & 9.15 &
            9.37 $\pm$ 0.02 & 11.01 $\pm$ 0.08 \\
HBC\,358\,B & J & 5.\,Oct\,1993 & & 11.33 $\pm$ 0.06 & & \\
            & H & 8.\,Jan\,1993 & & 10.75 $\pm$ 0.06 & & \\
            & K & 16.\,Feb\,1992 & & 10.44 $\pm$ 0.09 & & \\
LkCa\,3     & J & 5.\,Oct\,1993 & 0.62 $\pm$ 0.02 & 8.47 & 
            8.99 $\pm$ 0.01 & 9.51 $\pm$ 0.02 \\
            & H & 27.\,Sep\,1991 & 0.60 $\pm$ 0.03 &  7.74 $\pm$ 0.01 &
            8.25 $\pm$ 0.03 & 8.80 $\pm$ 0.04 \\
            &   & 29.\,Sep\,1996 & 0.885 $\pm$ 0.011 & &
            8.43 $\pm$ 0.02 & 8.56 $\pm$ 0.02 \\
            & K & 6.\,Dec\,1990 & 0.5 & 7.52 $\pm$ 0.05 & 7.96 $\pm$ 0.05 &
            8.71 $\pm$ 0.05 \\
            &   & 19.\,Nov\,1997 & 0.755 $\pm$ 0.007 & &
            8.13 $\pm$ 0.05 & 8.44 $\pm$ 0.06 \\ 
V\,773\,Tau & J & 5.\,Oct\,1993 & 0.113 $\pm$ 0.005 & 7.65 $\pm$ 0.05 &
            7.77 $\pm$ 0.05 & 10.13 $\pm$ 0.09 \\
            & H & 8.\,Jan\,1993 & 0.15 $\pm$ 0.02 & 6.85 $\pm$ 0.09  &
            7.00 $\pm$ 0.11 & 9.06 $\pm$ 0.22 \\
FO Tau      & J & 7.\,Oct\,1993 & 0.30 $\pm$ 0.01 & 9.70 $\pm$ 0.07 &
             9.99 $\pm$ 0.08 & 11.28 $\pm$ 0.09 \\
            &   & 29.\,Nov\,1996 & 0.55 $\pm$ 0.05 & &
            10.18 $\pm$ 0.11 & 10.83 $\pm$ 0.14 \\
            &   & 16.\,Nov\,1997 & 0.63 $\pm$ 0.02 & & 
            10.23 $\pm$ 0.08 & 10.73 $\pm$ 0.09 \\
            & H & 7.\,Jan\,1993 & 0.93 $\pm$ 0.03 & 8.71 $\pm$ 0.05 &
            9.42 $\pm$ 0.07 & 9.50 $\pm$ 0.07 \\
            &   & 16.\,Nov\,1997 & 0.698 $\pm$ 0.014 & &
            9.28 $\pm$ 0.06 & 9.68 $\pm$ 0.06 \\
            & K & 19.\,Sep\,1991 & 0.92 $\pm$ 0.04 & 8.18 $\pm$ 0.03 &
            8.89 $\pm$ 0.05 & 8.98 $\pm$ 0.05 \\
            &   & 14.\,Dec\,1994 & 0.72 $\pm$ 0.02 & &
            8.76 $\pm$ 0.04 & 9.13 $\pm$ 0.05 \\
            &   & 9.\,Oct\,1995 & 0.648 $\pm$ 0.018 & &
            8.72 $\pm$ 0.04 & 9.19 $\pm$ 0.05 \\
            &   & 27.\,Sep\,1996 & 0.628 $\pm$ 0.019 & &
            8.71 $\pm$ 0.04 & 9.21 $\pm$ 0.05 \\     
DD Tau      & J & 5.\,Dec\,1990 & 0.79 $\pm$ 0.01 & 9.47 $\pm$ 0.04 &
            10.10 $\pm$ 0.05 & 10.36 $\pm$ 0.05 \\
            & H & 5.\,Dec\,1990 & 0.79 $\pm$ 0.01 & 8.52 $\pm$ 0.07 &
            9.15 $\pm$ 0.08 & 9.41 $\pm$ 0.08 \\
            & K  & 5.\,Dec\,1990 & 0.64 $\pm$ 0.01 &  7.87 $\pm$ 0.07 &
            8.41 $\pm$ 0.08 & 8.89 $\pm$ 0.08 \\
CZ Tau      & J & 29.\,Nov\,1996 & 0.120 $\pm$ 0.005 & 10.51 $\pm$ 0.12 &
            10.63 $\pm$ 0.12 & 12.94 $\pm$ 0.16 \\
            & H & 9.\,Jan\,1993 & 0.23 $\pm$ 0.01 & 9.74 $\pm$ 0.03 &
            9.96 $\pm$ 0.04 & 11.56 $\pm$ 0.07 \\
            & K & 19.\,Mar\,1992 & 0.46 $\pm$ 0.03 & 9.28 $\pm$ 0.03 &
            9.69 $\pm$ 0.05 & 10.53 $\pm$ 0.08 \\
            &   & 28.\,Sep\,1996 & 0.183 $\pm$ 0.004 &  &
            9.46 $\pm$ 0.03 & 11.30 $\pm$ 0.04 \\
FQ Tau      & J & 16.\,Nov\,1997 & 1.06 $\pm$ 0.01 & 10.61 $\pm$ 0.02 &
            11.39 $\pm$ 0.03 & 11.33 $\pm$ 0.02 \\
            & H & 27.\,Sep\,1991 & 1.23 & 9.90 $\pm$ 0.07 &
            10.77 $\pm$ 0.07 & 10.55 $\pm$ 0.07 \\
            &   & 16.\,Nov\,1997 & 1.109 $\pm$ 0.016 &  &
            10.71 $\pm$ 0.08 & 10.60 $\pm$ 0.08 \\
            & K & 22.\,Sep\,1991 & 0.90 $\pm$ 0.01 & 9.47 $\pm$ 0.31 &
            10.17 $\pm$ 0.32 & 10.28 $\pm$ 0.32 \\
V\,819\,Tau & J & 2.\,Oct\,1993 & & & 9.45 $\pm$ 0.03 & 12.96 $\pm$ 0.06 \\
            & H & 2.\,Oct\,1993 & & & 8.76 $\pm$ 0.08 & 12.39 $\pm$ 0.08 \\
LkCa\,7     & J & 5.\,Oct\,1993 & 0.414 $\pm$ 0.008 & 9.25 &
            9.63 $\pm$ 0.01 & 10.58 $\pm$ 0.01 \\
            & H & 27.\,Sep\,1991 & 0.44 $\pm$ 0.02 & 8.58 &
            8.98 $\pm$ 0.02 & 9.87 $\pm$ 0.03 \\
            & K & 19.\,Sep\,1991 & 0.56 $\pm$ 0.02 & 8.36 $\pm$ 0.03 &
            8.84 $\pm$ 0.04 & 9.47 $\pm$ 0.05 \\
FS\,Tau     & J & 29.\,Nov\,1996 & 0.188 $\pm$ 0.007 & 10.66 $\pm$ 0.13 &
            10.85 $\pm$ 0.14 & 12.66 $\pm$ 0.16 \\
            & H & 28.\,Sep\,1996 & 0.183 $\pm$ 0.008 & 9.14 $\pm$ 0.09 &
            9.32 $\pm$ 0.10 & 11.17 $\pm$ 0.13 \\
            & K & 19.\,Nov\,1997 & 0.138 $\pm$ 0.005 & 7.74 $\pm$ 0.26 &
            7.88 $\pm$ 0.26 & 10.03 $\pm$ 0.29 \\
FW\,Tau     & H & 27.\,Sep\,1996 & 0.76 $\pm$ 0.10 & 9.78 &
            10.39 $\pm$ 0.06 & 10.69 $\pm$ 0.08 \\
            & K & 17.\,Oct\,1989 & 1.00 $\pm$ 0.01 & 9.37 &
            10.12 $\pm$ 0.01 & 10.12 $\pm$ 0.01 & \\
            &   & 13.\,Dec\,1994 & 0.61 $\pm$ 0.10 &      &
            9.89 $\pm$ 0.07 & 10.42 $\pm$ 0.11 & \\
            &   & 9.\,Oct\,1995 & 1.00 $\pm$ 0.05 &       &
            10.12 $\pm$ 0.03 & 10.12 $\pm$ 0.03 & \\
FV\,Tau     & J & 1.\,Sep\,1990 & 0.71 $\pm$ 0.05 & 9.51 $\pm$ 0.10 &
            10.09 $\pm$ 0.13 & 10.46 $\pm$ 0.14 \\
            &   & 30.\,Nov\,1996 & 0.38 $\pm$ 0.04 & &
            9.86 $\pm$ 0.13 & 10.91 $\pm$ 0.18 \\
            & H & 1.\,Sep\,1990 & 0.68 $\pm$ 0.02 & 8.22 $\pm$ 0.13 &
            8.78 $\pm$ 0.14 & 9.20 $\pm$ 0.15 \\
            & K & 1.\,Sep\,1990 & 0.83 $\pm$ 0.01 & 7.37 $\pm$ 0.10 &
            8.03 $\pm$ 0.11 & 8.23 $\pm$ 0.11 \\
            &   & 9.\,Oct\,1995 & 0.695 $\pm$ 0.007 & &
            7.94 $\pm$ 0.10 & 8.34 $\pm$ 0.11 \\
FV\,Tau\,/c & H & 9.\,Jan\,1991 & 0.03 & 9.42 $\pm$ 0.02 &
            9.45 $\pm$ 0.02 & 13.26 $\pm$ 0.03 \\
            & K & 9.\,Oct\,1995 & 0.076 $\pm$ 0.004 & 8.80 $\pm$ 0.02 &
            8.88 $\pm$ 0.02 & 11.68 $\pm$ 0.03 \\
UX\,Tau\,AC & J & 16.\,Nov\,1997 & & &
            8.97 $\pm$ 0.09 & 11.85 $\pm$ 0.09 \\
FX\,Tau     & J & 20.\,Mar\,1991 & 0.93 $\pm$ 0.03 & 9.16 $\pm$ 0.06 &
            9.87 $\pm$ 0.08 & 9.95 $\pm$ 0.08 \\
            &   & 30.\,Nov\,1996 & 0.934 $\pm$ 0.004 & &
            9.88 $\pm$ 0.06 & 9.95 $\pm$ 0.06 \\
            & H & 20.\,Mar\,1991 & 0.78 $\pm$ 0.01 & 8.75 $\pm$ 0.14 &
            9.37 $\pm$ 0.15 & 9.65 $\pm$ 0.15 \\
            &   & 16.\,11.\,1997 & 0.775 $\pm$ 0.012 &  &
            9.37 $\pm$ 0.15 & 9.65 $\pm$ 0.15 \\
\hline
\end{tabular}
\end{table*}
\clearpage
\setcounter{table}{0}
\begin{table*}
\caption{continued}
\begin{tabular}{llllllll}
& & & & & & & \\ \hline
System & Filter & Date & $I_2/I_1$ & $m_{\mathrm{Sys}}$ &
 $m_{\mathrm{Prim}}$ & $m_{\mathrm{Sec}}$ \\ \hline
FX\,Tau     & K & 4.\,Dec\,1990 & 0.54 $\pm$ 0.05 & 8.14 $\pm$ 0.14 &
            8.61 $\pm$ 0.18 & 9.28 $\pm$ 0.21 \\ 
            &   & 20.\,Mar\,1991 & 0.56 $\pm$ 0.01 & &
            8.63 $\pm$ 0.15 & 9.25 $\pm$ 0.15 \\
            &   & 18.\,Oct\,1991 & 0.509 $\pm$ 0.001 & &
            8.59 $\pm$ 0.14 & 9.32 $\pm$ 0.14 \\
DK\,Tau     & J & 16.\,Nov\,1997 & & &
            9.15 $\pm$ 0.09 & 10.52 $\pm$ 0.10 \\
Lk\,H$\alpha$\,331 & J & 26.\,Jan\,1994 & 0.706 $\pm$ 0.023 & 9.85 &
            10.43 $\pm$ 0.01 & 10.81 $\pm$ 0.02 \\
            & H & 6.\,Jan\,1993 & 0.91 $\pm$ 0.02 & 8.99 &
            9.69 $\pm$ 0.01 & 9.79 $\pm$ 0.01 \\
            &   & 29.\,Sep\,1996 & 0.70 $\pm$ 0.01 & &
            9.57 $\pm$ 0.01 & 9.95 $\pm$ 0.01 \\
            & K & 29,\,Oct\,1991 & 0.73 $\pm$ 0.04 & 8.68 &
            9.28 $\pm$ 0.03 & 9.62 $\pm$ 0.03 \\
            &   & 9.\,Oct\,1995 & 0.66 $\pm$ 0.03 &  &
            9.23 $\pm$ 0.02 & 9.68 $\pm$ 0.03 \\
XZ\,Tau     & J & 27.\,Jan\,1994 & 1.51 $\pm$ 0.03 & 9.91 $\pm$ 0.32 &
            10.91 $\pm$ 0.33 & 10.46 $\pm$ 0.33 \\
            &   & 30.\,Nov\,1996 & 3.54 $\pm$ 0.31 &  &
            11.55 $\pm$ 0.39 & 10.18 $\pm$ 0.34 \\
            & K & 28.\,Jan\,1994 & 0.41 $\pm$ 0.01 & 8.05 $\pm$ 0.56 &
            8.42 $\pm$ 0.57 & 9.39 $\pm$ 0.58 \\
            &   & 22.\,Nov\,1997 & 0.316 $\pm$ 0.007 & &
            8.35 $\pm$ 0.57 & 9.60 $\pm$ 0.58 \\
HK\,Tau\,G2 & J & 26.\,Jan\,1994 & 0.764 $\pm$ 0.023 & 9.41 $\pm$ 0.09 &
            10.03 $\pm$ 0.10 & 10.32 $\pm$ 0.11 \\
            & H & 6.\,Jan\,1993 & 0.69 $\pm$ 0.07 & 8.44 $\pm$ 0.06 &
            9.01 $\pm$ 0.10 & 9.41 $\pm$ 0.13 \\
            &   & 27.\,Jan\,1994 & 0.85 $\pm$ 0.12 & &
            9.11 $\pm$ 0.13 & 9.28 $\pm$ 0.14 \\
            &   & 27.\,Sep\,1996 & 0.76 $\pm$ 0.05 & &
            9.05 $\pm$ 0.09 & 9.35 $\pm$ 0.10 \\
            & K & 28.\,Sep\,1991 & 0.88 $\pm$ 0.03 & 8.05 $\pm$ 0.02 &
            8.74 $\pm$ 0.04 & 8.87 $\pm$ 0.04 \\
            &   & 9.\,Oct\,1995 & 0.85 $\pm$ 0.06 & &
            8.72 $\pm$ 0.03 & 8.89 $\pm$ 0.04 \\
            &   & 19.\,Nov\,1997 & 0.587 $\pm$ 0.017 & &
            8.55 $\pm$ 0.03 & 9.13 $\pm$ 0.04 \\
GG\,Tau\,Aa & J & 27.\,Jan\,1994 & 0.543 $\pm$ 0.004 & 9.01 $\pm$ 0.07 &
            9.48 $\pm$ 0.01 & 10.14 $\pm$ 0.01 \\
            & H & 2.\,Nov\,1991 & 0.549 $\pm$ 0.009 & 7.83 $\pm$ 0.05 &
            8.31 $\pm$ 0.06 & 8.96 $\pm$ 0.06 \\
            &   & 24.\,Sep\,1994 & 0.417 $\pm$ 0.017 & &
            8.31 $\pm$ 0.05 & 8.95 $\pm$ 0.06 \\
            & K & 2.\,Nov\,1990 & 0.64 $\pm$ 0.01 & 7.30 $\pm$ 0.03 &
            7.84 $\pm$ 0.04 & 8.32 $\pm$ 0.04 \\
            &   & 21.\,Oct\,1991 & 0.32 $\pm$ 0.05 & &
            7.77 $\pm$ 0.04 & 8.44 $\pm$ 0.06 \\
            &   & 16.\,Nov\,1997 & 0.564 $\pm$ 0.004 & &
            7.79 $\pm$ 0.03 & 8.41 $\pm$ 0.03 \\
            &   & 10.\,Oct\,1998 & 0.476 $\pm$ 0.005 & &
            7.72 $\pm$ 0.03 & 8.53 $\pm$ 0.04 \\
GG\,Tau\,Bb & J & 27.\,Jan\,1994 & & &
            11.44 $\pm$ 0.06 & 13.12 $\pm$ 0.09 \\
            & H & 10.\,Jan\,1993 & & &
            10.26 $\pm$ 0.06 & 12.58 $\pm$ 0.06 \\
            & K  & 2.\,Nov\,1990  & & &
            9.99 $\pm$ 0.10 & 11.79 $\pm$ 0.10 \\
UZ\,Tau\,w  & J & 26.\,Jan\,1994 & 0.76 $\pm$ 0.07 & 9.64 &
            10.25 $\pm$ 0.04 & 10.55 $\pm$ 0.06 \\
            & H & 9.\,Jan\,1993 & 0.69 $\pm$ 0.02 & 8.67 &
            9.24 $\pm$ 0.01 & 9.64 $\pm$ 0.02 \\
            &   & 29.\,Sep\,1996 & 0.66 $\pm$ 0.02 & &
            9.22 $\pm$ 0.01 & 9.67 $\pm$ 0.02 \\
GH\,Tau     & J & 26.\,Jan\,1994 & 0.89 $\pm$ 0.01 & 9.22 $\pm$ 0.08 &
            9.91 $\pm$ 0.09 & 10.04 $\pm$ 0.09 \\
            & H & 6.\,Jan\,1993 & 1.30 $\pm$ 0.10 & 8.34 $\pm$ 0.07 &
            9.24 $\pm$ 0.12 & 8.96 $\pm$ 0.11 \\
            &   & 29.\,Sep\,1996 & 1.03 $\pm$ 0.10 & &
            9.11 $\pm$ 0.12 & 9.08 $\pm$ 0.12 \\
            &   & 27.\,Oct\,1991 & 0.91 $\pm$ 0.04 & &
            8.48 $\pm$ 0.15 & 8.58 $\pm$ 0.15 \\
Elias\,12   & J & 27.\,Jan\,1994 & 0.579 $\pm$ 0.022 & 8.22 $\pm$ 0.04 &
            8.72 $\pm$ 0.06 & 9.31 $\pm$ 0.07 \\
            & H & 5.\,Jan\,1993 & 0.54 $\pm$ 0.01 & 7.41 $\pm$ 0.02 &
            7.88 $\pm$ 0.03 & 8.55 $\pm$ 0.03 \\
            &   & 29.\,Sep\,1996 & 0.58 $\pm$ 0.04 & &
            7.91 $\pm$ 0.05 & 8.50 $\pm$ 0.07 \\
            & K  & 25.\,Sep\,1991 & 0.460 $\pm$ 0.008 & 6.95 $\pm$ 0.02 &
            7.36 $\pm$ 0.03 & 8.20 $\pm$ 0.03 \\
            &   & 26.\,Oct\,1991 & 0.45 $\pm$ 0.02 & &
            7.35 $\pm$ 0.03 & 8.22 $\pm$ 0.05 \\
IS\,Tau     & J & 26.\,Jan\,1994 & 0.32 $\pm$ 0.01 & 10.26 &
            10.56 $\pm$ 0.01 & 11.80 $\pm$ 0.03 \\
            &   & 29.\,Nov\,1996 & 0.21 $\pm$ 0.01 & &
            10.47 $\pm$ 0.01 & 12.16 $\pm$ 0.04 \\
            & H & 9.\,Jan\,1993 & 0.36 $\pm$ 0.03 & 9.25 &
            9.58 $\pm$ 0.02 & 10.69 $\pm$ 0.07 \\
            &   & 29.\,Sep\,1996 & 0.20 $\pm$ 0.01 & &
            9.45 $\pm$ 0.01 & 11.20 $\pm$ 0.05 \\
            & K & 9.\,Jan\,1993 & 0.21 $\pm$ 0.02 & 8.68 &
            8.87 $\pm$ 0.02 & 10.58 $\pm$ 0.09 \\
            &   & 9.\,Oct\,1995 & 0.165 $\pm$ 0.004 & &
            8.85 $\pm$ 0.01 & 10.80 $\pm$ 0.01 \\
CoKu\,Tau\,3 & J & 16.\,Nov\,1997 & & &
            11.19 $\pm$ 0.09 & 12.30 $\pm$ 0.09 \\
            & H & 27.\,Sep\,1991 & & &
            9.34 & 10.89 \\
HBC\,412    & J & 27.\,Jan\,1994 & 0.82 $\pm$ 0.11 & 10.06 &
            10.71 $\pm$ 0.07 & 10.93 $\pm$ 0.08 \\
            & H & 6.\,Jan\,1993 & 0.90 $\pm$ 0.02 & 9.33 &
            10.03 $\pm$ 0.01 & 10.14 $\pm$ 0.01 \\
            & K & 19.\,Mar\,1992 & 1.00 $\pm$ 0.02 & 9.10 &
            9.85 $\pm$ 0.01 & 9.85 $\pm$ 0.01 \\
Haro 6-28   & J & 27.\,Jan\,1994 & 0.20 $\pm$ 0.01 & 11.08 &
            11.28 $\pm$ 0.01 & 13.03 $\pm$ 0.05 \\
            & H & 6.\,Jan\,1993 & 0.44 $\pm$ 0.01 & 10.01 $\pm$ 0.06 &
            10.41 $\pm$ 0.07 & 11.30 $\pm$ 0.08 \\
            & K & 16.\,Feb\,1992 & 0.63 $\pm$ 0.03 & 9.27 $\pm$ 0.02 &
            9.80 $\pm$ 0.04 & 10.30 $\pm$ 0.05 \\
VY\,Tau     & J & 29.\,Nov\,1996 & 0.306 $\pm$ 0.006 & 9.86 $\pm$ 0.29 &
            10.15 $\pm$ 0.29 & 11.44 $\pm$ 0.31 \\
\hline
\end{tabular}
\end{table*}
\clearpage
\setcounter{table}{0}
\begin{table*}
\caption{continued}
\begin{tabular}{llllllll}
& & & & & & & \\ \hline
System & Filter & Date & $I_2/I_1$ & $m_{\mathrm{Sys}}$ &
$m_{\mathrm{Prim}}$ & $m_{\mathrm{Sec}}$ \\ \hline
            & H & 22.\,Sep\,1991 & 0.24 $\pm$ 0.01 & 9.26 $\pm$ 0.17 &
            9.49 $\pm$ 0.18 & 11.04 $\pm$ 0.18 \\
            &   & 28.\,Sep\,1996 & 0.26 $\pm$ 0.01 & &
            9.51 $\pm$ 0.18 & 10.97 $\pm$ 0.20 \\
            & K & 5.\,Dec\,1990 & 0.26 $\pm$ 0.02 & 8.97 $\pm$ 0.16 &
            9.22 $\pm$ 0.18 & 10.68 $\pm$ 0.23 \\
IW\,Tau     & J & 27.\,Jan\,1994 & 1.24 $\pm$ 0.04 & 9.33 &
            10.21 $\pm$ 0.02 & 9.97 $\pm$ 0.02 \\
            & H & 8.\,Jan\,1993 & 0.93 $\pm$ 0.02 & 8.61 &
            9.32 $\pm$ 0.01 & 9.40 $\pm$ 0.01 \\
            &   & 28.\,Sep\,1996 & 1.30 $\pm$ 0.15 & &
            9.51 $\pm$ 0.07 & 9.23 $\pm$ 0.05 \\
            & K & 28.\,Oct\,1991 & 0.91 $\pm$ 0.04 & 8.35 $\pm$ 0.02 &
            9.05 $\pm$ 0.04 & 9.15 $\pm$ 0.04 \\
Lk\,H$\alpha$\,332\,G1 & J & 29,\,Nov\,1996 & 0.511 $\pm$ 0.026
            & 9.64 $\pm$ 0.01 & 10.09 $\pm$ 0.03 & 10.82 $\pm$ 0.05 \\
            & H & 10.\,Nov\,1992 & 0.62 $\pm$ 0.01 & 8.68 $\pm$ 0.03 &
            9.20 $\pm$ 0.04 & 9.72 $\pm$ 0.04 \\
            &   & 28.\,Sep\,1996 & 0.560 $\pm$ 0.009 & &
            9.16 $\pm$ 0.04 & 9.79 $\pm$ 0.04 \\
            & K & 27.\,Oct\,1991 & 0.58 $\pm$ 0.03 & 8.18 $\pm$ 0.05 &
            8.68 $\pm$ 0.07 & 9.27 $\pm$ 0.09 \\
            &   & 12.\,Dec\,1994 & 0.555 $\pm$ 0.006 & &
            8.66 $\pm$ 0.05 & 9.30 $\pm$ 0.06 \\
            &   & 22.\,Nov\,1997 & 0.51 $\pm$ 0.02 &  &
            8.63 $\pm$ 0.06 & 9.36 $\pm$ 0.08 \\
Lk\,H$\alpha$\,332\,G2 & J & 27.\,Jan\,1994 & 0.458 $\pm$ 0.003
            & 9.44 $\pm$ 0.02 & 9.85 $\pm$ 0.02 & 10.70 $\pm$ 0.02 \\
            & H & 10.\,Nov\,1992 & 0.509 $\pm$ 0.023 & 8.35 $\pm$ 0.04 &
            8.80 $\pm$ 0.06 & 9.53 $\pm$ 0.07 \\
            &   & 28.\,Sep\,1996 & 0.411 $\pm$ 0.012 & &
            8.72 $\pm$ 0.05 & 9.69 $\pm$ 0.06 \\
            & K & 27.\,Oct\,1991 & 0.60 $\pm$ 0.05 & 7.88 $\pm$ 0.07 &
            8.39 $\pm$ 0.10 & 8.94 $\pm$ 0.13 \\
            &   & 15.\,Dec\,1994 & 0.530 $\pm$ 0.017 & &
            8.34 $\pm$ 0.08 & 9.03 $\pm$ 0.09 \\
            &   & 17.\,Nov\,1997 & 0.465 $\pm$ 0.022 & &
            8.29 $\pm$ 0.09 & 9.13 $\pm$ 0.11 \\
Lk\,H$\alpha$\,332 & J & 27.\,Jan\,1994 & 0.80 $\pm$ 0.04 & 9.81 $\pm$ 0.02 &
             10.45 $\pm$ 0.05 & 10.69 $\pm$ 0.05 \\
            & H & 10.\,Nov\,1992 & 0.56 $\pm$ 0.06 & 8.61 $\pm$ 0.06 &
            9.09 $\pm$ 0.10 & 9.72 $\pm$ 0.13 \\
            &   & 27.\,Jan\,1994 & 0.65 $\pm$ 0.04 & &
            9.15 $\pm$ 0.09 & 9.62 $\pm$ 0.10 \\
            &   & 28.\,Sep\,1996 & 0.485 $\pm$ 0.015 & &
            9.04 $\pm$ 0.07 & 9.82 $\pm$ 0.08 \\
            & K & 27.\,Oct\,1991 & 0.23 $\pm$ 0.03 & 7.83 $\pm$ 0.08 &
            8.05 $\pm$ 0.11 & 9.65 $\pm$ 0.20 \\
            &   & 22.\,Nov\,1997 & 0.46 $\pm$ 0.01 & &
            8.24 $\pm$ 0.09 & 9.08 $\pm$ 0.10 \\
Haro\,6-37\,/c  & H & 8.\,Jan\,1998 & & 9.51 $\pm$ 0.06 & & & \\
            & K & 8.\,Jan\,1998 & & 8.97 $\pm$ 0.08 & & & \\
RW\,Aur     & J & 30\,Sep\,1993 & 0.30 $\pm$ 0.02 & 8.38 $\pm$ 0.21
            & 8.66 $\pm$ 0.23 & 9.97 $\pm$ 0.27 \\
            & H & 9\,Jan\,1993 & 0.37 $\pm$ 0.02 & 7.53 $\pm$ 0.06 
            & 7.87 $\pm$ 0.08 & 8.95 $\pm$ 0.10 \\
            & K & 28\,Nov\,1991 & 0.23 $\pm$ 0.01 & 6.83 $\pm$ 0.15
            & 7.05 $\pm$ 0.20 & 8.65 $\pm$ 0.23 \\
NTTS\,155203-2338 & J & 5.\,May\,1998 & 0.102 $\pm$ 0.006 & 7.56 $\pm$ 0.02 &
                 7.67 $\pm$ 0.03 & 10.14 $\pm$ 0.08 \\         
                 & H & 5.\,May\,1998 & 0.121 $\pm$ 0.007 & 7.15 $\pm$ 0.02 &
                 7.27 $\pm$ 0.03 & 9.57 $\pm$ 0.08 & \\
NTTS\,155808-2219 & J & 7.\,May\,1998 & 0.585 $\pm$ 0.019 & 9.73 $\pm$ 0.02 &
                 10.23 $\pm$ 0.03 & 10.81 $\pm$ 0.04 \\
                 & H & 7.\,May\,1998 & 0.575 $\pm$ 0.008 & 9.02 $\pm$ 0.02 &
                 9.51 $\pm$ 0.03 & 10.11 $\pm$ 0.03 \\
NTTS\,155808-2219 & J & 7.\,May\,1998 & 0.585 $\pm$ 0.019 & 9.73 $\pm$ 0.02 &
                 10.23 $\pm$ 0.03 & 10.81 $\pm$ 0.04 \\
                 & H & 7.\,May\,1998 & 0.575 $\pm$ 0.008 & 9.02 $\pm$ 0.02 &
                 9.51 $\pm$ 0.03 & 10.11 $\pm$ 0.03 \\
NTTS\,155913-2233 & J & 7.\,May\,1998 & 0.548 $\pm$ 0.014 & 8.84 $\pm$ 0.02 &
                 9.31 $\pm$ 0.03 & 9.97 $\pm$ 0.04 \\
                 & H & 7.\,May\,1998 & 0.577 $\pm$ 0.016 & 8.23 $\pm$ 0.02 &
                 8.72 $\pm$ 0.03 & 9.32 $\pm$ 0.04 \\
NTTS\,160735-1857 & J & 8.\,May\,1998 & 0.656 $\pm$ 0.019 & 9.72 $\pm$ 0.02 &
                 10.27 $\pm$ 0.03 & 10.73 $\pm$ 0.04 \\
                 & H & 8.\,May\,1998 & 0.584 $\pm$ 0.021 & 8.98 $\pm$ 0.02 &
                 9.48 $\pm$ 0.03 & 10.06 $\pm$ 0.05 \\
NTTS\,160946-1851 & J & 8.\,May\,1998 & 0.210 $\pm$ 0.007 & 8.27 $\pm$ 0.02 &
                 8.48 $\pm$ 0.03 & 10.17 $\pm$ 0.05 \\
                 & H & 8.\,May\,1998 & 0.194 $\pm$ 0.009 & 7.65 $\pm$ 0.02 &
                 7.84 $\pm$ 0.03 & 9.62 $\pm$ 0.06 \\
RXJ\,1546.1-2804 & H & 7.\,May\,1998 & 0.537 $\pm$ 0.016 & 7.51 &
                 7.98 $\pm$ 0.01 & 8.65 $\pm$ 0.02 \\
RXJ\,1549.3-2600 & H & 7.\,May\,1998 & 0.443 $\pm$ 0.003 & 8.15 &
                 8.55 $\pm$ 0.01 & 9.43 $\pm$ 0.01 \\
RXJ\,1601.7-2049 & H & 13.\,May\,1998 & 0.504 $\pm$ 0.012 & 8.82 &
                 9.26 $\pm$ 0.01 & 10.01 $\pm$ 0.02 \\
RXJ\,1600.5-2027 & H & 7.\,May\,1998 & 0.735 $\pm$ 0.021 & 9.13 &
                 9.73 $\pm$ 0.01 & 10.06 $\pm$ 0.02 \\
RXJ\,1601.8-2445 & H & 7.\,May\,1998 & 0.595 $\pm$ 0.030 & 8.71 &
                 9.22 $\pm$ 0.02 & 9.78 $\pm$ 0.03 \\
RXJ\,1603.9-2031B & H & 7.\,May\,1998 & 0.666 $\pm$ 0.022 & 8.94 &
                 9.49 $\pm$ 0.01 & 9.94 $\pm$ 0.02 \\
RXJ\,1604.3-2130B & H & 7.\,May\,1998 & 0.471 $\pm$ 0.004 & 9.82 &
                 10.24 $\pm$ 0.01 & 11.06 $\pm$ 0.01 \\
WX\,Cha          & J & 7.\,May\,1998 & 0.226 $\pm$ 0.017 & 10.00 &
                 10.22 $\pm$ 0.02 & 11.84 $\pm$ 0.07 \\
                 & H & 7.\,May\,1998 & 0.188 $\pm$ 0.007 & 9.01 &
                 9.20 $\pm$ 0.01 & 11.01 $\pm$ 0.03 \\
VW\,Cha\,AB      & J & 5.\,May\,1998 & 0.573 $\pm$ 0.022 & 8.73 &
                 9.22 $\pm$ 0.02 & 9.83 $\pm$ 0.03 \\
                 & H & 5.\,May\,1998 & 0.432 $\pm$ 0.019 & 7.66 &
                 8.05 $\pm$ 0.01 & 8.96 $\pm$ 0.03 \\
HM\,Anon         & J & 5.\,May\,1998 & 0.205 $\pm$ 0.002 & 8.79 &
                 8.99 $\pm$ 0.01 & 10.71 $\pm$ 0.01 \\
                 & H & 5.\,May\,1998 & 0.240 $\pm$ 0.006 & 8.15 &
                 8.38 $\pm$ 0.01 & 9.93 $\pm$ 0.02 \\
HN\,Lup          & J & 7.\,May\,1998 & 0.831 $\pm$ 0.044 & 9.24 $\pm$ 0.10 &
                 9.90 $\pm$ 0.13 & 10.10 $\pm$ 0.13 \\
                 & H & 7.\,May\,1998 & 0.628 $\pm$ 0.013 & 8.04 $\pm$ 0.10 &
                 8.57 $\pm$ 0.11 & 9.07 $\pm$ 0.11 \\
\hline
\end{tabular}
\end{table*}
\clearpage
%



\begin{table*}
\caption{\label{jhk-phot} NIR photometry for components of young binary
 systems. The values given here are the mean of all spatially resolved
 photometric observations, i.\,e. our data from Table \ref{new-obs} and 
 measurements published by Chelli et al.\,(\cite{Chelli95}),
 Duch\^{e}ne\,(\cite{Duchene99a}), Ghez et al.\,(\cite{Ghez93},
 \cite{Ghez97a}, \cite{Ghez97b}), Haas et al.\,(\cite{Haas90}),
 Hartigan et al.\,(\cite{Hartigan94}), K\"ohler et al.\,(\cite{Koehler99}),
 Moneti \& Zinnecker\,(\cite{Moneti91}), Richichi et al.\,(\cite{Richichi99}),
 Roddier et al.\,(\cite{Roddier96}) and  Simon et al.\,(\cite{Simon92}).
 The values for $A_{\mathrm{V}}$ and the spectral type given for the primaries,
 belong to the systems (see Sect.\,\ref{sysprops} for references).
 The numbers in the last column denote Taurus-Auriga (1), Upper Scorpius (2),
 Chamaeleon\,I (3) and Lupus (4).}
\begin{tabular}{llllllll}
& & & & & & \\ \hline
System & Component & J & H & K & $A_{\mathrm{V}}$ & Spectral type & SFR\\
 \hline
HBC 351 & A & 10.09 $\pm$ 0.05 & 9.49 $\pm$ 0.03 & 9.37 $\pm$ 0.02 &
           0.00 & K5 & 1 \\
        & B & 12.04 $\pm$ 0.07 & 11.29 $\pm$ 0.09 & 11.01 $\pm$ 0.08 & & & \\
HBC 352\,/ & & 10.12 $\pm$ 0.04 & 9.76 $\pm$ 0.03 & 9.62 $\pm$ 0.02  &
         0.87 & G0 & 1 \\
HBC 353    & & 10.47 $\pm$ 0.02 & 10.04 $\pm$ 0.03 & 9.91 $\pm$ 0.03 & & & \\
HBC 358 & A & 11.51 $\pm$ 0.06 & 10.87 $\pm$ 0.05 & 10.60 $\pm$ 0.06 &
         0.21 & M2 & 1 \\
        & a & 11.56 $\pm$ 0.06 & 10.94 $\pm$ 0.05 & 10.69 $\pm$ 0.06 & & & \\
        & B & 11.33 $\pm$ 0.06 & 10.75 $\pm$ 0.06 & 10.44 $\pm$ 0.09 & & & \\ 
HBC 360\,/ & & 10.83 $\pm$ 0.08 & 10.24 $\pm$ 0.08 & 9.98 $\pm$ 0.08 &
  0.28 & M3 & 1 \\
HBC 361 & & 10.96 & 10.37 & 10.11 & & & \\
LkCa\,3 & A & 8.99 $\pm$ 0.01 & 8.34 $\pm$ 0.09 & 8.11 $\pm$ 0.08 &
  0.42 & M1 & 1 \\
        & B & 9.51 $\pm$ 0.02 & 8.68 $\pm$ 0.12 & 8.48 $\pm$ 0.12 & & & \\
V\,773\,Tau & A & 7.77 $\pm$ 0.05 & 7.03 $\pm$ 0.03 & 6.77 $\pm$ 0.09 &
  1.32 & K3 & 1 \\
            & B & 10.13 $\pm$ 0.09 & 8.91 $\pm$ 0.15 & 8.09 $\pm$ 0.31 & & & \\
FO\,Tau & A & 10.13 $\pm$ 0.07 & 9.35 $\pm$ 0.07 & 8.76 $\pm$ 0.03 &
   1.87 & M2 & 1 \\
        & B & 10.95 $\pm$ 0.17 & 9.59 $\pm$ 0.09 & 9.14 $\pm$ 0.04 & & & \\
DD\,Tau & A & 10.10 $\pm$ 0.05 & 9.15 $\pm$ 0.08 & 8.37 $\pm$ 0.04 &
   0.76 & M1 & 1 \\
        & B & 10.36 $\pm$ 0.05 & 9.41 $\pm$ 0.08 & 8.96 $\pm$ 0.07 & & & \\
CZ\,Tau & A & 10.63 $\pm$ 0.12 & 9.96 $\pm$ 0.04 & 9.58 $\pm$ 0.12 &
   1.32 & M1.5 & 1 \\
        & B & 12.94 $\pm$ 0.16 & 11.56 $\pm$ 0.07 & 10.92 $\pm$ 0.39 & & & \\
FQ\,Tau & A & 11.39 $\pm$ 0.03 & 10.74 $\pm$ 0.03 & 10.17 $\pm$ 0.32 &
   1.87 & M2 & 1 \\
        & B & 11.33 $\pm$ 0.02 & 10.58 $\pm$ 0.03 & 10.28 $\pm$ 0.32 & & & \\
V\,819\,Tau & A & 9.45 $\pm$ 0.03 & 8.76 $\pm$ 0.08 & 8.50 &
   1.35 & K7 & 1 \\
            & B & 12.96 $\pm$ 0.06 & 12.39 $\pm$ 0.08 & 12.14 & & & \\
LkCa\,7 & A & 9.63 $\pm$ 0.01 & 8.98 $\pm$ 0.02 & 8.84 $\pm$ 0.04 &
   0.59 & K7 & 1 \\
        & B & 10.58 $\pm$ 0.01 & 9.87 $\pm$ 0.03 & 9.47 $\pm$ 0.05 & & & \\
FS\,Tau & A & 10.85 $\pm$ 0.14 & 9.32 $\pm$ 0.10 & 7.88 $\pm$ 0.26 &
   1.84 & M1 & 1 \\
        & B & 12.66 $\pm$ 0.16 & 11.17 $\pm$ 0.13 & 10.03 $\pm$ 0.29 & & & \\
T\,Tau  & A & 7.26 $\pm$ 0.22 & 6.36 $\pm$ 0.14 & 5.56 $\pm$ 0.18 &
   1.39 & K0 & 1 \\
        & B &   &  8.81 $\pm$ 0.11 & 6.52 $\pm$ 0.11 & & & \\
FV\,Tau & A & 9.98 $\pm$ 0.12 & 8.78 $\pm$ 0.14 & 7.96 $\pm$ 0.04 &
   4.72 & K5 & 1 \\
        & B & 10.69 $\pm$ 0.23 & 9.20 $\pm$ 0.15 & 8.32 $\pm$ 0.05 & & & \\
FV\,Tau/c & A & & 9.45 $\pm$ 0.02 & 8.93 $\pm$ 0.05 & 3.40 & M3.5 & 1 \\
          & B & & 13.26 $\pm$ 0.02 & 11.29 $\pm$ 0.40 & & & \\
Haro\,6-10 & A & 9.88 $\pm$ 0.03 & 9.13 $\pm$ 0.31 & 7.89 $\pm$ 0.04 &
   4.78 & K3 & 1 \\
        & B & & 12.02 $\pm$ 0.44 & 10.71 $\pm$ 0.56 & & & \\
FW\,Tau & A & & 10.39 $\pm$ 0.06 & 10.04 $\pm$ 0.08 & 0.35 & M4 & 1 \\
        & B & & 10.69 $\pm$ 0.08 & 10.22 $\pm$ 0.10 & & & \\
UX\,Tau & A & 9.00 $\pm$ 0.10 & 8.06 $\pm$ 0.10 & 7.54 $\pm$ 0.10 &
    0.21 & K2 & 1 \\
        & B & 11.14 $\pm$ 0.10 & 10.10 $\pm$ 0.03 & 9.42 $\pm$ 0.04 & & & \\
        & b & 11.42 $\pm$ 0.10 & 10.27 $\pm$ 0.03 & 9.55 $\pm$ 0.04 & & & \\
        & C & 11.85 $\pm$ 0.09 & 10.84 & 10.43 & & & \\
FX\,Tau & A & 9.88 $\pm$ 0.01 & 9.37 $\pm$ 0.01 & 8.61 $\pm$ 0.02 &
    1.08 & M1 & 1 \\
        & B & 9.95 $\pm$ 0.01 & 9.65 $\pm$ 0.01 & 9.28 $\pm$ 0.02 & & & \\
DK\,Tau & A & 9.15 $\pm$ 0.09 & 8.10 $\pm$ 0.02 & 7.32 $\pm$ 0.11 &
    0.76 & K7 & 1 \\
        & B & 10.52 $\pm$ 0.10 & 9.43 $\pm$ 0.03 & 8.73 $\pm$ 0.04 & & & \\
LkH$\alpha$\,331 & A & 10.43 $\pm$ 0.01 & 9.63 $\pm$ 0.06 & 9.26 $\pm$ 0.03 &
 0.38 & M5.5 & 1 \\
                 & B & 10.81 $\pm$ 0.02 & 9.87 $\pm$ 0.08 & 9.65 $\pm$ 0.03 &
 & & \\
XZ\,Tau & A & 11.04 $\pm$ 0.27 & 9.44 $\pm$ 0.17 & 8.41 $\pm$ 0.03 &
  2.91 & M3 & 1 \\
        & B & 10.43 $\pm$ 0.14 & 10.07 $\pm$ 0.20 & 9.44 $\pm$ 0.08 & & & \\
\hline
\end{tabular}
\end{table*}
\clearpage
\setcounter{table}{1}
\begin{table*}
\caption{continued}
\begin{tabular}{llllllll}
& & & & & & \\ \hline
System & Component & J & H & K & $A_{\mathrm{V}}$ & Spectral type & SFR \\
 \hline
HK\,Tau &  & 10.41 $\pm$ 0.23 & 9.13 $\pm$ 0.22 & 8.42 $\pm$ 0.20 &
  2.32 & M0.5 & 1 \\
HK\,Tau/c  & & 13.66 $\pm$ 0.20 & 12.18 $\pm$ 0.20 & 11.77 $\pm$ 0.20 & & & \\
V\,710\,Tau & A & 9.82 $\pm$ 0.16 & 9.06 $\pm$ 0.16 & 8.69 $\pm$ 0.16 &
   0.87 & M1 & 1 \\
            & B & 10.20 $\pm$ 0.27 & 9.21 $\pm$ 0.27 & 8.82 $\pm$ 0.27 &
   &  & \\
HK\,Tau\,G2 & A & 10.03 $\pm$ 0.10 & 9.05 $\pm$ 0.03 & 8.64 $\pm$ 0.05 &
   1.87 & M0.5 & 1 \\
            & B & 10.32 $\pm$ 0.11 & 9.35 $\pm$ 0.04 & 9.00 $\pm$ 0.07 & & \\
GG\,Tau & A & 9.39 $\pm$ 0.09 & 8.29 $\pm$ 0.09 & 7.79 $\pm$ 0.05 &
   0.76 & M0 & 1 \\
        & a & 10.16 $\pm$ 0.02 & 9.05 $\pm$ 0.08 & 8.58 $\pm$ 0.08 & & & \\
        & B & 11.41 $\pm$ 0.16 & 10.55 $\pm$ 0.15 & 10.11 $\pm$ 0.12 &
   0.76 & M4.5 & \\
        & b & 13.24 $\pm$ 0.12 & 12.64 $\pm$ 0.06 & 12.01 $\pm$ 0.22 & & & \\
UZ\,Tau\,w & A & 10.25 $\pm$ 0.04 & 9.23 $\pm$ 0.01 & 8.80 $\pm$ 0.02 &
   0.83 & M3 & 1 \\
           & B & 10.55 $\pm$ 0.06 & 9.66 $\pm$ 0.02 & 9.41 $\pm$ 0.03 & & & \\
UZ\,Tau\,e &   & 9.83 $\pm$ 0.02 & 8.46 $\pm$ 0.02 & 7.59 $\pm$ 0.02 &
   1.49 & M1 & \\
GH\,Tau & A & 9.91 $\pm$ 0.09 & 9.18 $\pm$ 0.07 & 8.69 $\pm$ 0.21 &
   0.52 & M2 & 1 \\
        & B & 10.04 $\pm$ 0.09 & 9.02 $\pm$ 0.06 & 8.42 $\pm$ 0.16 & & & \\
Elias\,12 & A & 8.72 $\pm$ 0.06 & 7.90 $\pm$ 0.02 & 7.33 $\pm$ 0.02 &
   2.87 & K7 & 1 \\
          & B & 9.31 $\pm$ 0.07 & 8.53 $\pm$ 0.03 & 8.26 $\pm$ 0.05 & & & \\
IS\,Tau & A & 10.52 $\pm$ 0.05 & 9.52 $\pm$ 0.07 & 8.85 $\pm$ 0.01 &
   4.17 & K2 & 1 \\
        & B & 11.98 $\pm$ 0.18 & 10.95 $\pm$ 0.03 & 10.74 $\pm$ 0.08 & & & \\
GK\,Tau\,/ & & 9.02 $\pm$ 0.16 & 8.02 $\pm$ 0.07 & 7.31 $\pm$ 0.23 &
   0.87 & K7 & 1 \\
GI\,Tau    & & 9.42 $\pm$ 0.08 & 8.46 $\pm$ 0.07 & 7.79 $\pm$ 0.18 & & & \\
HN\,Tau    & A & 10.82 $\pm$ 0.35 & 9.49 $\pm$ 0.34 & 8.44 $\pm$ 0.33 &
   0.89 & K5 & 1 \\
           & B & 12.62 & 12.05 & 11.62 & & & \\
CoKu\,Tau\,3 & A & 11.19 $\pm$ 0.09 & 9.34 & 8.53 & 3.26 & M1 & 1 \\
             & B & 12.30 $\pm$ 0.09 & 10.89 & 9.87 & & & \\
HBC\,412 & A & 10.71 $\pm$ 0.07 & 10.03 $\pm$ 0.01 & 9.85 $\pm$ 0.01 &
   0.69 & M2 & 1 \\
         & B & 10.93 $\pm$ 0.08 & 10.14 $\pm$ 0.01 & 9.85 $\pm$ 0.01 & & & \\
HP\,Tau\,G2\,/ & & 8.19 $\pm$ 0.09 & 7.49 $\pm$ 0.08 & 7.28 $\pm$ 0.07 &
   0.67 & G0 & 1 \\
HP\,Tau\,G3    & & 10.13 $\pm$ 0.10 & 9.18 $\pm$ 0.06 & 8.85 $\pm$ 0.06 & 
  & & \\
Haro 6-28 & A & 11.28 $\pm$ 0.01 & 10.41 $\pm$ 0.07 & 9.80 $\pm$ 0.04 &
   1.77 & M5 & 1 \\
          & B & 13.03 $\pm$ 0.05 & 11.30 $\pm$ 0.08 & 10.30 $\pm$ 0.05 & & & \\
VY\,Tau & A & 10.15 $\pm$ 0.29 & 9.50 $\pm$ 0.01 & 9.22 $\pm$ 0.18 &
   0.38 & M0 & 1 \\
        & B & 11.44 $\pm$ 0.31 & 11.01 $\pm$ 0.04 & 10.68 $\pm$ 0.23 & & & \\
IW\,Tau & A & 10.21 $\pm$ 0.02 & 9.42 $\pm$ 0.10 & 9.05 $\pm$ 0.04 &
   0.83 & K7 & 1 \\
        & B & 9.97 $\pm$ 0.02 & 9.32 $\pm$ 0.09 & 9.15 $\pm$ 0.04 & & & \\
LkH$\alpha$\,332\,G1 & A & 10.09 $\pm$ 0.03 & 9.18 $\pm$ 0.02 &
 8.66 $\pm$ 0.01 & 2.98 & M1 & 1 \\
                 & B & 10.82 $\pm$ 0.05 & 9.76 $\pm$ 0.04 & 9.31 $\pm$ 0.02
 & & & \\
LkH$\alpha$\,332\,G2 & A & 9.85 $\pm$ 0.02 & 8.76 $\pm$ 0.04 & 8.34 $\pm$ 0.03
 & 3.16 & K7 & 1 \\
                 & B & 10.70 $\pm$ 0.02 & 9.61 $\pm$ 0.08 & 9.03 $\pm$ 0.05
 & & & \\
LkH$\alpha$\,332 & A & 10.45 $\pm$ 0.05 & 9.09 $\pm$ 0.03 & 8.15 $\pm$ 0.10 &
 2.67 & K7 & 1 \\
                 & B & 10.69 $\pm$ 0.05 & 9.72 $\pm$ 0.06 & 9.37 $\pm$ 0.29 
 & & & \\
Haro\,6-37 & A & 9.99 & 8.70 $\pm$ 0.06 & 8.15 $\pm$ 0.08 & 2.12 & K6 & 1 \\
           & a & 11.79 & 10.39 $\pm$ 0.06 & 9.70 $\pm$ 0.08 & & & \\
           & B & 10.65 & 9.51 $\pm$ 0.06 & 8.97 $\pm$ 0.08 & & & \\
UY\,Aur & A & 9.01 $\pm$ 0.06 & 8.26 $\pm$ 0.07 & 7.42 $\pm$ 0.05 &
   1.35 & K7 & 1 \\
        & B & 11.20 $\pm$ 0.07 & 9.85 $\pm$ 0.06 & 8.53 $\pm$ 0.05 & & & \\
\hline
\end{tabular}
\end{table*}
\clearpage
\setcounter{table}{1}
\begin{table*}
\caption{continued}
\begin{tabular}{llllllll}
& & & & & & \\ \hline
System & Component & J & H & K & $A_{\mathrm{V}}$ & Spectral type & SFR \\
 \hline
RW\,Aur & A & 8.66 $\pm$ 0.23 & 7.87 $\pm$ 0.08 & 6.97 $\pm$ 0.19 &
   0.53 & K3 & 1 \\
        & B & 9.97 $\pm$ 0.27 & 8.95 $\pm$ 0.10 & 8.90 $\pm$ 0.23 & & & \\ 
 & & & \\
NTTS\,155203-2338 & A & 7.67 $\pm$ 0.03 & 7.27 $\pm$ 0.03 & 7.21 $\pm$ 0.01 &
 0.2 & G2 & 2 \\
                  & B & 10.14 $\pm$ 0.08 & 9.57 $\pm$ 0.08 & 9.24 $\pm$ 0.04
 & & & \\
NTTS\,155219-2314 & A & 10.66 $\pm$ 0.04 & 9.94 $\pm$ 0.04 & 9.69 $\pm$ 0.04 &
 0.2 & M4 & 2 \\
                  & B & 11.58 $\pm$ 0.06 & 11.02 $\pm$ 0.08 & 10.78 $\pm$ 0.08
 & & & \\
NTTS\,155808-2219 & A & 10.23 $\pm$ 0.03 & 9.51 $\pm$ 0.03 & 9.29 $\pm$ 0.03
 & 0.3 & M3 & 2 \\
                  & B & 10.81 $\pm$ 0.04 & 10.11 $\pm$ 0.03 & 9.90 $\pm$ 0.03
 & & & \\
NTTS\,155913-2233 & A & 9.31 $\pm$ 0.03 & 8.72 $\pm$ 0.03 & 8.54 $\pm$ 0.02 &
 0.2 & K5 & 2 \\
                  & B & 9.97 $\pm$ 0.04 & 9.32 $\pm$ 0.04 & 9.24 $\pm$ 0.05 
 & & & \\
NTTS\,160735-1857 & A & 10.27 $\pm$ 0.03 & 9.48 $\pm$ 0.03 & 9.29 $\pm$ 0.04 & 
 1.1 & M3 & 2 \\
                  & B & 10.73 $\pm$ 0.04 & 10.06 $\pm$ 0.05 & 9.71 $\pm$ 0.05
 & & & \\
NTTS\,160946-1851 & A & 8.48 $\pm$ 0.03 & 7.84 $\pm$ 0.03 & 7.72 $\pm$ 0.01 &
 1.3 & K0 & 2 \\
                  & B & 10.17 $\pm$ 0.05 & 9.62 $\pm$ 0.06 & 9.28 $\pm$ 0.05
 & & & \\
RXJ\,1546.1-2804 & A & & 7.98 $\pm$ 0.01 & 8.01 $\pm$ 0.01 & & & 2 \\
                 & B & & 8.65 $\pm$ 0.02 & 8.39 $\pm$ 0.01 & \\
RXJ\,1549.3-2600 & A & & 8.55 $\pm$ 0.01 & 8.43 $\pm$ 0.02 & & & 2 \\
                 & B & & 9.43 $\pm$ 0.01 & 9.19 $\pm$ 0.03 & \\
RXJ\,1600.5-2027 & A & & 9.73 $\pm$ 0.01 & 9.46 $\pm$ 0.02 & & & 2 \\
                 & B & & 10.06 $\pm$ 0.02 & 9.89 $\pm$ 0.04 & \\
RXJ\,1601.7-2049 & A & & 9.26 $\pm$ 0.01 & 9.08 $\pm$ 0.02 & & & 2 \\
                 & B & & 10.01 $\pm$ 0.02 & 9.65 $\pm$ 0.03 & \\
RXJ\,1601.8-2445 & A & & 9.22 $\pm$ 0.02 & 8.87 $\pm$ 0.05 & & & 2 \\
                 & B & & 9.78 $\pm$ 0.03 & 9.89 $\pm$ 0.13 & \\
RXJ\,1603.9-2031B & A & & 9.49 $\pm$ 0.01 & 9.20 $\pm$ 0.02 & & & 2 \\
                 & B & & 9.94 $\pm$ 0.02 & 9.73 $\pm$ 0.04 & \\
RXJ\,1604.3-2130B & A & & 10.24 $\pm$ 0.01 & 10.08 $\pm$ 0.08 & & & 2 \\
                 & B &  & 11.06 $\pm$ 0.01 & 10.46 $\pm$ 0.11 & \\
WX\,Cha       & A & 10.22 $\pm$ 0.02 & 9.20 $\pm$ 0.01 & 8.43 $\pm$ 0.03 &
 2.14 & K7-M0 & 3 \\
              & B & 11.84 $\pm$ 0.07 & 11.01 $\pm$ 0.03 & 10.94 $\pm$ 0.30
 & & & \\
VW\,Cha       & A & 9.22 $\pm$ 0.02 & 8.05 $\pm$ 0.01 & 7.25 $\pm$ 0.03 &
 2.39 & K5 & 3 \\
              & B & 9.83 $\pm$ 0.03 & 8.96 $\pm$ 0.03 & 8.89 $\pm$ 0.14
  & & & \\
              & C & $\ge$ 11.65 & $\ge$ 11.35 & 9.75 $\pm$ 0.49 & & & \\
HM\,Anon & A &  8.99 $\pm$ 0.01 & 8.38 $\pm$ 0.01 & 8.11 $\pm$ 0.02 &
 1.21 & G8 & 3 \\
              & B & 10.71 $\pm$ 0.01 & 9.93 $\pm$ 0.02 & 10.25 $\pm$ 0.14
 & & & \\
LkH$\alpha$ 332-17 & A & 7.88 & 7.01 & 6.28 & 2.35 & G2 & 3 \\
              & B & 11.58 & 10.82 & 10.35 & & & \\
IK\,Lup       & A & 9.46 $\pm$ 0.09 & 8.71 $\pm$ 0.05 & 8.33 $\pm$ 0.02 &
 0.20 & M0 & 4 \\
              & B & 10.82 $\pm$ 0.09 & 9.86 $\pm$ 0.05 & 9.43 $\pm$ 0.02
 & & & \\
HT\,Lup       & A & 7.66 $\pm$ 0.04 & 6.95 $\pm$ 0.02 & 6.55 $\pm$ 0.03 &
 1.45 & K2 & 4 \\
              & B & 10.39 & 9.72 & 8.96 & & & \\       
HN\,Lup       & A & 9.90 $\pm$ 0.13 & 8.57 $\pm$ 0.11 & 7.76 $\pm$ 0.06 &
 2.30 & M1.5 & 4 \\
              & B & 10.10 $\pm$ 0.13 & 9.07 $\pm$ 0.11 & 8.71 $\pm$ 0.08
 & & & \\
HBC\,603      & A & 9.54 &  8.72 & 8.69 $\pm$ 0.10 & 0.79 & M0 & 4 \\
              & B & $\ge$ 12.43 & $\ge$ 11.61 & 9.77 $\pm$ 0.25 & & & \\
HBC\,604      & A & 10.23 $\pm$ 0.08 & 9.55 $\pm$ 0.05 & 9.18 $\pm$ 0.06 &
 0.12 & M5.5 & 4 \\
              & B & 11.58 & 11.07 & 11.14 & & & \\
HO\,Lup       & A & 10.04 $\pm$ 0.06 & 9.14 $\pm$ 0.06 & 8.60 $\pm$ 0.03 &
  1.25 & M1 & 4 \\
              & B & 11.11 & 10.38 & 9.82 & & & \\
\hline
\end{tabular}
\end{table*}
\clearpage


\section{Color-magnitude diagrams}  
\label{fhd-multiplotz}

\begin{figure*} 
 \resizebox{\hsize}{!}{\includegraphics{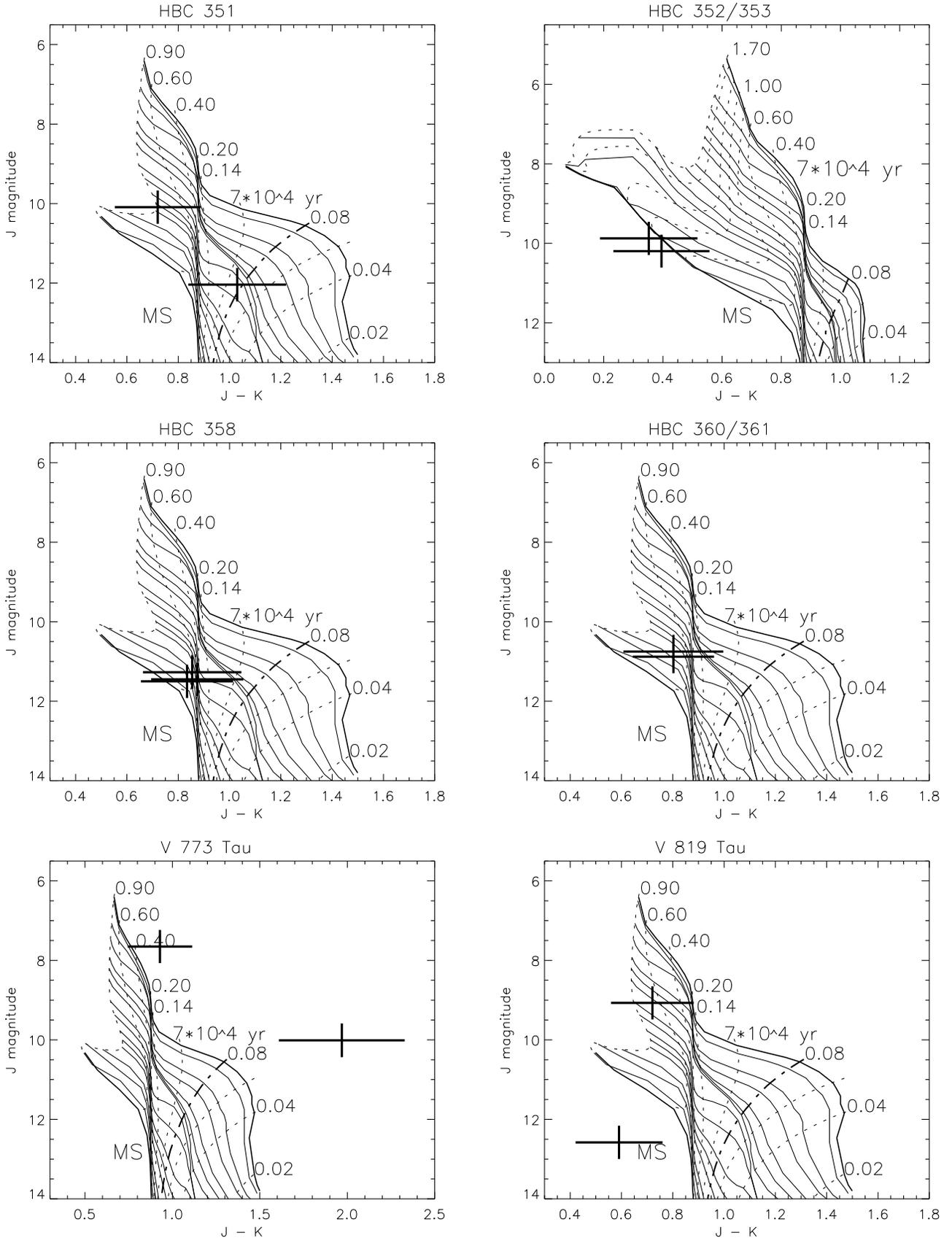}}
  \caption{\label{fhd-multiplots} . Components of WTTS systems placed into NIR
    color-magnitude diagrams together with the PMS model by
    D'Antona \& Mazzitelli\,(\cite{dm98}). The dashed lines denote
    evolutionary tracks for masses from 0.02 to $0.9\,M_{\sun}$, the
    solid lines are isochrones for ages $7\cdot 10^4$, $10^5$, $2\cdot 10^5$,
    $3\cdot 10^5$, $5\cdot 10^5$, $7\cdot 10^5$, $10^6$, $2\cdot 10^6$,
    $3\cdot 10^6$, $5\cdot 10^6$, $7\cdot 10^6$, $10^7$, $2\cdot 10^7$,
    $3\cdot 10^7$, $5\cdot 10^7$ and  $10^8\,\mathrm{yr}$ (MS).}
\end{figure*}
\clearpage

\setcounter{figure}{0}

\begin{figure*}
  \resizebox{\hsize}{!}{\includegraphics{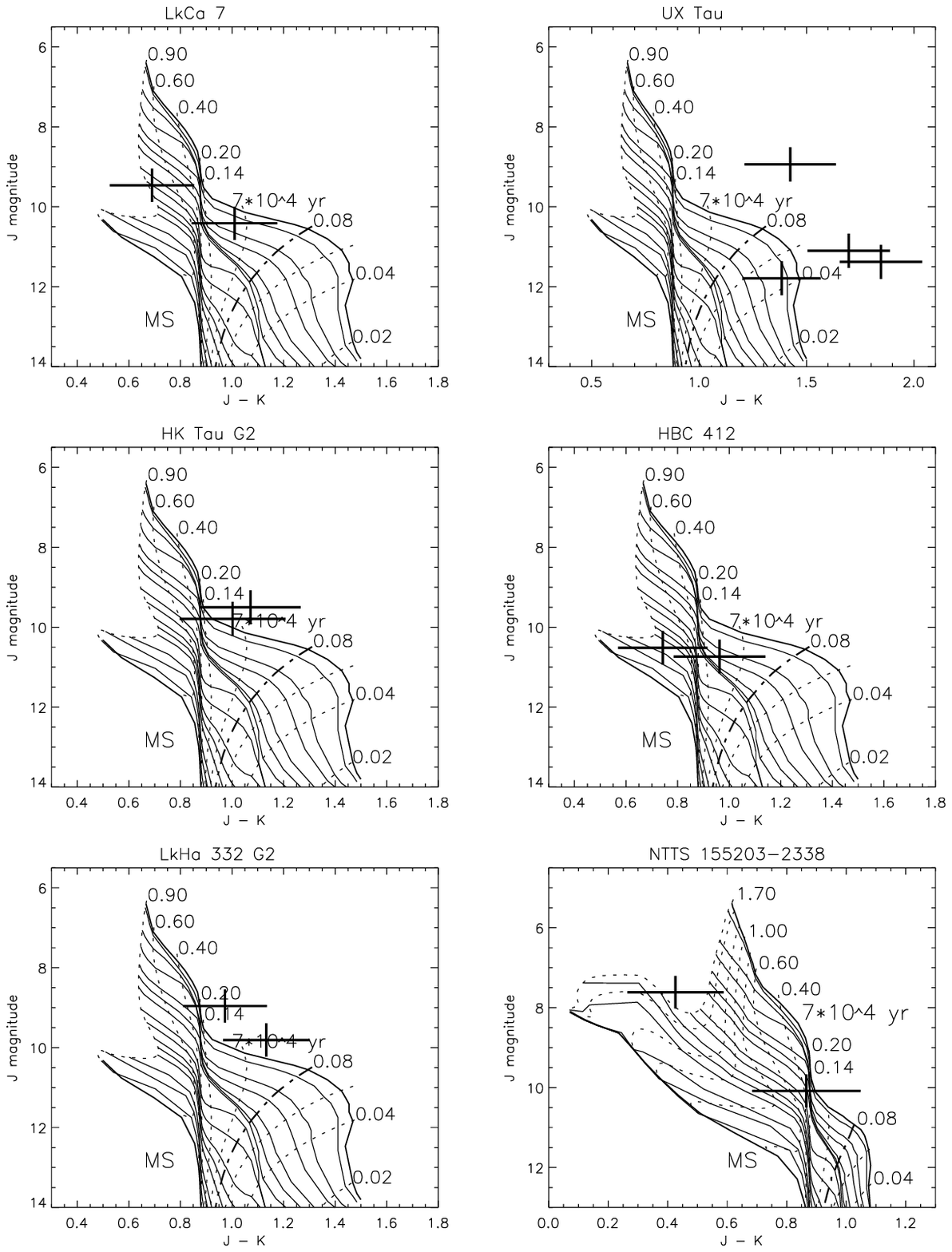}}
  \caption{continued}
\end{figure*}
\clearpage

\setcounter{figure}{0}

\begin{figure*}
  \resizebox{\hsize}{!}{\includegraphics{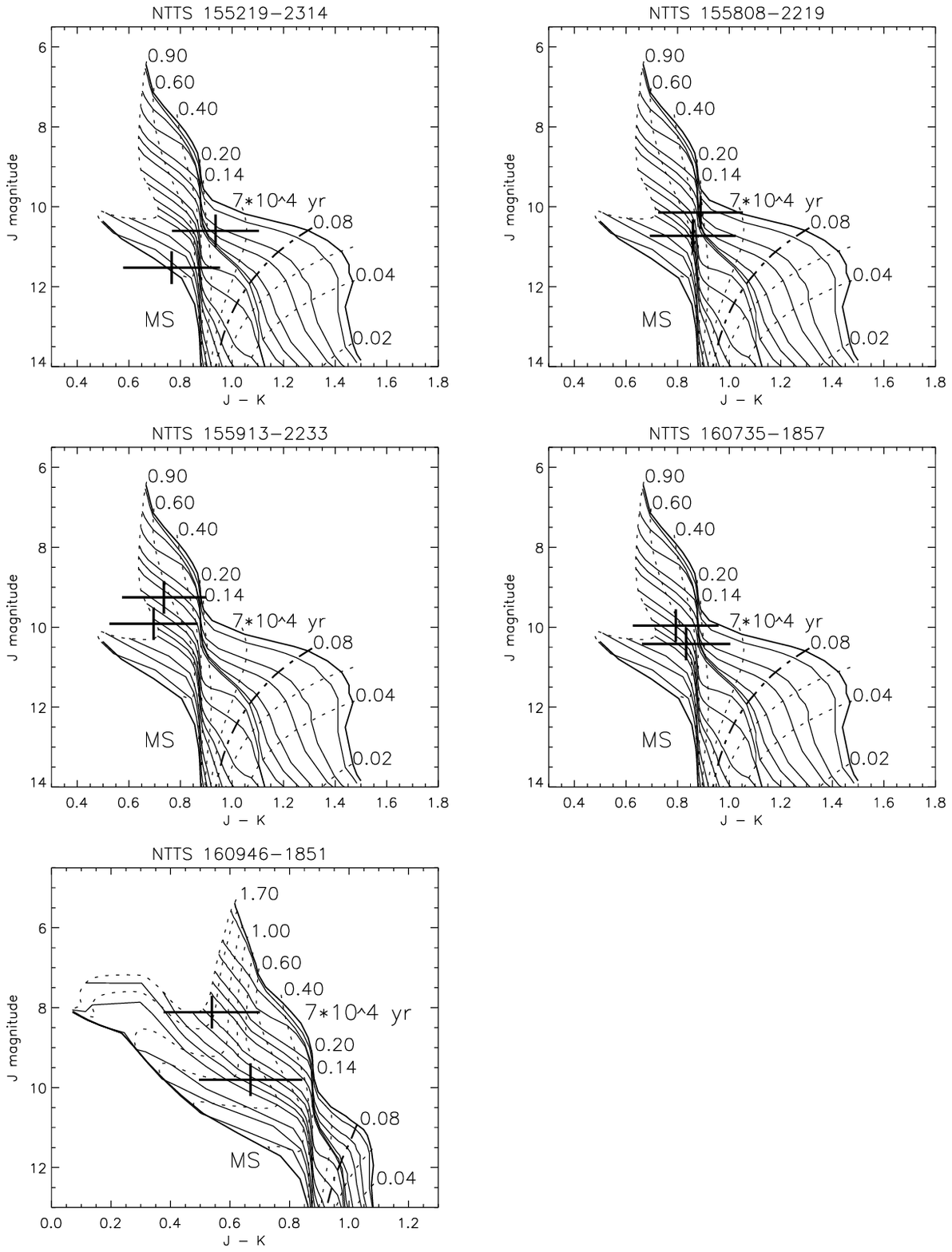}}
  \caption{continued}
\end{figure*}
\clearpage

\section{Hertzsprung-Russell diagrams} 
\label{hrd-multiplotz}

\begin{figure*}
  \resizebox{\hsize}{!}{\includegraphics{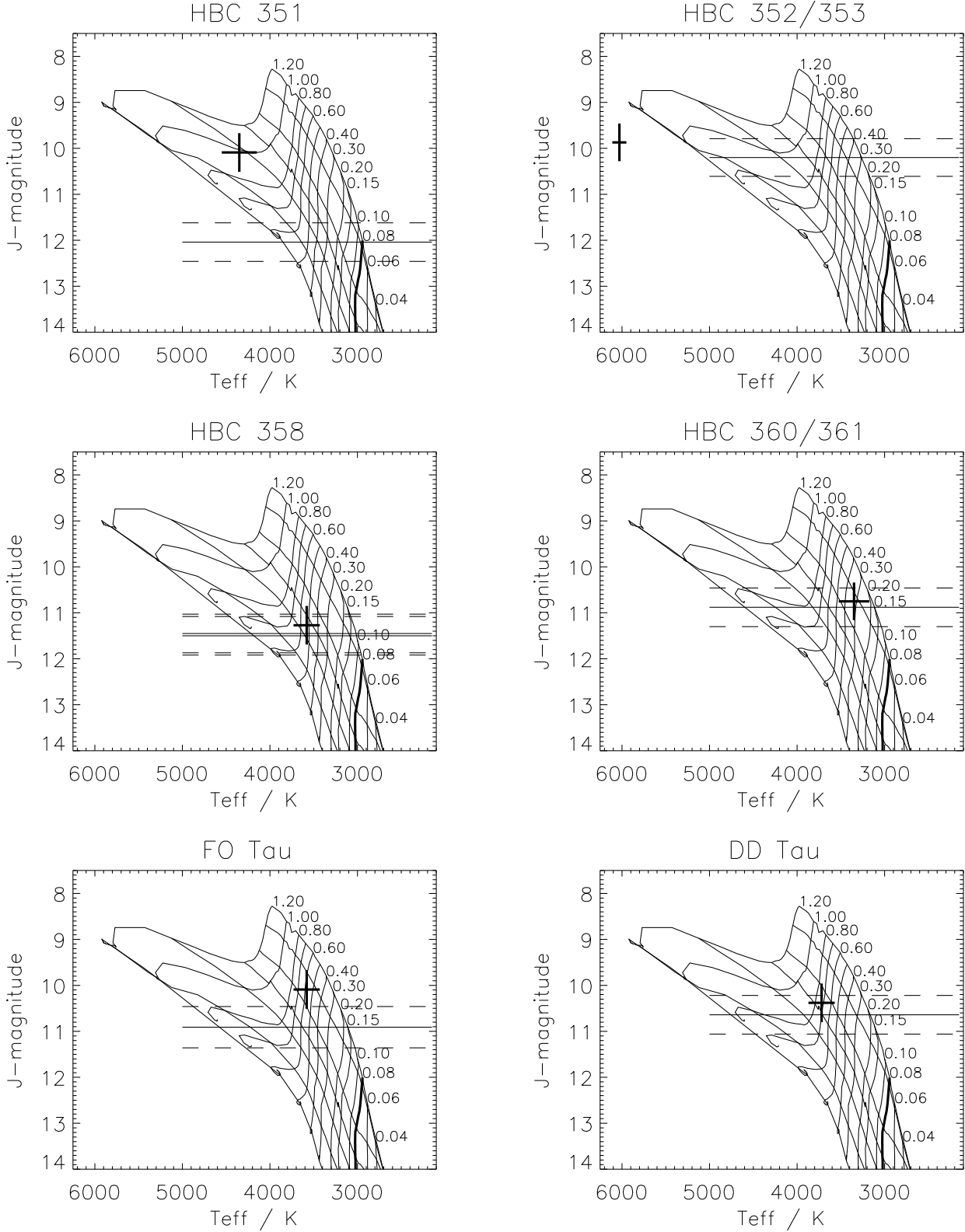}}
  \caption{\label{hrd-appendix} Components of T~Tauri binary systems placed
  into the HRD as
  described in Sect.\,\ref{hrd}. The cross gives the position of the
  primary, the horizontal dashed lines give the locus for the companion
  and the respective error. The PMS model from Baraffe et al.\,
  (\cite{Baraffe98}) is also indicated. Evolutionary tracks are plotted
  for masses of 0.04, 0.06, 0.08 (bold), 0.10, 0.15, 0.20, 0.30, 0.40, 0.50,
  0.60, 0.70, 0.80, 1.0 and 1.2 $M_{\sun}$, the isochrones denote ages of
  1, 2, 5, 10, 20, 50 and 90 Myr. }
\end{figure*}

\setcounter{figure}{0}

\begin{figure*}
  \resizebox{\hsize}{!}{\includegraphics{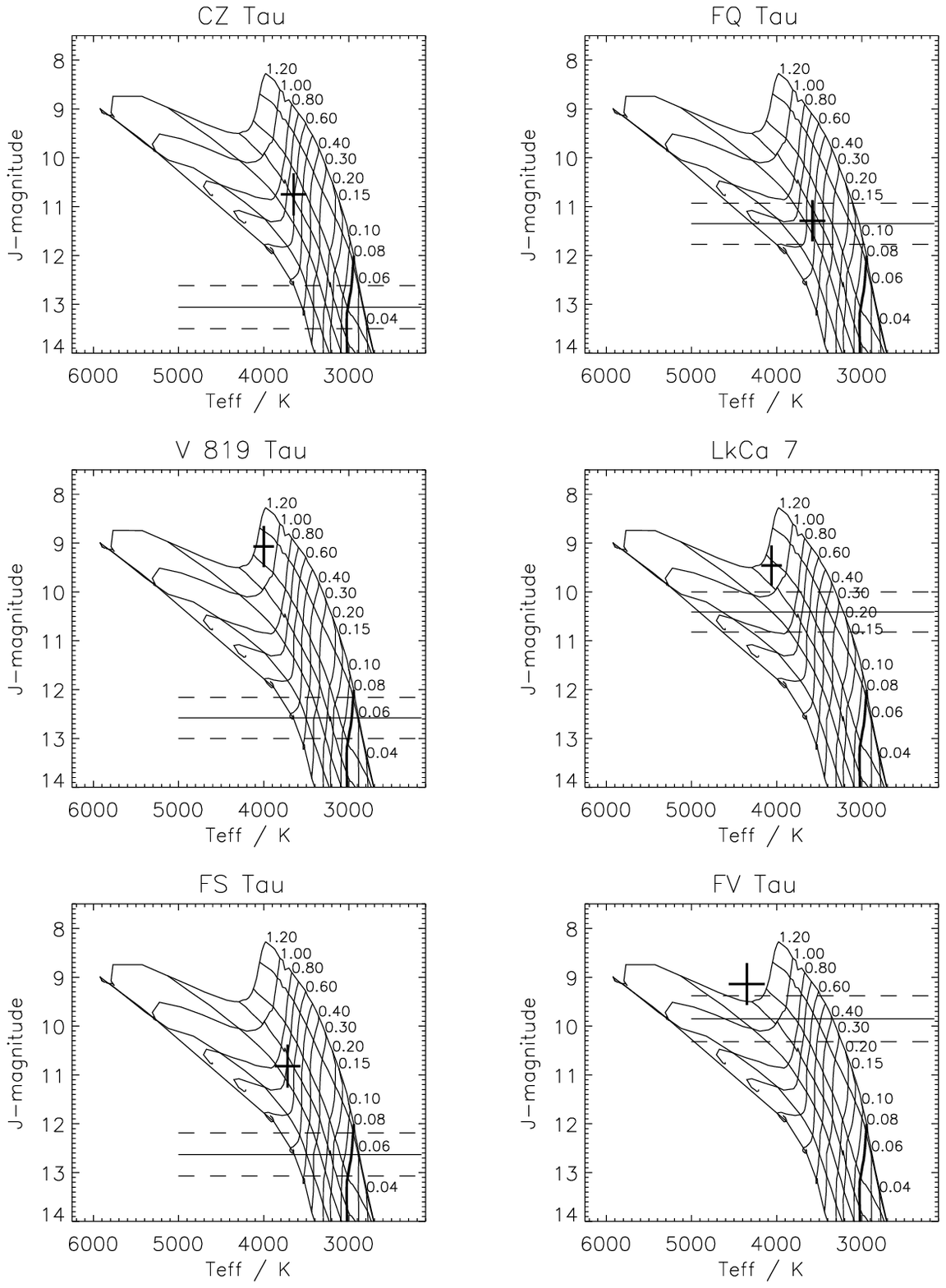}}
  \caption{continued}
\end{figure*}

\setcounter{figure}{0}

\begin{figure*}
  \resizebox{\hsize}{!}{\includegraphics{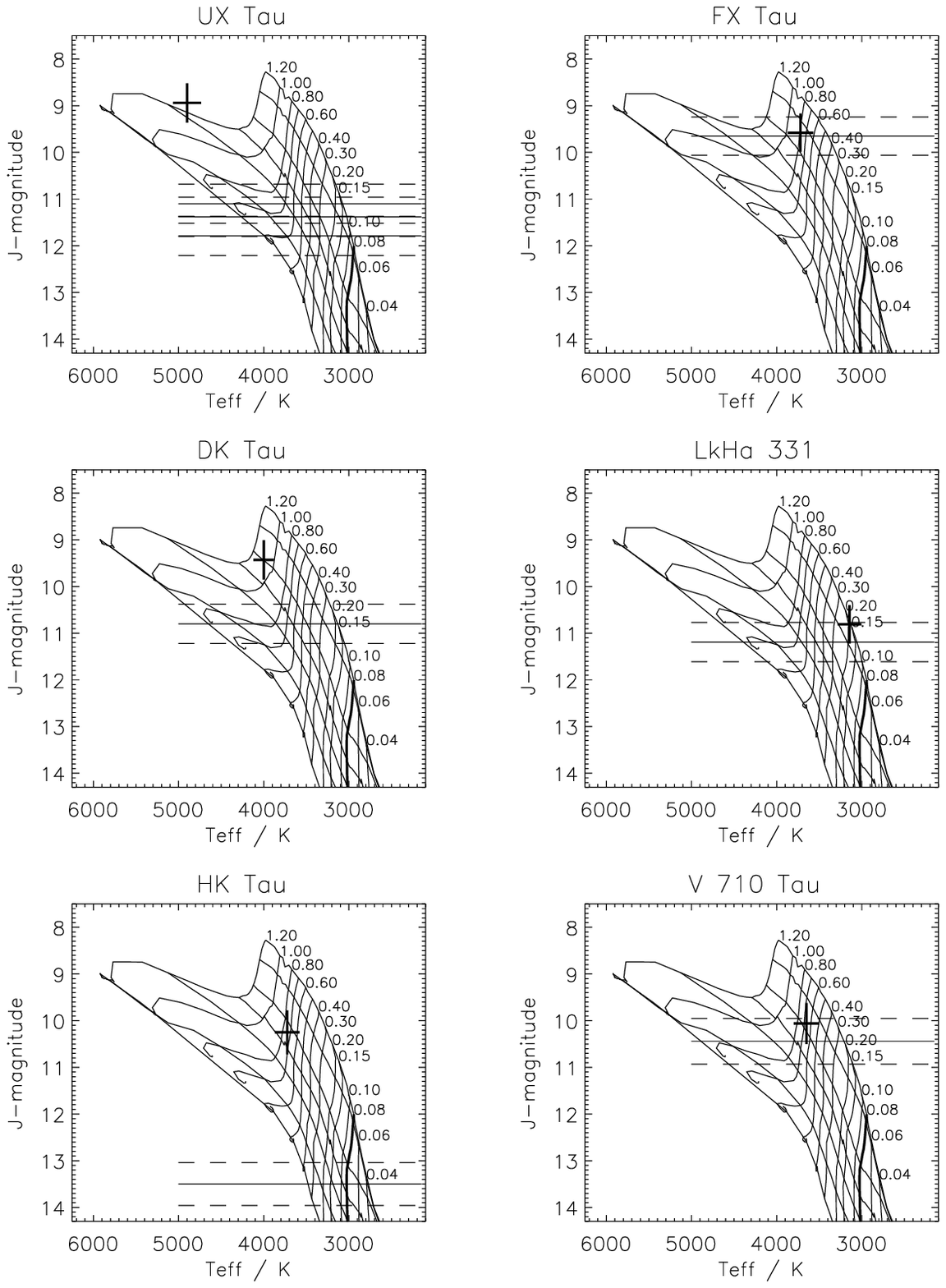}}
  \caption{continued}
\end{figure*}

\setcounter{figure}{0}

\begin{figure*}
  \resizebox{\hsize}{!}{\includegraphics{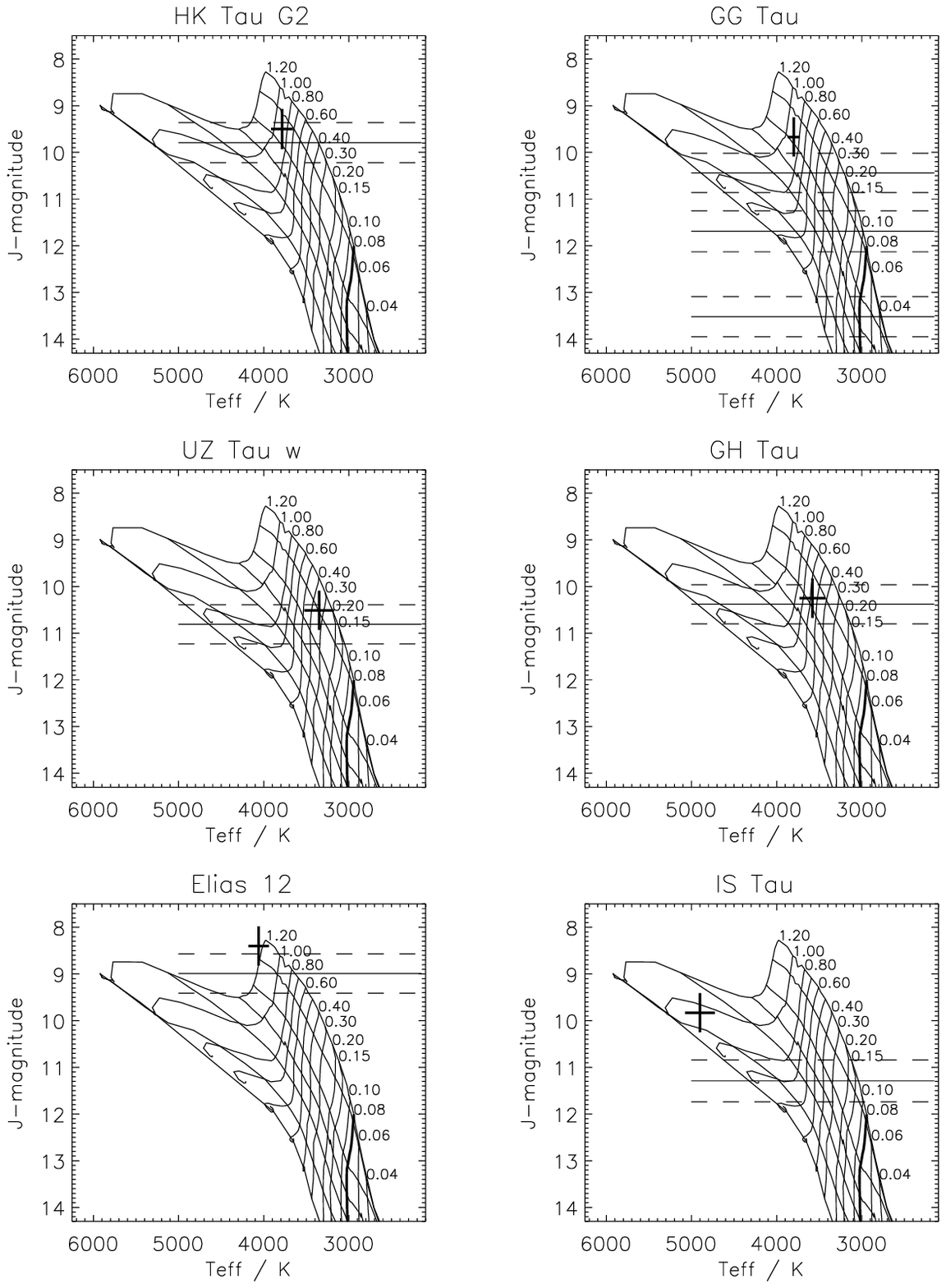}}
  \caption{continued}
\end{figure*}

\setcounter{figure}{0}

\begin{figure*}
 \resizebox{\hsize}{!}{\includegraphics{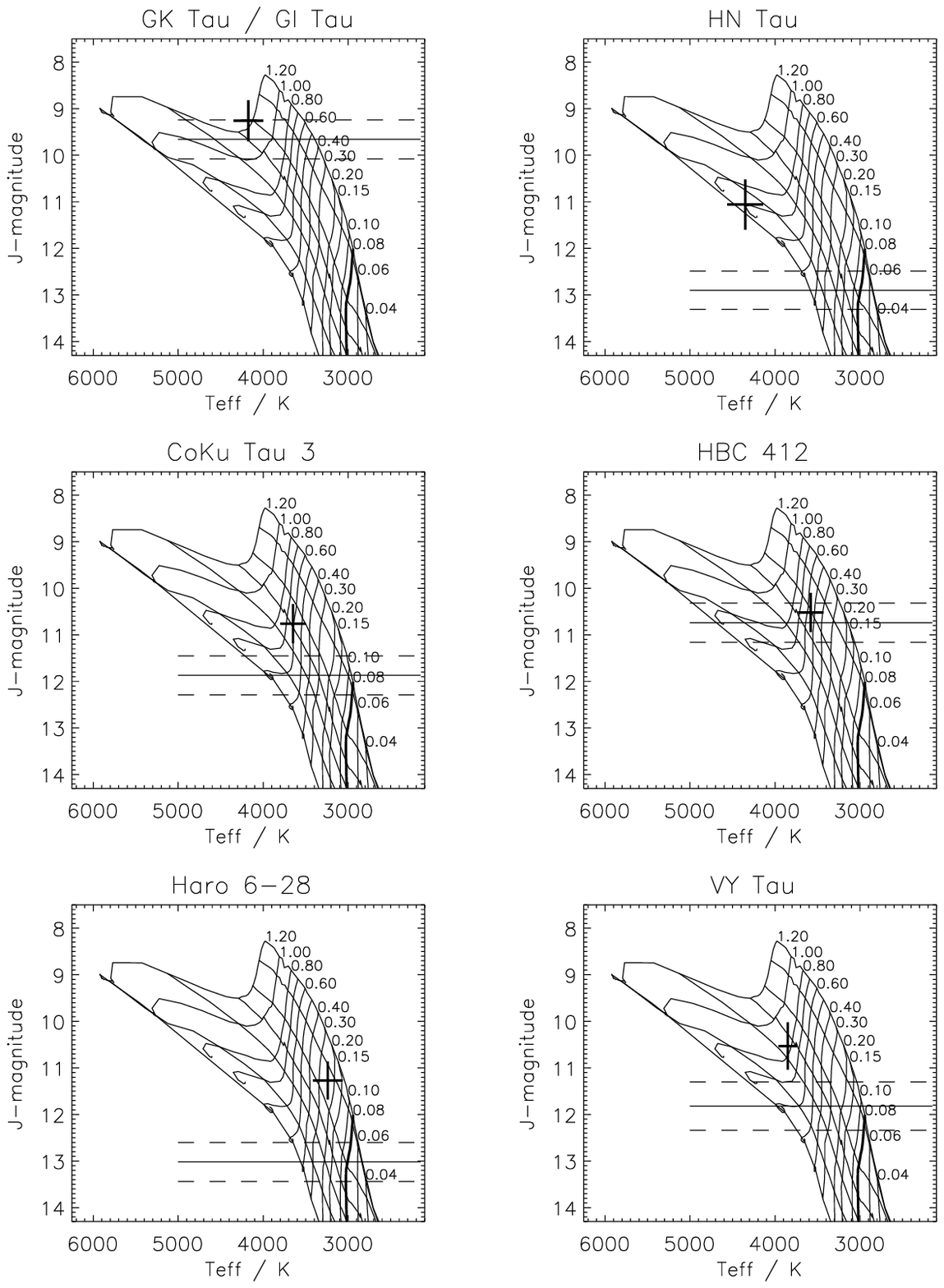}}
  \caption{continued}
\end{figure*}

\setcounter{figure}{0}

\begin{figure*}
 \resizebox{\hsize}{!}{\includegraphics{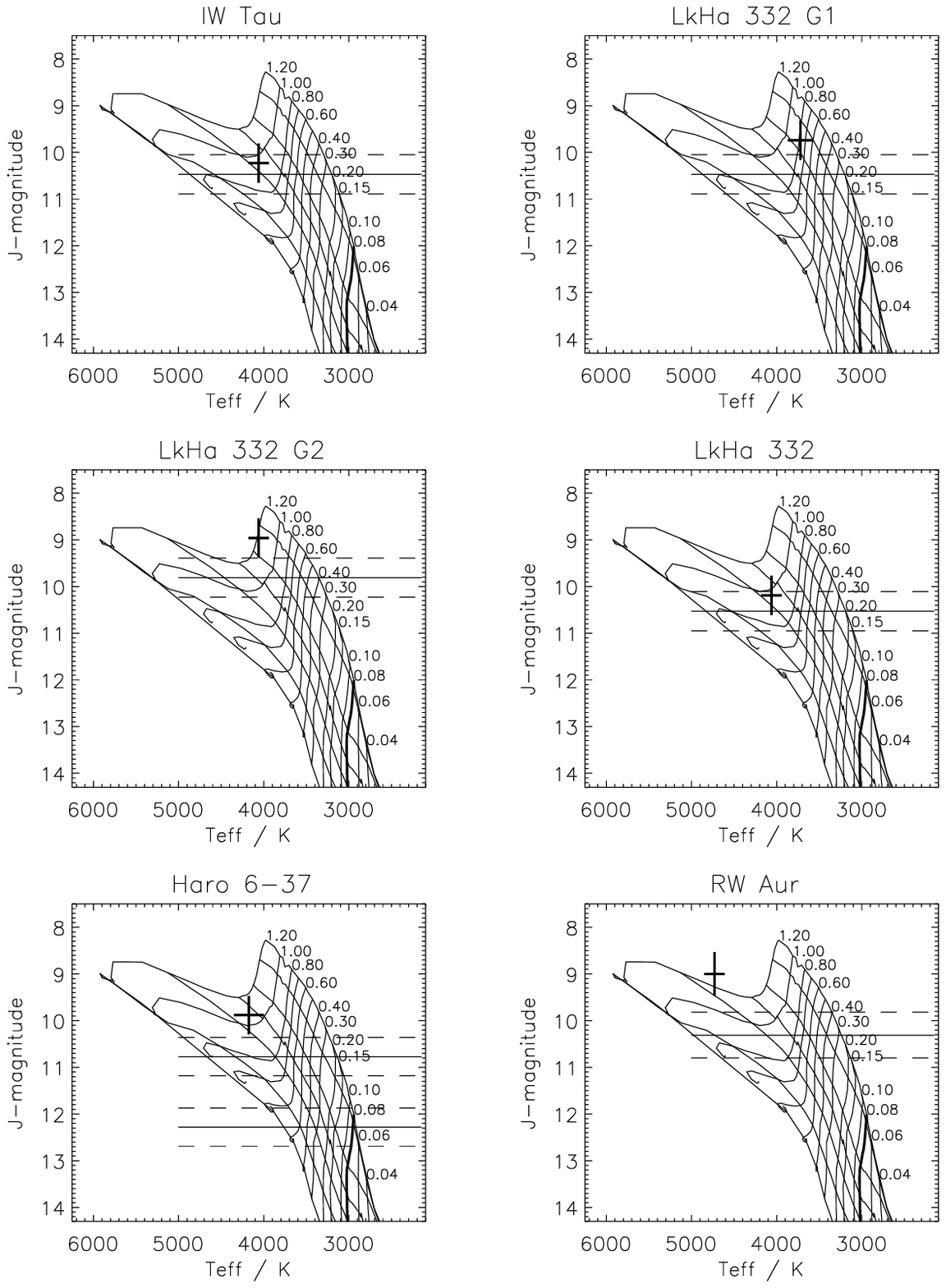}}
  \caption{continued}
\end{figure*}

\setcounter{figure}{0}

\begin{figure*}
  \resizebox{\hsize}{!}{\includegraphics{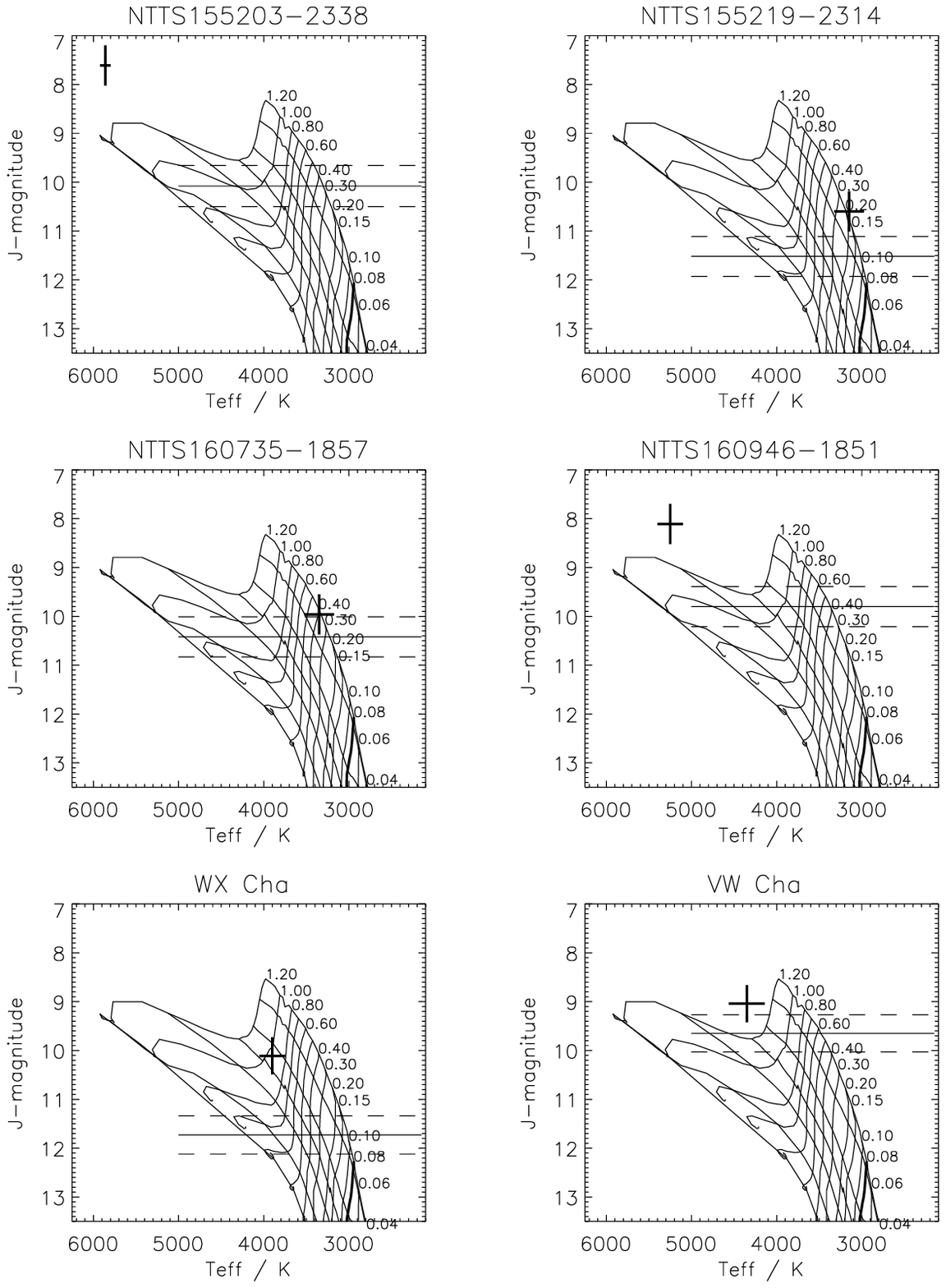}}
  \caption{continued}
\end{figure*}

\setcounter{figure}{0}

\begin{figure*}
  \resizebox{\hsize}{!}{\includegraphics{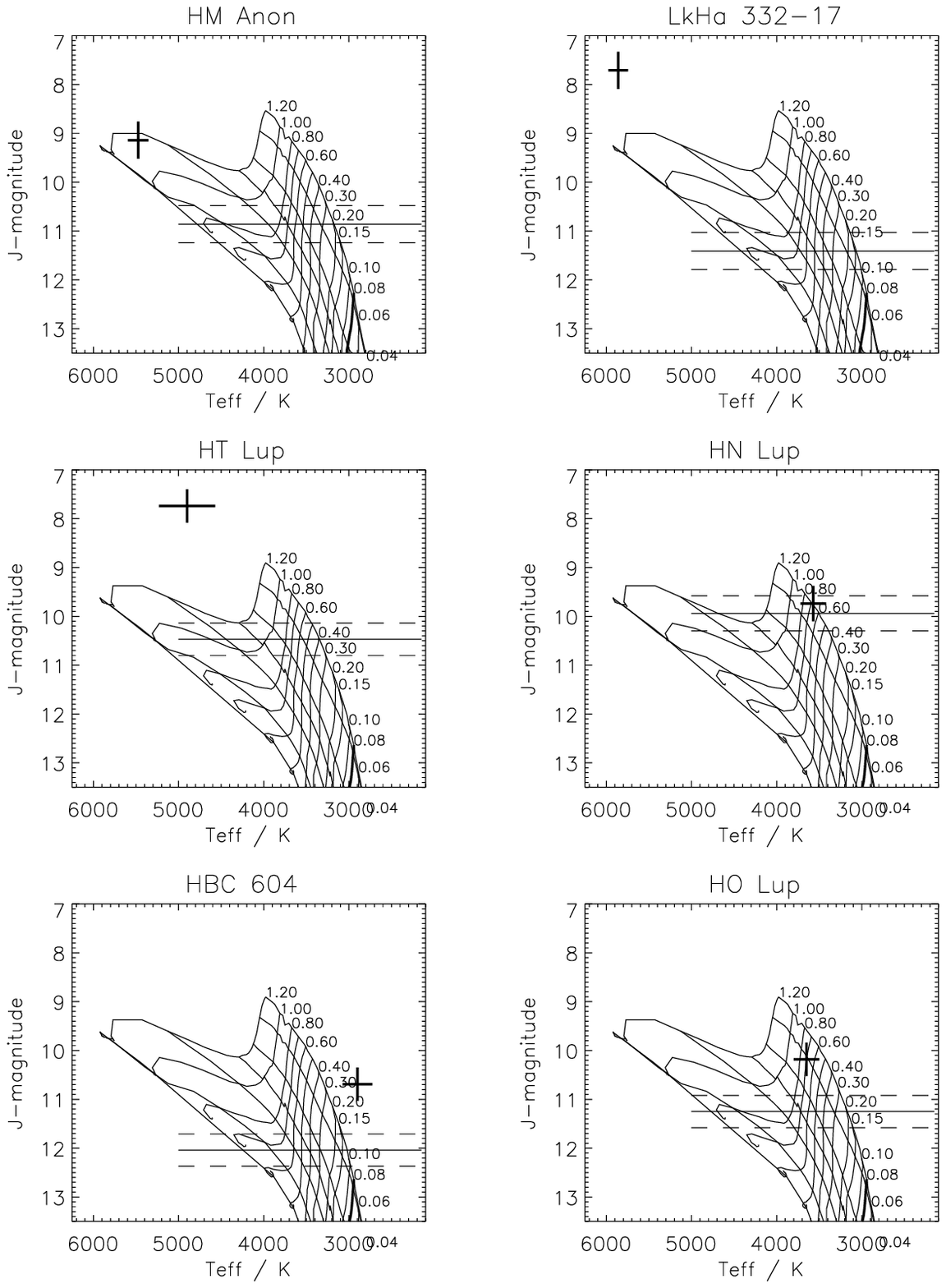}}
  \caption{continued}
\end{figure*}

\end{document}